\documentclass[floatfix,twocolumn,showpacs,preprintnumbers,amsmath,amssymb,pre]{revtex4-1}
\usepackage{color}
\usepackage[usenames,dvipsnames,svgnames,table]{xcolor}
\usepackage[colorlinks=true,linkcolor=blue,urlcolor=blue,citecolor=blue]{hyperref}

\usepackage{mathtools}
\usepackage{graphicx}
\usepackage{dcolumn}
\usepackage{array}
\usepackage{lipsum}
\usepackage{bm}
\usepackage{subfigure}
\usepackage{amssymb}
\usepackage{multirow}
\usepackage{tabularx}
\usepackage{amsmath}
\usepackage{braket}
\graphicspath{{plots/}}

\begin{document}
\title{ The Fermion Sign Problem in Path Integral Monte Carlo Simulations:\\ Quantum Dots, Ultracold Atoms, and Warm Dense Matter}

\author{ T.~Dornheim}
 \email{t.dornheim@hzdr.de}

\affiliation{
Center for Advanced Systems Understanding (CASUS), G\"orlitz, Germany
}

\begin{abstract}
The \textit{ab initio} thermodynamic simulation of correlated Fermi systems is of central importance for many applications, such as warm dense matter, electrons in quantum dots, and ultracold atoms. Unfortunately, path integral Monte Carlo (PIMC) simulations of fermions are severely restricted by the notorious fermion sign problem (FSP). In this work, we present a hands-on discussion of the FSP and investigate in detail its manifestation with respect to temperature, system size, interaction-strength and -type, and the dimensionality of the system. Moreover, we analyze the probability distribution of fermionic expectation values, which can be non-Gaussian and fat-tailed when the FSP is severe. As a practical application, we consider electrons and dipolar atoms in a harmonic confinement, and the uniform electron gas in the warm dense matter regime. In addition, we provide extensive PIMC data, which can be used as a reference for the development of new methods and as a benchmark for approximations.
\end{abstract}

\maketitle

\section{Introduction}

The numerical solution of the well-known, but highly complex equations that govern quantum mechanics using modern high-performance computers has emerged as one of the most active and successful fields in theoretical physics and chemistry. A particularly useful approach to accomplish this goal for a correlated quantum system in thermodynamic equilibrium (i.e., at finite temperature) was already outlined by Feynman~\cite{feynman}, who proposed to map the complicated quantum system of interest onto a classical ensemble of interacting ring-polymers~\cite{chandler} via the path-integral formalism~\cite{kleinert}. 
The basic idea of the path-integral Monte-Carlo (PIMC) method~\cite{berne3,imada,pollock,cep,review} is to stochastically evaluate the resulting high-dimensional integrals using the Metropolis algorithm~\cite{metropolis}, which, remarkably, does not suffer from the \emph{curse of dimensionality}~\cite{curse} that renders standard quadrature methods unfeasible in this case~\cite{filinov_chapter,liu_monte_carlo}.

Since its first application to He$^4$ in the late sixties~\cite{pimc_original,pimc_original2}, PIMC has emerged as one of the most successful tools in statistical physics and has allowed for profound insights into exciting physical phenomena such as superfluidity~\cite{ceperley_superfluid, sindzingre,kwon,dornheim_superfluid} and Bose-Einstein-condensation~\cite{BEC1,BEC2}. Moreover, PIMC provides exact simulations at strong coupling, which makes it possible to study crystallization in real quantum systems~\cite{jones_crystal, filinov_PRL, clark_casula}, and the direct access to imaginary-time correlation functions~\cite{berne1,berne2,dynamic_folgepaper} can be used as input for an \textit{analytic continuation}~\cite{jarrell,schoett}, which makes even possible the computation of dynamic properties such as collective excitations~\cite{dornheim_dynamic,vitali,supersolid}. In fact, novel Monte-Carlo sampling techniques allow for exact calculations of up to $N\sim10^4$ bosons and boltzmannons (i.e., distinguishable particles obeying Boltzmann statistics)~\cite{boninsegni1,boninsegni2}.

On the other hand, the situation is entirely different in the case of fermions. More specifically, the antisymmetry of the fermionic density matrix under the exchange of particles [cf.~Eq.~(\ref{eq:Z})] leads to a near cancellation of positive and negative terms both with decreasing temperature and increasing the system size~\cite{ceperley_fermions}. This issue is commonly known as the fermion sign problem (FSP)~\cite{loh,troyer,lyubartsev,vozn}, and generally prevents fermionic PIMC simulations once quantum degeneracy effects start to get important~\cite{fsp_note}. This is very unfortunate, as correlated Fermi systems offer a wealth of interesting effects such as the BCS-BEC crossover~\cite{fermi1,fermi2,fermi3} in ultracold atoms and the formation of Wigner molecules in quantum dots~\cite{wigner_molecule1,wigner_molecule2,wigner_molecule3}.

Of particular importance is the so-called warm dense matter (WDM) regime~\cite{review,ross,koenig,fortov_review}, an extreme state of matter with high temperatures ($T\sim10^4-10^8$K) and extreme densities ($n\sim10^{21}-10^{27}$cm$^{-3}$).
These conditions occur in astrophysical objects such as giant planet interiors~\cite{militzer1,militzer2,knudson} and brown dwarfs~\cite{saumon1,saumon2,becker}, and are expected to play an important role on the pathway towards inertial confinement fusion~\cite{hu1,hu2}. Moreover, WDM is now routinely realized in the lab (see Ref.~\cite{falk_wdm} for a topical review article) and constitutes one of the most active frontiers in plasma science~\cite{wdm_book}.
The theoretical description of WDM, however, is notoriously difficult due to the nontrivial interplay of (i) Coulomb coupling, (ii) quantum degeneracy effects, and (iii) thermal excitations. This regime is typically characterized by two parameters, which are both of the order of one: the density parameter $r_s=\overline{r}/a_\textnormal{B}$ (with $\overline{r}$ and $a_\textnormal{B}$ being the mean interparticle distance and first Bohr radius) and the degeneracy parameter~\cite{torben_eur} $\theta=k_\textnormal{B}T/E_\textnormal{F}$ (with $E_\textnormal{F}$ being the usual Fermi energy~\cite{quantum_theory}). Therefore, both perturbation theory and ground state methods are not applicable, which leaves \textit{ab initio} PIMC simulations as one of the most promising options~\cite{brown_chapter}.

Consequently, there has been a spark of new developments in the field of fermionic quantum Monte-Carlo simulations at finite temperature over the last years~\cite{cpimc_original,brown_ethan,blunt1,schoof_prl,malone1,blunt2,malone2,dornheim,dornheim2,vladimir_UEG,groth,dornheim3,dornheim_prl,dornheim_pop,groth_prl,dubois,dornheim_pre,groth_jcp,claes,dornheim_cpp,brenda,universe,dornheim_neu}. Despite this exciting progress, a thorough study of the fermion sign problem itself seems to be missing. In this work, we aim to fill this gap by presenting a detailed practical investigation of the FSP within standard PIMC simulations of i) electrons in quantum dots, ii) ultracold dipolar atoms in a harmonic confinement, and iii) the uniform electron gas at warm dense matter conditions~\cite{review,groth_prl}. More specifically, we study the manifestation of the FSP regarding different parameters (e.g., system size, coupling strength, etc.) and discuss the probability distribution of Monte-Carlo expectation values, which, in the presence of a sign problem, is not necessarily given by a simple Gaussian. In addition, we provide extensive benchmark data, which will aid the development of new methods and can be used to gauge the accuracy of novel approximations.

The paper is organized as follows: In Sec.~\ref{sec:theory}, we introduce the required theory, in particular the standard path integral Monte Carlo approach (\ref{sec:PIMC}), followed by the fermion sign problem (\ref{sec:FSP}) and the considered system types and Hamiltonians (\ref{sec:hamiltonian}). In Sec.~\ref{sec:results}, we present our simulation results, starting with a detailed discussion of the Monte-Carlo sampling and the probability distribution of fermionic expectation values (\ref{sec:histograms}). In addition, we study the manifestation of the FSP with respect to temperature (\ref{sec:temperature}), system-size (\ref{sec:N}), interaction-strength and -type (\ref{sec:coupling}), and the dimensionality (\ref{sec:dimensionality}), all for electrons and ultracold atoms in a harmonic confinement.
Lastly, we extend our considerations to the uniform electron gas in the warm dense matter regime, where the FSP exhibits a somewhat different manifestation regarding system size.
The paper is concluded by a concise summary and discussion in Sec.~\ref{sec:summary}.

\section{Theory\label{sec:theory}}

\subsection{Path Integral Monte Carlo\label{sec:PIMC}}

\begin{figure}
\includegraphics[width=0.41\textwidth]{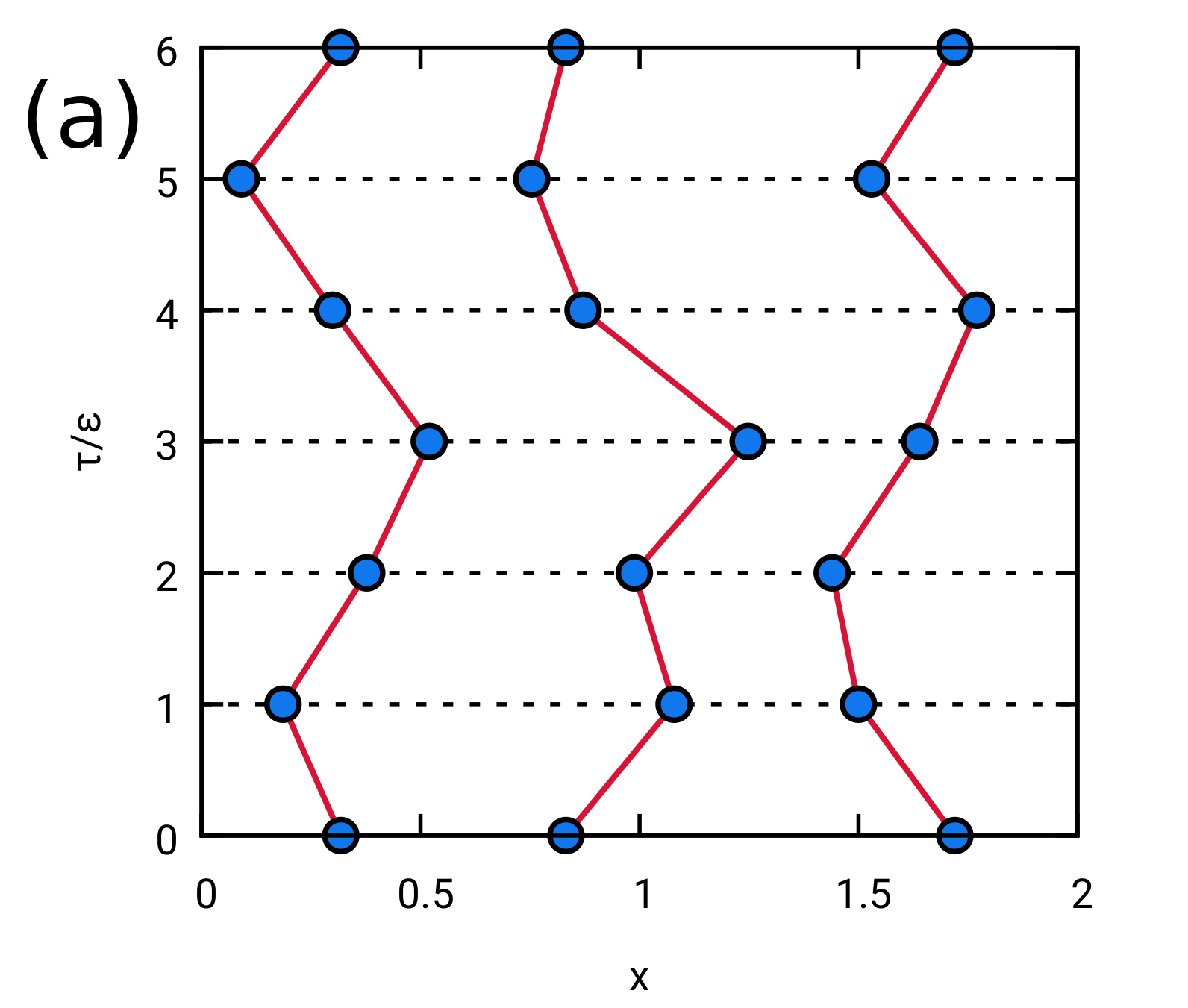}
\includegraphics[width=0.41\textwidth]{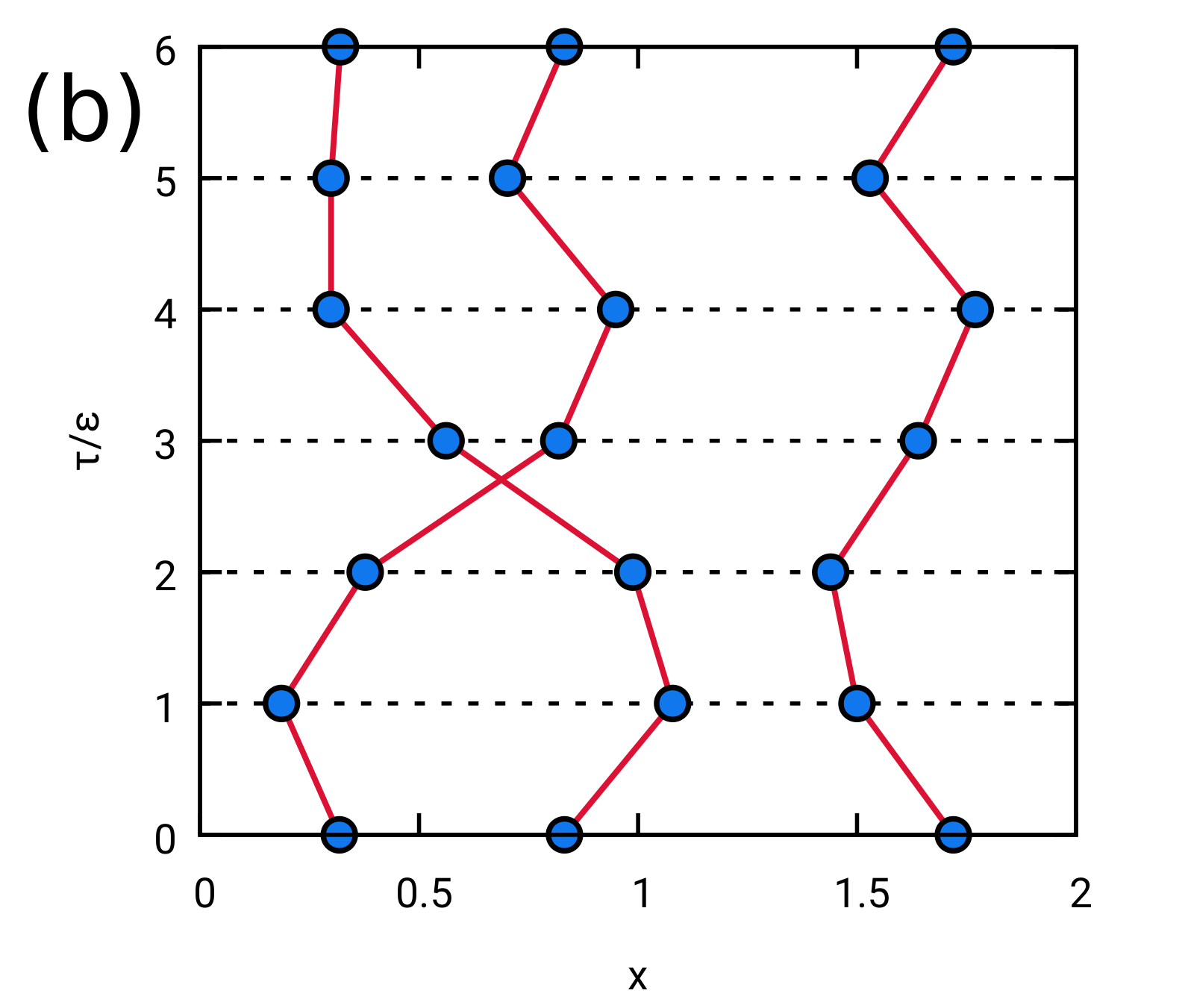}
\caption{\label{fig:paths}
Example configurations from a PIMC simulation of $N=3$ spin-polarized fermions: Each particle is represented by an entire path in the imaginary time $\tau\in[0,\beta]$ (with $\epsilon=\beta/P$ being the so-called time-step). In panel (a), there is no exchange of particle coordinates, and the weight function $W(\mathbf{X})$ is positive. In panel (b), the two particles on the left side form a combined exchange-cycle, and the weight is negative. Taken from Ref.~\cite{dornheim_permutation_cycles} with the permission of the authors.
}
\end{figure}

Throughout this work, we restrict ourselves to the discussion of $N$ spin-polarized fermions in the canonical ensemble, i.e., the inverse temperature $\beta=1/k_\textnormal{B}T$, volume $V$ (or trap frequency $\Omega$ in case of a harmonic confinement, see Eq.~(\ref{eq:Hamiltonian_trap}) below) and particle number $N$ are fixed.
The central quantity in statistical physics is the partition function, which can be written in coordinate space as
\begin{eqnarray}\label{eq:Z}
Z = \frac{1}{N!} \sum_{\sigma\in S_N} \textnormal{sgn}(\sigma) \int \textnormal{d}\mathbf{R}\ \bra{\mathbf{R}} e^{-\beta\hat H} \ket{\hat\pi_\sigma \mathbf{R}} \quad ,
\end{eqnarray}
where $\mathbf{R}=(\mathbf{r}_1,\dots,\mathbf{r}_N)^T$ contains the coordinates of all particles. Since we are interested in fermions, we have to explicitly evaluate the sum over all possible permutations of particle coordinates $\sigma$, with $S_N$ denoting the permutation group and $\hat\pi_\sigma$ being the corresponding permutation operator. Note that the sign $\textnormal{sgn}(\sigma)$ is positive (negative) for an even (odd) number of pair exchanges. For completeness, we mention that we restrict ourselves to the spin-polarized case (i.e., only one species of fermions, like spin-up electrons) throughout this work, but the generalization to multiple particle species is straightforward and does not affect the manifestation of the FSP.
To make the evaluation the matrix elements of the density operator $\hat\rho = e^{-\beta\hat H}$ in Eq.~(\ref{eq:Z}) possible, one typically performs a Trotter decomposition~\cite{trotter} and finds that $Z$ can be expressed as the sum over all closed paths $\mathbf{X}$ in the imaginary time $\tau$. However, since both the derivation and final formulas have already been presented elsewhere~\cite{cep,review}, they need not be repeated here. For the present purposes, it is fully sufficient to work with the abstract expression
\begin{eqnarray}
Z = \int \textnormal{d}\mathbf{X}\ W(\mathbf{X}) \quad ,
\end{eqnarray}
which can be interpreted as follows: The $PND$-dimensional (with $P$ denoting the number of so-called imaginary time-slices, cf.~Fig.~\ref{fig:paths}, and $D$ being the dimensionality of the system) variable $\mathbf{X}$ constitutes a so-called configuration, and each configuration contributes to $Z$ with the appropriate configuration weight $W(\mathbf{X})$, which is a function that can be readily evaluated.
This is illustrated in Fig.~\ref{fig:paths}, where we show example configurations from a PIMC simulation of $N=3$ fermions. First and foremost, we note that each particle is now represented by an entire path in the imaginary time $\tau$, with $P=6$ imaginary time-slices. In panel (a), there is no exchange of particle coordinates and, consequently, the configuration weight $W(\mathbf{X})$ is positive. In contrast, the second depicted configuration contains an exchange-cycle comprised of two fermions. Due to this pair-exchange, the corresponding $W$ is negative.

The basic idea of the path integral Monte Carlo method~\cite{berne3,cep} is to generate a Markov chain of path-configurations $\mathbf{X}$ that are distributed according to $P(\mathbf{X}) = W(\mathbf{X})/Z$. Although the normalization $Z$ is not known, this can be accomplished efficiently using the celebrated Metropolis algorithm~\cite{metropolis}. Indeed, simulations of up to $N\sim10^4$ bosons and boltzmannons (i.e., distinguishable particles obeying classical Boltzmann statistics~\cite{clark_casula,dornheim_analyzing}) are feasible using novel Monte-Carlo sampling techniques~\cite{boninsegni1,boninsegni2} without the introduction of any approximation. Unfortunately, since the weight function $W(\mathbf{X})$ is not strictly positive in the case of fermions, it cannot be interpreted as a probability distribution, which, as we shall see in the next section, is the origin of the infamous fermion sign problem.

For completeness, we mention that it is, at least in principle, possible to recast Eq.~(\ref{eq:Z}) into a sum over only positive terms by exploiting the nodal structure of the density matrix~\cite{ceperley_fermions,fermion_nodes}. However, since the exact nodes are a-priori unknown, this simplification comes at the cost of an uncontrolled approximation~\cite{node_note}.

\subsection{The fermion sign problem\label{sec:FSP}}

The first task at hand is to find a way to generate the paths $\mathbf{X}$ using the Metropolis algorithm, although their weight function is negative. In practice, we switch to the modified partition function 
\begin{eqnarray}\label{eq:Z_modified}
Z' = \int \textnormal{d}\mathbf{X}\ |W(\mathbf{X})| \quad ,
\end{eqnarray}
where the paths $\mathbf{X}$ are now generated according to the absolute value of the weight function. We note that in the case of standard PIMC, as it has been introduced above, Eq.~(\ref{eq:Z_modified}) coincides with the (symmetrized) bosonic partion function, which has some interesting implications that are discussed later on.
To calculate the fermionic expectation value of an observable $\hat A$, we then have to evaluate the ratio
\begin{eqnarray}\label{eq:fermionic_expectation_value}
\braket{\hat A} = \frac{\braket{\hat A \hat S}'}{\braket{\hat S}'} \quad ,
\end{eqnarray}
where the operator $\hat S$ measures the sign of the configuration weight, i.e., $S(\mathbf{X})=W(\mathbf{X})/|W(\mathbf{X})|$.
The problem with this approach is that both the enumerator and the denominator in Eq.~(\ref{eq:fermionic_expectation_value}) vanish simultaneously both towards low temperature (i.e., large $\beta$) and with increasing system size $N$.
This is captured by the average value of $\hat S$, which is given by the ratio of the fermionic and bosonic partition function
\begin{eqnarray}\label{eq:sign}
S \coloneqq \braket{\hat S}' &=& \frac{1}{Z'}\int \textnormal{d}\mathbf{X}\ |W(\mathbf{X})|S(\mathbf{X}) \\ \nonumber
 &=&\frac{Z}{Z'} = e^{-\beta N (f-f')} \quad ,
\end{eqnarray}
and which we will simply refer to as the \emph{average sign}
throughout this work.
In fact, Eq.~(\ref{eq:sign}) constitutes a direct measure for the amount of cancellations within a fermionic PIMC simulation, and exponentially decays both with $N$ and $\beta$ (with $f$ and $f'$ being the free energy density of the fermionic and modified system, respectively). This is bad news, because a small sign (typically $S\sim10^{-3}$) means that simulations are no longer feasible.
This can be understood by considering the relative Monte-Carlo error of Eq.~(\ref{eq:fermionic_expectation_value}), which is given by~\cite{ceperley_fermions}
\begin{eqnarray}\label{eq:FSP}
\frac{\Delta A}{A} \sim \frac{1}{S\sqrt{M}} \sim \frac{ e^{\beta N (f-f')} }{\sqrt{M}}\quad .
\end{eqnarray}
Evidently, the statistical error exponentially increases with $N$ and $\beta$, which can only be compensated by increasing the number of Monte-Carlo samples as $~1/\sqrt{M}$. In practice, one thus quickly runs into an \emph{exponential wall}, which is nothing else than the fermion sign problem.

\subsection{System types and Hamiltonians\label{sec:hamiltonian}}
\subsubsection{Harmonic confinement}
The most widely used model system that is considered in this work are fermions in a harmonic confinement, which is governed by the Hamiltonian 
\begin{eqnarray}\label{eq:Hamiltonian_trap}
\hat H = - \frac{1}{2} \sum_{k=1}^N \nabla_k^2 + \frac{1}{2} \sum_{k=1}^N \mathbf{\hat r}_k^2 + \sum_{k>l}^N \frac{ \lambda }{ |\mathbf{\hat r}_l - \mathbf{\hat r}_k|^\alpha } \quad ,
\end{eqnarray}
where we assume oscillator units, i.e., the characteristic length $l_0=\sqrt{\hbar/m\Omega}$ (with $\Omega$ being the trap frequency) and energy scale $E_0=\hbar\Omega$.
Of particular importance is the exponent $\alpha\in\{1,3\}$, which distinguishes between Coulomb interaction ($\alpha=1$, corresponding to electrons in a quantum dot~\cite{dornheim,dornheim_analyzing,reimann}) and dipole interaction ($\alpha=3$, corresponding to ultracold atoms~\cite{stuhler,dynamic_alex1,jan_willem}).
In addition, the coupling constant $\lambda$ is defined as the ratio of the interaction energy to $E_0$,
\begin{eqnarray}
\lambda = \begin{cases} \frac{e^2}{4\pi\epsilon_0l_0E_0}\ , &\mbox{if } \alpha = 1 \\
\frac{D}{4\pi l_0^3 E_0} & \mbox{if } \alpha=3 \end{cases} \quad ,
\end{eqnarray}
with $D$ being the usual dipole-dipole interaction constant, see, e.g., Ref.~\cite{dynamic_alex1}.
Finally, we consider two- and three-dimensional systems in this work, and the dimensionality of the harmonic confinement is always equal to the overall number of dimensions.

\subsubsection{Uniform electron gas}
The second type of model system that we consider in this work is the uniform electron gas (see Refs.~\cite{review,loos} for topical review articles), which is defined as an ensemble of $N$ electrons in a periodic box of length $L$ and volume $V=L^3$. The corresponding Hamiltonian is given by
\begin{eqnarray}\label{eq:Hamiltonian_UEG}
\hat H = - \frac{1}{2} \sum_{k=1}^N \nabla^2_k + \sum_{k>l}^N w(\mathbf{\hat r}_l, \mathbf{\hat r}_k ) \quad .
\end{eqnarray}
Note that we always assume Hartree atomic units (i.e., energies in Hartree and distances in units of the first Bohr radius $a_0$) when discussing the UEG.
Let us briefly turn our attention to the pair interaction potential $w(\mathbf{\hat r}_l, \mathbf{\hat r}_k )$ in Eq.~(\ref{eq:Hamiltonian_UEG}). In order to mitigate finite-size effects, one typically employs the Ewald summation, which takes into account both the interaction between the two electrons $l$ and $k$ (and the respective positive background) and the infinite array of periodic images~\cite{fraser}. In this work, we use a pre-averaged (i.e., with respect to the orientation of the array of images) Ewald potential introduced by Yakub and Ronchi~\cite{yakub1,yakub2}, where the infinite sums both in real and reciprocal space are evaluated analytically beforehand. This leads to a significant saving of time, while the differences to the \emph{real} Ewald summation are expected to be small under the conditions considered in this work.

For completeness, we mention that a complete thermodynamic description of the UEG at warm dense matter conditions was achieved only recently~\cite{groth_prl} on the basis of novel configuration PIMC and permutation-blocking PIMC simulation data, see Ref.~\cite{review} for a comprehensive discussion.

\section{Results\label{sec:results}}
All results in this work have been obtained using a canonical adaption~\cite{mezza} of the worm algorithm~\cite{boninsegni1,boninsegni2}. Further, we use $P=200$ imaginary-time propagators based on the \textit{primitive action}, see Appendix~\ref{sec:convergence} for details, and Refs.~\cite{brualla,sakkos} for an accessible discussion. All fermionic results listed in Tabs.~\ref{tab:N_dependence_lambda0p5_Coulomb}, \ref{tab:lambda_dependence_N6_beta1}, and \ref{tab:beta_dependence_N6_lambda0.5} are conveged with respect to $P$ within the given statistical uncertainty.

\subsection{Monte-Carlo sampling and probability distribution of expectation values\label{sec:histograms}}

\begin{figure}
\includegraphics[width=0.41547\textwidth]{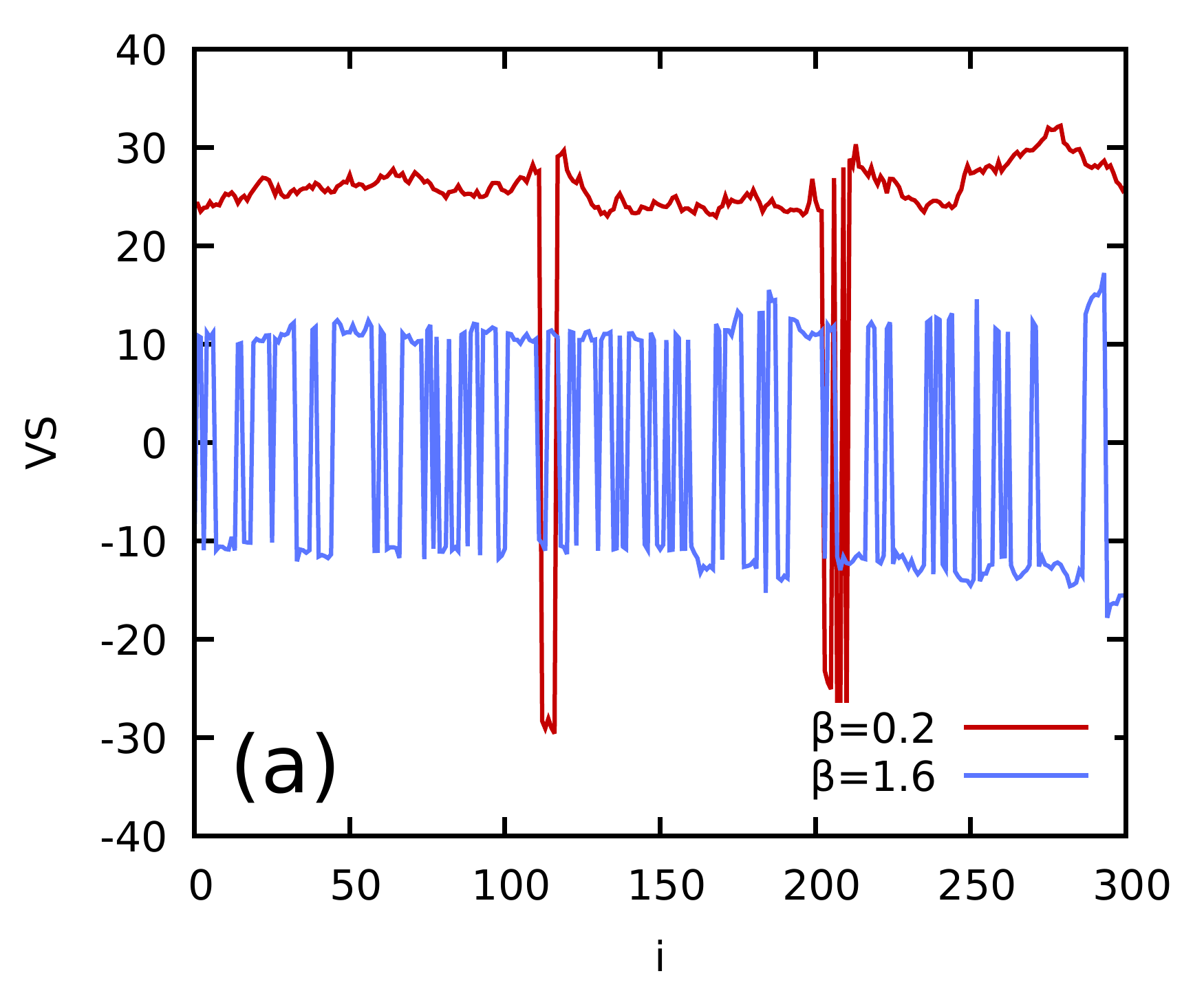}
\hspace*{-0.33cm}\includegraphics[width=0.434\textwidth]{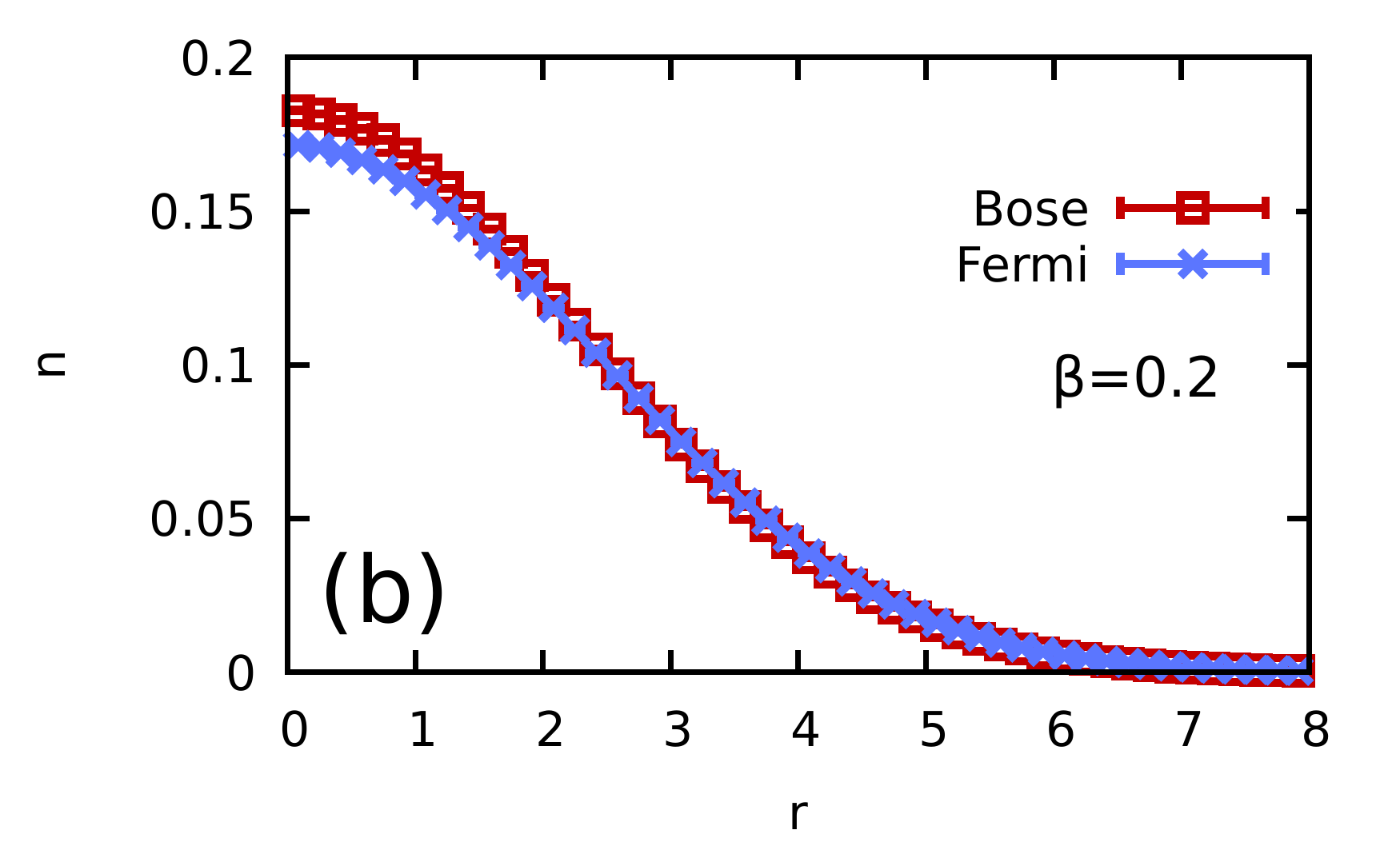}
\includegraphics[width=0.41547\textwidth]{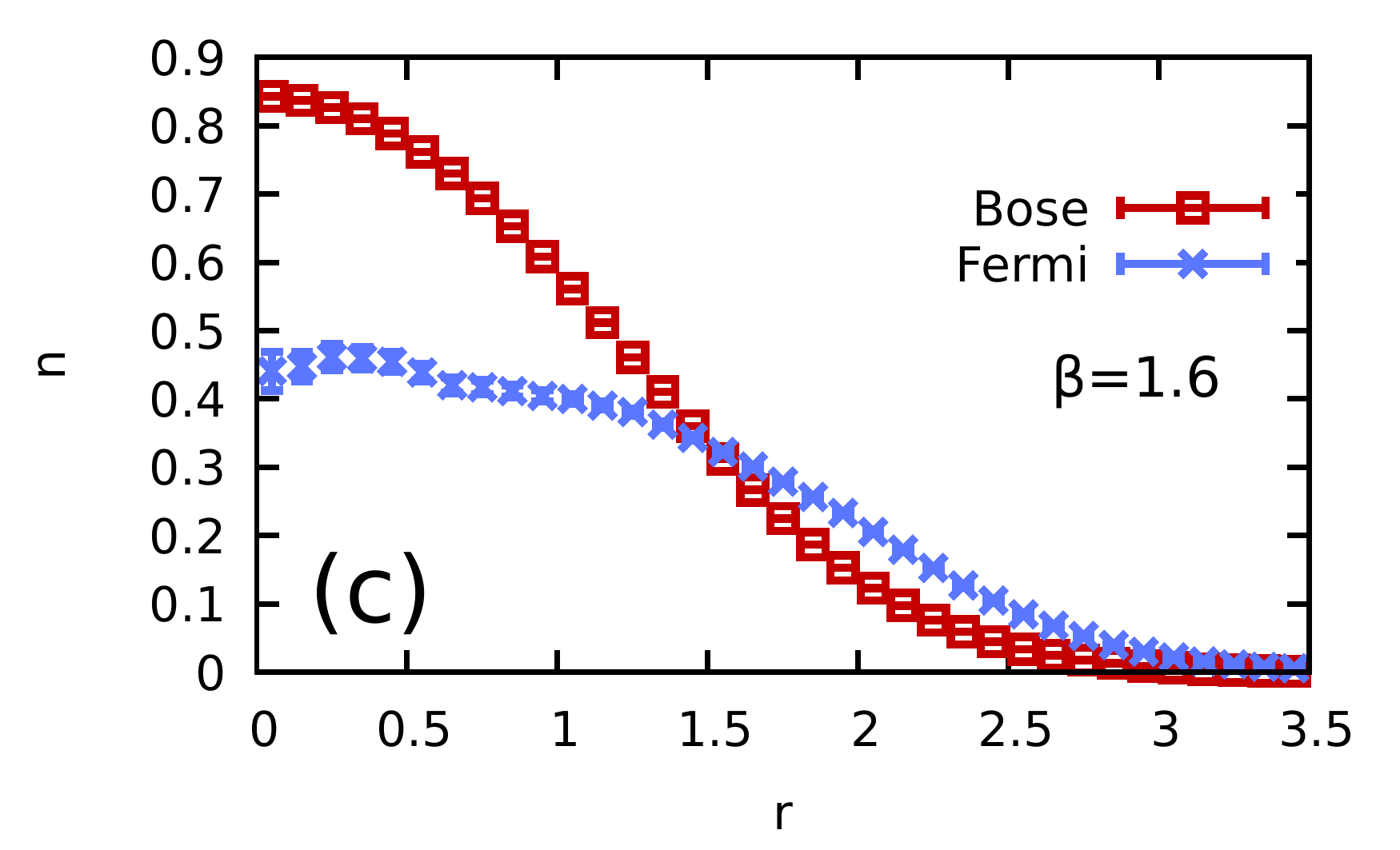}
\caption{\label{fig:Series_Coulomb_2D_N6_lambda0p5}
Manifestation of the fermion sign problem in the Monte Carlo sampling of an observable: panel (a) shows a series of $M=600$ consecutive measurements of the signed potential energy $VS$ [see Eq.~(\ref{eq:fermionic_expectation_value})] within a PIMC simulation of $N=6$ spin-polarized electrons in a $2D$ harmonic confinement with $\lambda=0.5$ and $\beta=0.2$ (red) and $\beta=1.6$ (blue). Panels (b) ($\beta=0.2$) and (c) ($\beta=1.6$) depict the corresponding radial densities $n(r)$ computed in the modified configuration space (i.e., for Bose statistics, red squares) and for electrons (i.e., Fermi statistics, blue crosses).
}
\end{figure}  

Let us start our discussion of the fermion sign problem with an illustration of the sampling of the expectation value of an observable. In Fig.~\ref{fig:Series_Coulomb_2D_N6_lambda0p5}, we show PIMC results for a simulation of $N=6$ electrons in a $2D$ harmonic trap [cf.~Eq.~(\ref{eq:Hamiltonian_trap})] at moderate coupling $\lambda=0.5$ and two inverse temperatures, $\beta=0.2$ (red) and $\beta=1.6$ (blue). Panel (a) shows a series of $M=300$ measurements for the signed potential energy $VS$, i.e., the enumerator from Eq.~(\ref{eq:fermionic_expectation_value}). The solid red line corresponds to $\beta=0.2$, which is a comparatively high temperature, where fermionic exchange-effects are not that important. Consequently, the sign stays mostly positive (with $S\approx0.82$), and sign-changes due to permutation cycles appear as brief negative spikes in the series of measurements. In stark contrast, the blue line corresponds to $\beta=1.6$, and the situation looks completely different: at this low temperature, positive and negative signs appear with a similar frequency and the average sign has decreased to $S\approx0.002$.

\begin{figure*}
\hspace*{-0.5cm}\includegraphics[width=0.41547\textwidth]{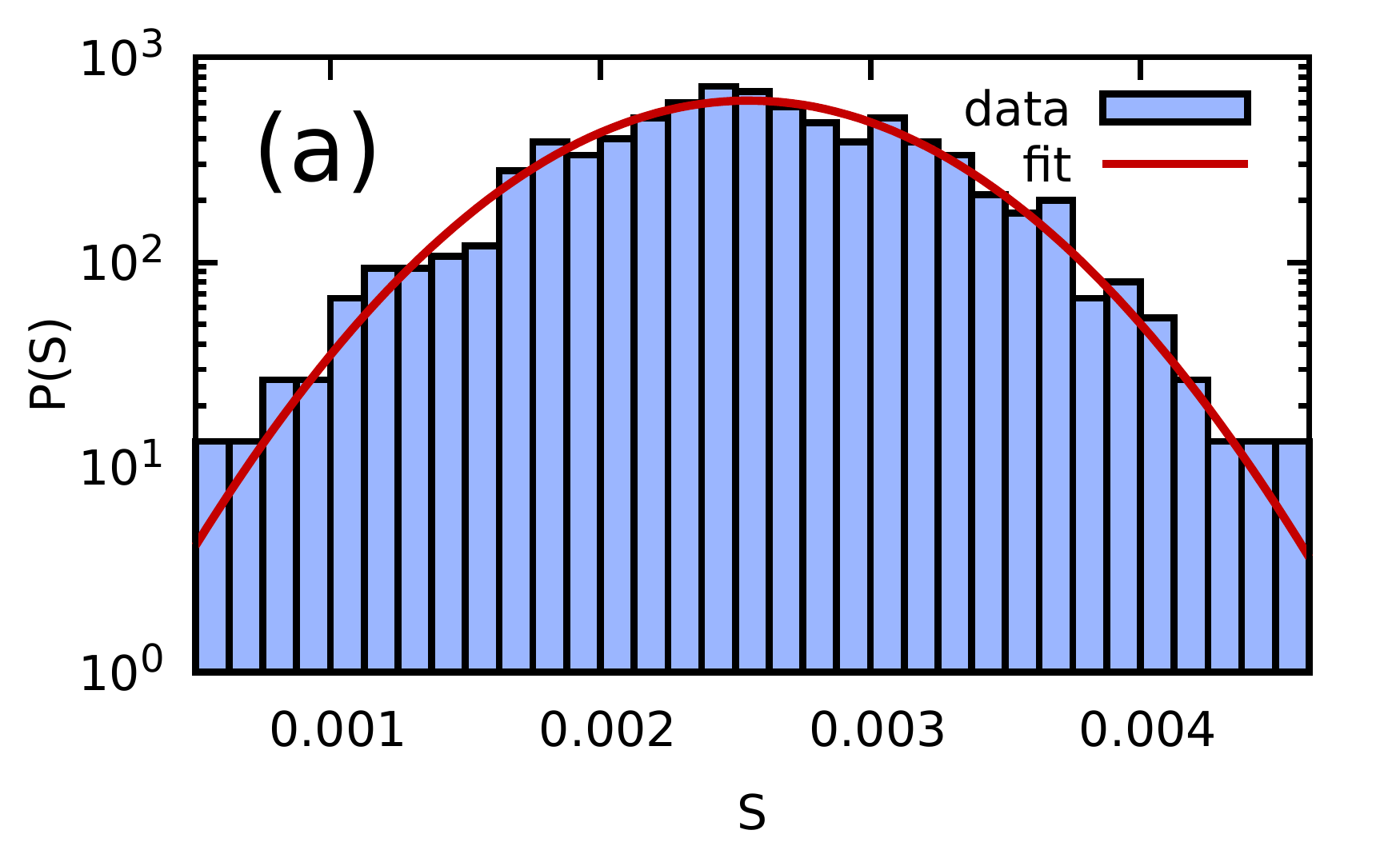}
\includegraphics[width=0.41547\textwidth]{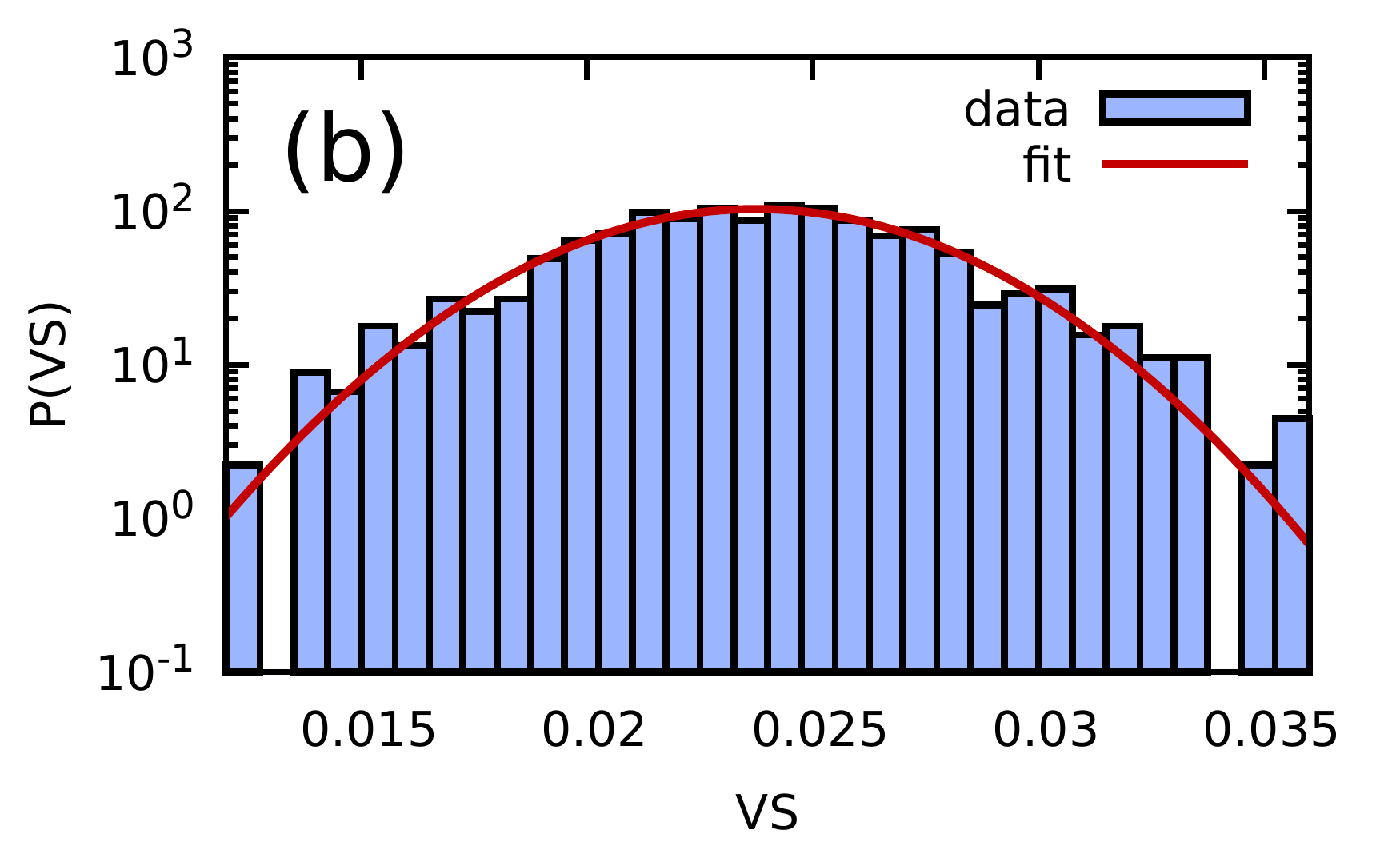}\\ \hspace*{0.79cm}
\includegraphics[width=0.41547\textwidth]{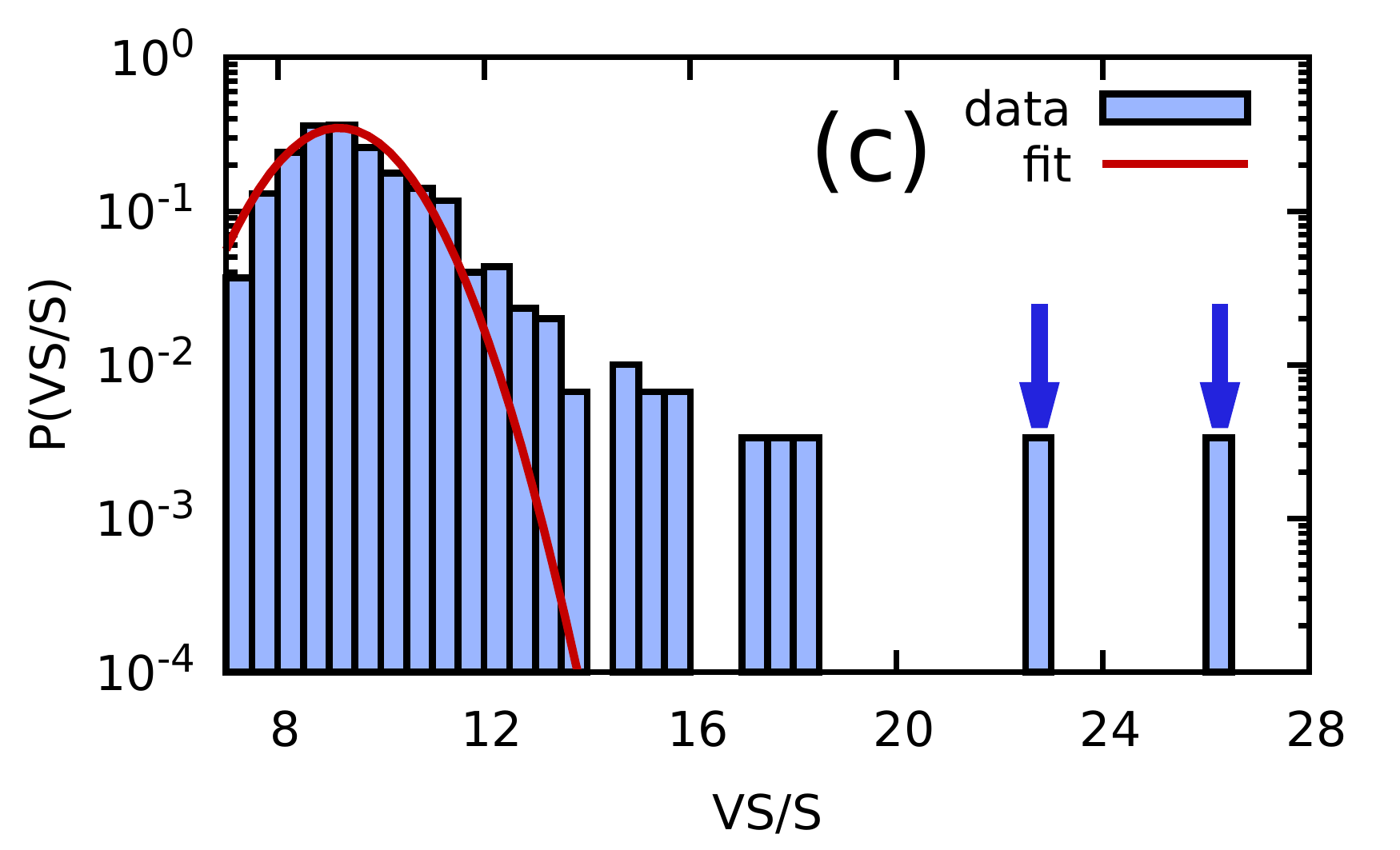} \hspace*{0.19cm}
\includegraphics[width=0.489\textwidth]{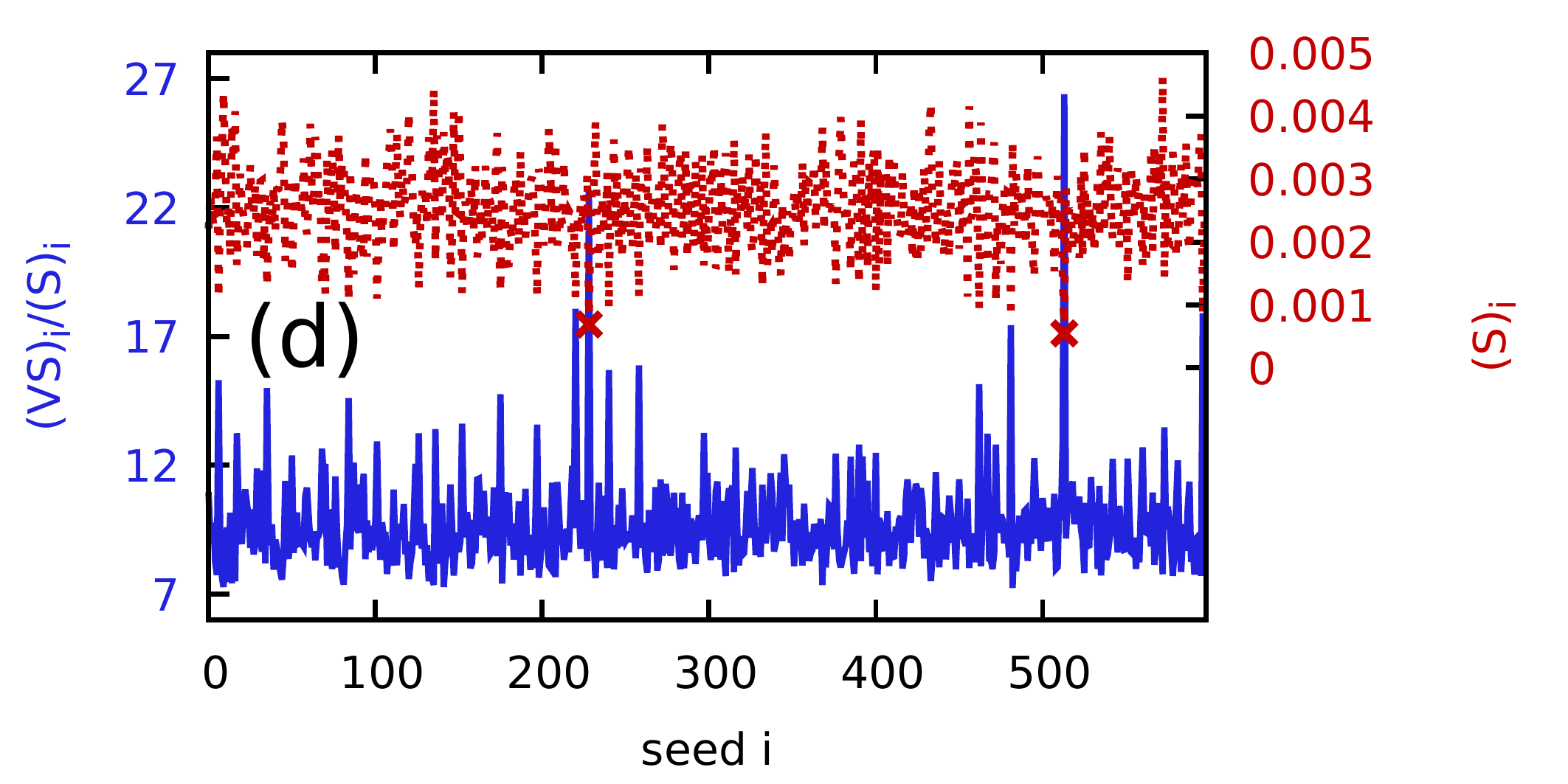}
\caption{\label{fig:histogram_2D_Coulomb_N6_beta1_lambda0}
PIMC simulation results with $N_\textnormal{s}=600$ independent seeds (and $M=5\cdot10^6$ measurements per seed) for $N=6$ spin-polarized noninteracting fermions in a $2D$ harmonic trap for $\beta=1$. Panels (a) and (b) depict the histograms of the average values per seed for the sign $\braket{S}'_i$ and the interaction energy times the sign $\braket{VS}'_i$. Panel (c) shows the corresponding histogram of fermionic expectation values $\braket{VS}'_i/\braket{S}'_i$ [cf.~Eq.~(\ref{eq:fermionic_expectation_value})], again evaluated for each seed, and the blue arrows indicate two extreme values at $V\approx22.7$ and $V\approx26.4$. The solid red lines depict Gaussian fits according to Eq.~(\ref{eq:Gaussian}). Finally, panel (d) shows the individual results for $\braket{VS}'_i/\braket{S}'_i$ (blue, left ordinate) and $\braket{S}'_i$ (red, right ordinate) for all $N_\textnormal{s}$ seeds. The two red crosses at $i=228$ and $i=513$ indicate the two smallest results for $\braket{S}'_i$, which, in turn, are responsible for the extreme values in  $\braket{VS}'_i/\braket{S}'_i$.
}
\end{figure*}

To further illustrate the origin of these cancellations, it is instructive to consider the modified (bosonic) probability distribution, which is used to generate the paths $\mathbf{X}$. To this end, we show in Fig.~\ref{fig:Series_Coulomb_2D_N6_lambda0p5}~(b) ($\beta=0.2$) and Fig.~\ref{fig:Series_Coulomb_2D_N6_lambda0p5}~(c) ($\beta=1.6$) the radial density distribution $n(r)$ both for Bose (red squares) and Fermi (blue crosses) statistics. At high temperature, the two data sets are very similar and the most significant deviations occur around the center of the trap, where the density is at the maximum. At $\beta=1.6$, on the other hand, the two densities exhibit severe discrepancies over the entire $r$-range. While bosons tend to cluster around the center of the trap, the fermions are pushed outward by the Pauli blocking. Since the paths in our simulations are distributed according to the bosonic density, the difference in the results for fermions at low temperature can only be accomplished by the cancellation and subsequent division by the small value for $S$ [cf.~Eq.~(\ref{eq:fermionic_expectation_value})].

Let us next consider the probability distribution of the expectation values within a fermionic PIMC simulation. According to the central limiting theorem~\cite{monte_carlo_book}, the average value of a Metropolis Monte-Carlo calculation of an expectation value $\braket{\hat A}$ with $M$ measurements (and $M$ being large) is normally distributed around the exact value, and the standard deviation decreases as $\sigma_M\sim1/\sqrt{M}$. This is verified in Fig.~\ref{fig:histogram_2D_Coulomb_N6_beta1_lambda0}, where we show histograms for the Monte-Carlo average of $S$ (a) and $VS$ (b) for $N_\textnormal{s}=600$ independent seeds with $M=5\cdot10^6$ measurements per seed, for a system of $N=6$ noninteracting fermions in a $2D$ harmonic trap at $\beta=1$. The blue bars correspond to our PIMC data, and the solid red curves to Gaussian fits according to
\begin{eqnarray}\label{eq:Gaussian}
P(A) = \frac{ e^{-\frac{(A-\mu)^2}{2\sigma^2}} }{ \sqrt{2\pi\sigma^2}} \quad ,
\end{eqnarray}
with $\sigma$ and $\mu$ being the free parameters. Evidently, we do indeed find the expected normal distribution for both cases, which means that the statistical uncertainty for a single seed can be straightforwardly estimated from the Monte-Carlo data via
\begin{eqnarray}\label{eq:naive_error}
\Delta \braket{V}' = \left(
\frac{1}{M} \sum_{i=1}^M (V_i - \braket{V}')^2
\right)^{1/2} \quad .
\end{eqnarray}
For completeness, we note that the error bars given in all tables and figures have been obtained by evaluating Eq.~(\ref{eq:naive_error}) for $N_\textnormal{s}$ statistically independent seeds, instead of $M$ measurements from a single seed, see Ref.~\cite{error_note} for details.

Let us next consider the distribution of the fermionic observable $\braket{V}=\braket{VS}'/\braket{S}'$, which is shown in Fig.~\ref{fig:histogram_2D_Coulomb_N6_beta1_lambda0}~(c). Remarkably, the histogram does not exhibit a Gaussian form, and the corresponding normal fit is not in agreement with the PIMC results. More specifically, the simulation results show a distinct tail towards large values of $VS/S$, with the two largest outliers (see the two blue arrows in the plot) being located around $VS/S=24$, i.e., around $16$ standard deviations (assuming the $\sigma$ value from the Gaussian fit) away from the mean.

\begin{figure*}
\hspace*{-0.5cm}\includegraphics[width=0.41547\textwidth]{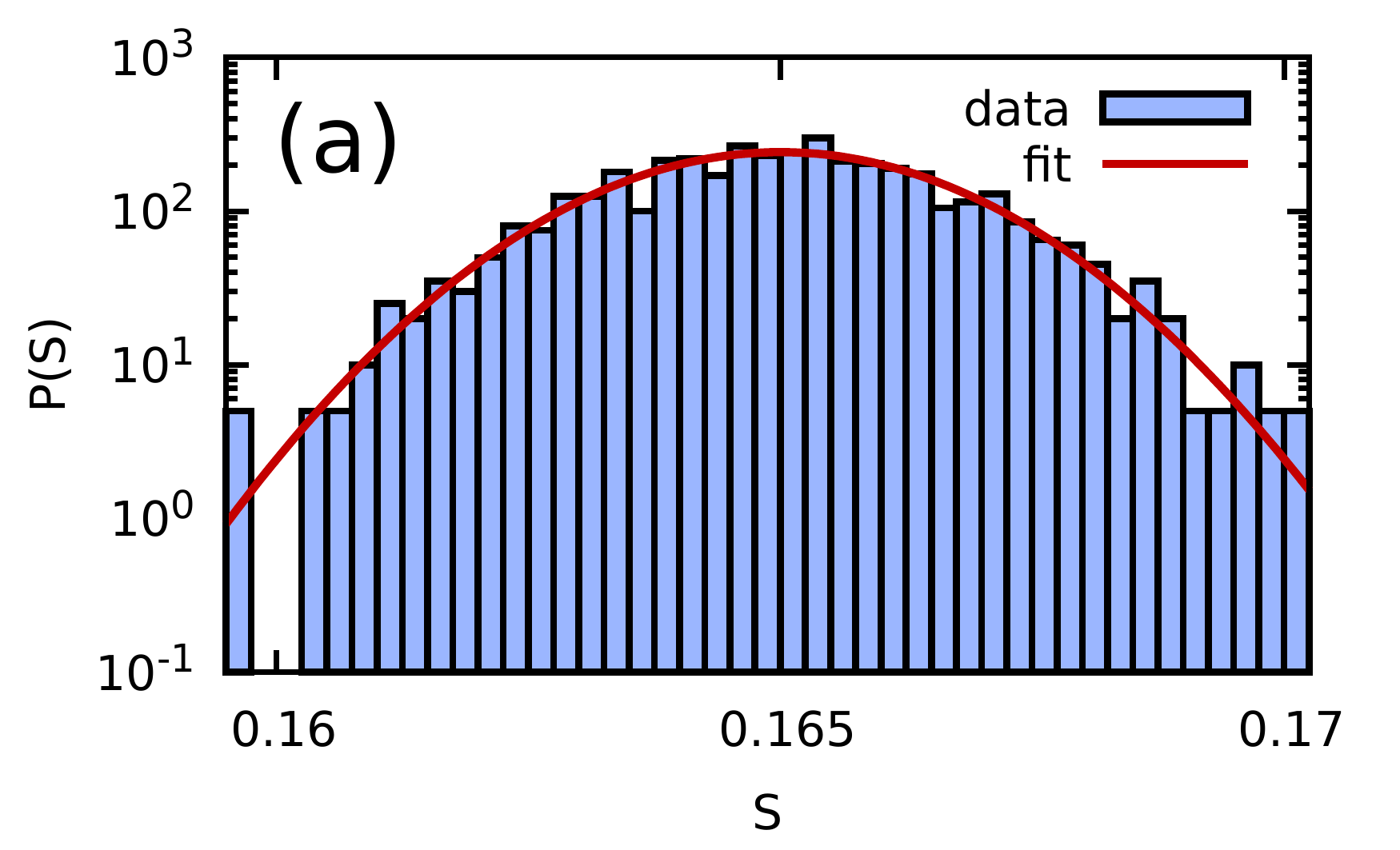}
\includegraphics[width=0.41547\textwidth]{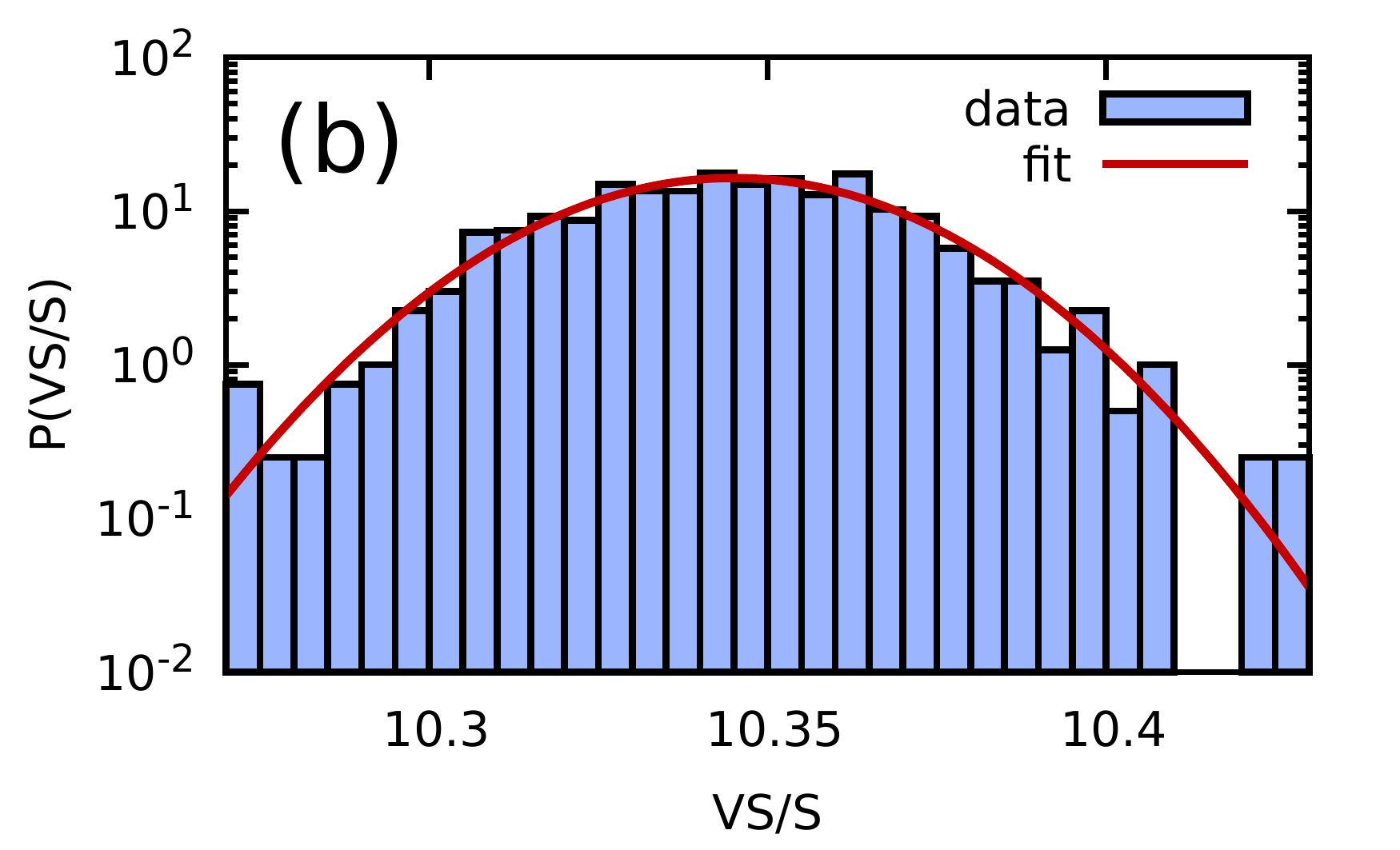}
\caption{\label{fig:histogram_2D_Dipole_N6_beta1_lambda0.1}
PIMC simulation results with $N_\textnormal{s}=800$ independent seeds (and $M=5\cdot10^6$ measurements per seed) for $N=6$ spin-polarized fermions with dipole interaction (i.e., ultracold atoms) and $\lambda=0.1$ in a $2D$ harmonic trap for $\beta=1$. Panel (a) depicts the histograms of the average values for the sign $\braket{S}'$, and panel (b) shows the corresponding histogram of fermionic expectation values $\braket{VS}'_i/\braket{S}'_i$ [cf.~Eq.~(\ref{eq:fermionic_expectation_value})], again evaluated for each seed.
}
\end{figure*}

The reason for this peculiar finding is the nonlinear nature of the fermionic expectation value from Eq.~(\ref{eq:fermionic_expectation_value}). In fact, it can be shown~\cite{hatano} that the probability distribution of the ratio of $\braket{VS}'$ and $\braket{S}'$ is the superposition of a Lorentzian (also known as Cauchy distribution) and a Gaussian, with the former one being responsible for the tail. Moreover, the sample deviation as defined in Eq.~(\ref{eq:naive_error})
actually diverges, and, therefore, does not constitute a good measure for the real uncertainty in the fermionic expectation value $\braket{V}$. For completeness, we mention that a similar behavior has been found in other fields, most notably financial modelling~\cite{mandelbrot}.

To further illustrate the occurrence of these tail events in our fermionic PIMC simulation, we show the series of the average values for all $N_\textnormal{s}=600$ seeds of both the ratio $VS/S$ (blue, left $y$-axis) and the sign $S$ (red, right $y$-axis) in Fig.~\ref{fig:histogram_2D_Coulomb_N6_beta1_lambda0}~(d). Let us first consider the blue curve: evidently, the expectation values of most seeds are located somewhere around the mean value, with a few spikes corresponding to the upward outliers. In contrast, the red curve does not exhibit any spikes, and we have already seen that $S_i$ follows a normal distribution, see Fig.~\ref{fig:histogram_2D_Coulomb_N6_beta1_lambda0}~(a).
A comparison of both curves reveals that the spikes in the ratio appear in those seeds with the smallest values of $S$, and the two smallest values, which are responsible for the blue arrows in Fig.~\ref{fig:histogram_2D_Coulomb_N6_beta1_lambda0}~(c), are highlighted by red crosses. Indeed, these $S_i$ are more than an order of magnitude smaller than the corresponding mean value of the distribution.

Up to this point, one might conclude that fermionic PIMC simulations appear to be doomed as 1) we do not have a good measure for the statistical uncertainty, which would make the Monte-Carlo expectation value an uncontrolled approximation, and 2) the distribution of the ratio $P(VS/S)$ is \emph{fat-tailed} and outliers exceeding $16$-times the standard deviation (often called \emph{black swan} events~\cite{taleb}) do appear with finite probability.
However, as we will see next, all is not lost.

In Fig.~\ref{fig:histogram_2D_Dipole_N6_beta1_lambda0.1}, we show PIMC results for the same conditions as in Fig.~\ref{fig:histogram_2D_Coulomb_N6_beta1_lambda0}, but with dipole-interaction and a coupling constant $\lambda=0.1$ (i.e., ultracold atoms). Due to the dipolar repulsion, fermionic exchange is suppressed, and we find an average sign of $S\approx0.165$, as compared to $S\approx0.0025$ for the noninteracting case. Panel (a) shows results for $P(S)$, and we again find the expected normal distribution. In contrast to the noninteracting case, this time $P(S)$ has a significantly smaller relative deviation $\sigma/S$, and no expectation values $S_i$ with a value that is an order of magnitude smaller than the average appear. Consequently, there are no spikes in the seed-averages of the ratio, and the corresponding distribution $P(VS/S)$ [Fig.~\ref{fig:histogram_2D_Dipole_N6_beta1_lambda0.1}~(b)] cannot be distinguished from a Gaussian.

In summary, the nonlinear nature of the fermionic expectation value Eq.~(\ref{eq:fermionic_expectation_value}) causes the distribution $P(VS/S)$ to be non-Gaussian, with a \emph{fat tail} towards larger values. To put it in another way, if the relative uncertainty of the denominator (i.e., $S$) is large, the smallest signs $S_i$ lead to spikes in the fermionic observable. In contrast, if the relative error of the sign is small (as in Fig.~\ref{fig:histogram_2D_Dipole_N6_beta1_lambda0.1}), these spikes do not appear (or are sufficiently unlikely), and the resulting distribution $P(VS/S)$ cannot be distinguished from a normal distribution in practice. Since $S$ itself does obey a normal distribution in any case, this condition can always be checked, and fermionic PIMC results can safely be labelled as \emph{quasi-exact} after all.
Thus, the statistical uncertainty for the fermionic expectation values given in all figures and data tables has been computed assuming the Gaussian form from Eq.~(\ref{eq:naive_error}), which is reliable for all presented cases.

\subsection{Temperature dependence\label{sec:temperature}}

\begin{figure}
\includegraphics[width=0.41547\textwidth]{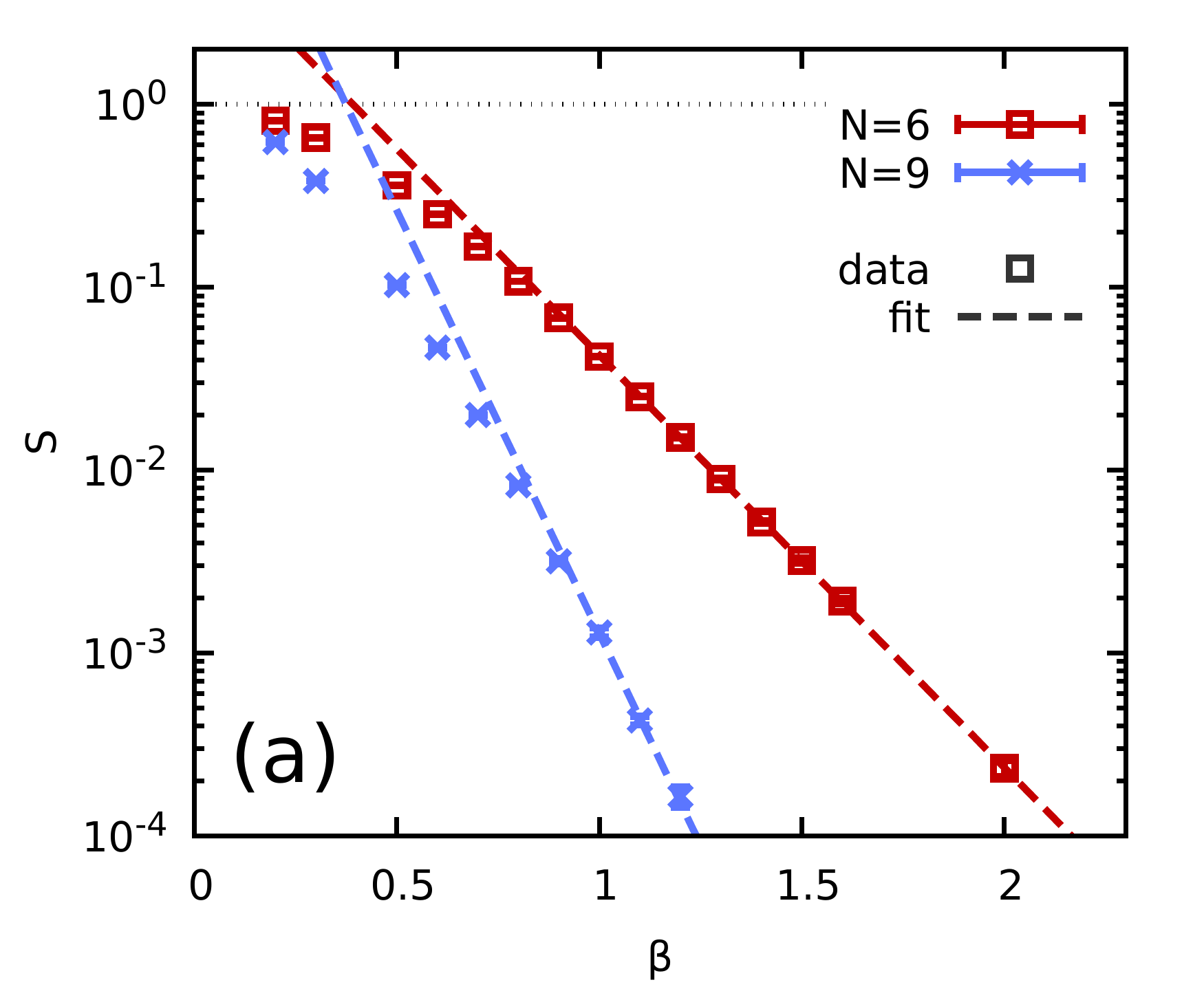}
\includegraphics[width=0.42547\textwidth]{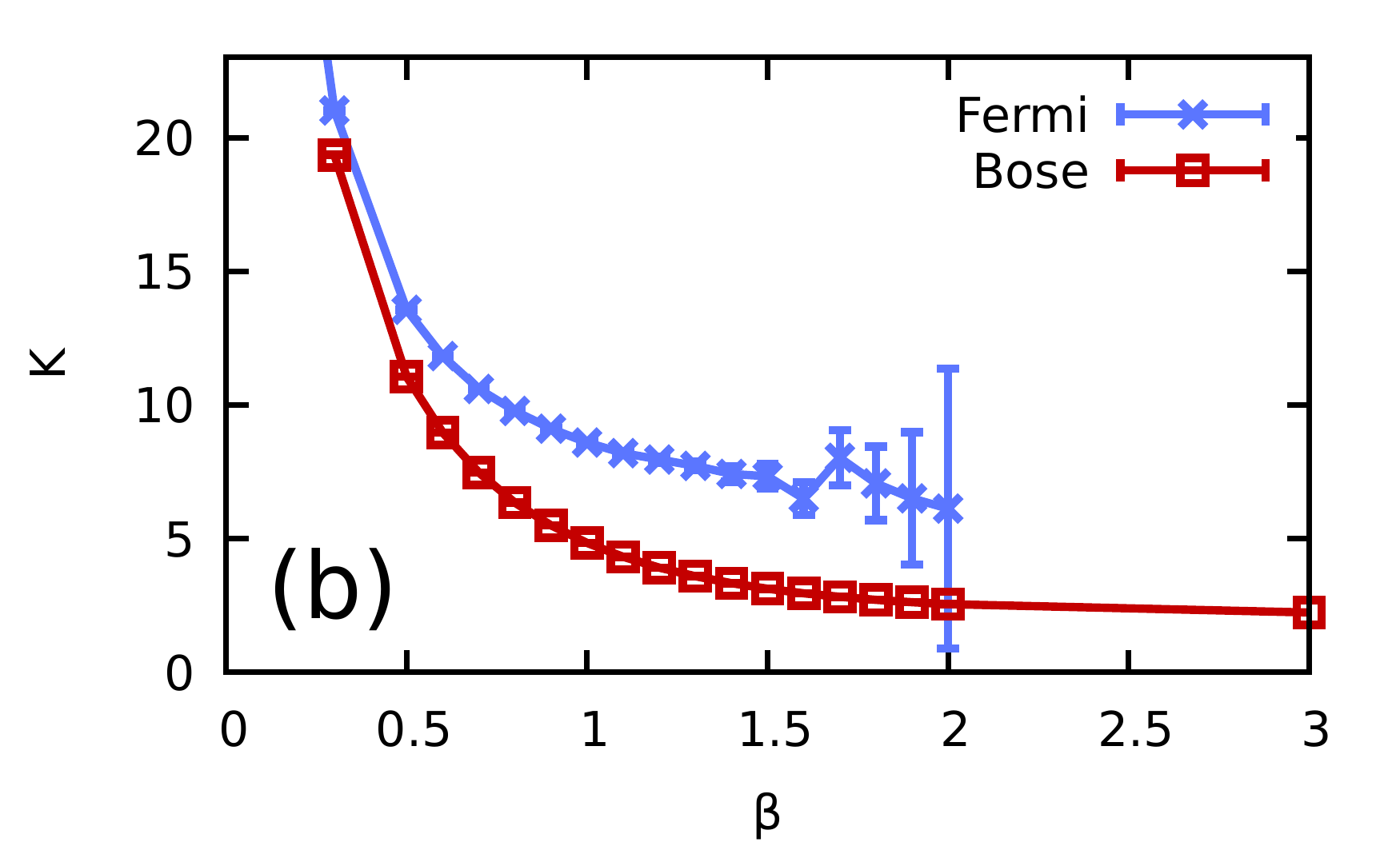}
\includegraphics[width=0.41547\textwidth]{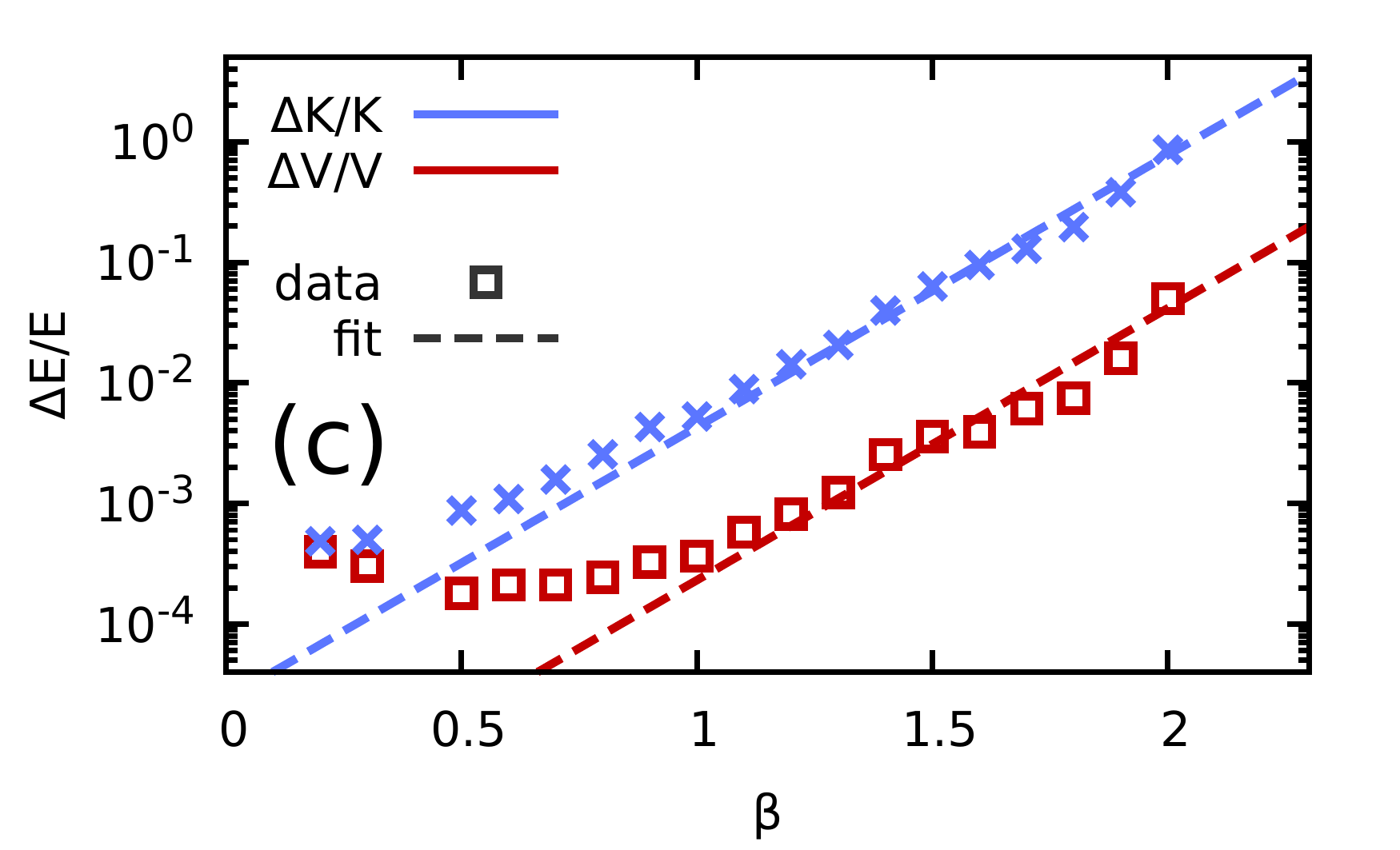}
\caption{\label{fig:2D_Coulomb_beta_dependence}
Temperature-dependence of the fermion sign problem for electrons in a $2D$ harmonic trap: Panel (a) shows the average sign for $N=6$ (red squares) and $N=9$ (blue crosses) electrons in a $2D$ harmonic trap with $\lambda=0.5$. In panel (b), we plot the corresponding kinetic energy for $N=6$ for both Fermi (blue crosses) and Bose statistics (red squares), and panel (c) depicts the relative statistical uncertainty for the case of fermions both for the kinetic energy (blue crosses) and the total potential energy (red squares), obtained for calculations with $N_\textnormal{s}=80$ seeds and $M=5\cdot10^6$ measurements per seed. The PIMC data for $S$ and different energies are given in Tab.~\ref{tab:beta_dependence_N6_lambda0.5} (for $N=6$).
}
\end{figure}

Let us next investigate the manifestiation of the FSP upon decreasing the temperature. To this end, we simulate spin-polarized electrons in a $2D$ harmonic trap at intermediate coupling $\lambda=0.5$. Fig.~\ref{fig:2D_Coulomb_beta_dependence}~(a) shows PIMC data for the $\beta$-dependence of the average sign $S$ for $N=6$ (red squares) and $N=9$ (blue crosses). 
First and foremost, we note that both data sets exhibit a qualitatively similar behavior: for small $\beta$, the system is nearly classical and $S$ is large, whereas it monotonically decreases with increasing $\beta$. In addition, the sign for $N=9$ is always smaller than for $N=6$, as it is expected. To verify the predicted exponential decrease of $S$ with $\beta$ (see Eq.~(\ref{eq:sign}) in Sec.~\ref{sec:FSP}), we perform fits (starting at $\beta\geq1$) of the form
\begin{eqnarray}\label{eq:beta_fit}
S_N(\beta)= a_N e^{-b_N \beta} \quad ,
\end{eqnarray}
with $a_N$ and $b_N$ being the free parameters. The results are shown as the dashed lines and are indeed in excellent agreement with the PIMC data for $\beta\geq1$. Note that at higher temperature, the free energy density $f$ [cf.~Eq.~(\ref{eq:sign} )] changes significantly with $\beta$, which leads to the deviation from Eq.~(\ref{eq:beta_fit}) in this regime.

Panel (b) shows the kinetic energy $K$ for the case of $N=6$ both for fermions (blue crosses) and bosons (red squares). Firstly, we mention that the relative deviations between Bose and Fermi statistics increase towards low temperature, as it is expected. Secondly, the red curve is very smooth over the entire depicted $\beta$-range, and the error bars cannot be seen with the naked eye. In contrast, the fermionic data are accurate for small $\beta$, but eventually the error bars markedly increase when $S$ becomes small.

To check if we are really running into the \emph{exponential wall} as predicted by Eq.~(\ref{eq:FSP}), we show the corresponding relative statistical uncertainty of $K$ (blue crosses) and the total potential energy $V$ (i.e., both interaction and external potential, red squares) in Fig.~\ref{fig:2D_Coulomb_beta_dependence}~(c). The dashed lines depict exponential fits (for $\beta\geq1$) of the form
\begin{eqnarray}
\frac{\Delta K}{K}(\beta) = a_6 e^{b_6\beta} c_K\ ,\
\frac{\Delta V}{V}(\beta) = a_6 e^{b_6\beta} c_V\ ,
\end{eqnarray}
with $C_K$ ($C_V$) being the only free parameter, as $a_6, b_6$ have already been determined by a fit to $S$. Evidently, the data and the fit are in excellent agreement both for $V$ and $K$, which (sadly) confirms the severity of the FSP.

Extensive PIMC data for the temperature-dependence of electrons in $2D$ and $3D$ (cf.~Sec.~\ref{sec:dimensionality}) are given in Tab.~\ref{tab:beta_dependence_N6_lambda0.5}.

\begin{figure}
\includegraphics[width=0.41547\textwidth]{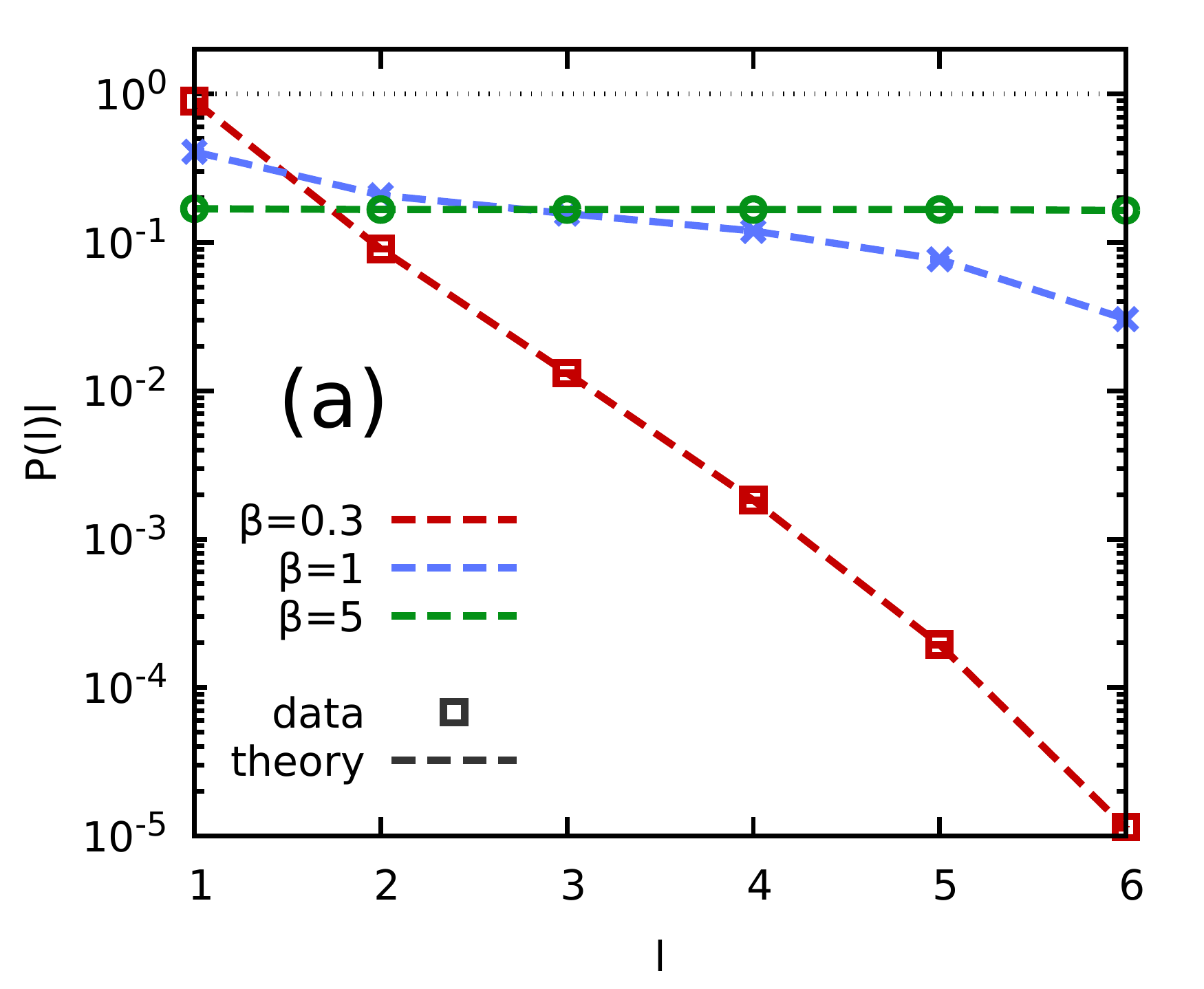}
\includegraphics[width=0.41547\textwidth]{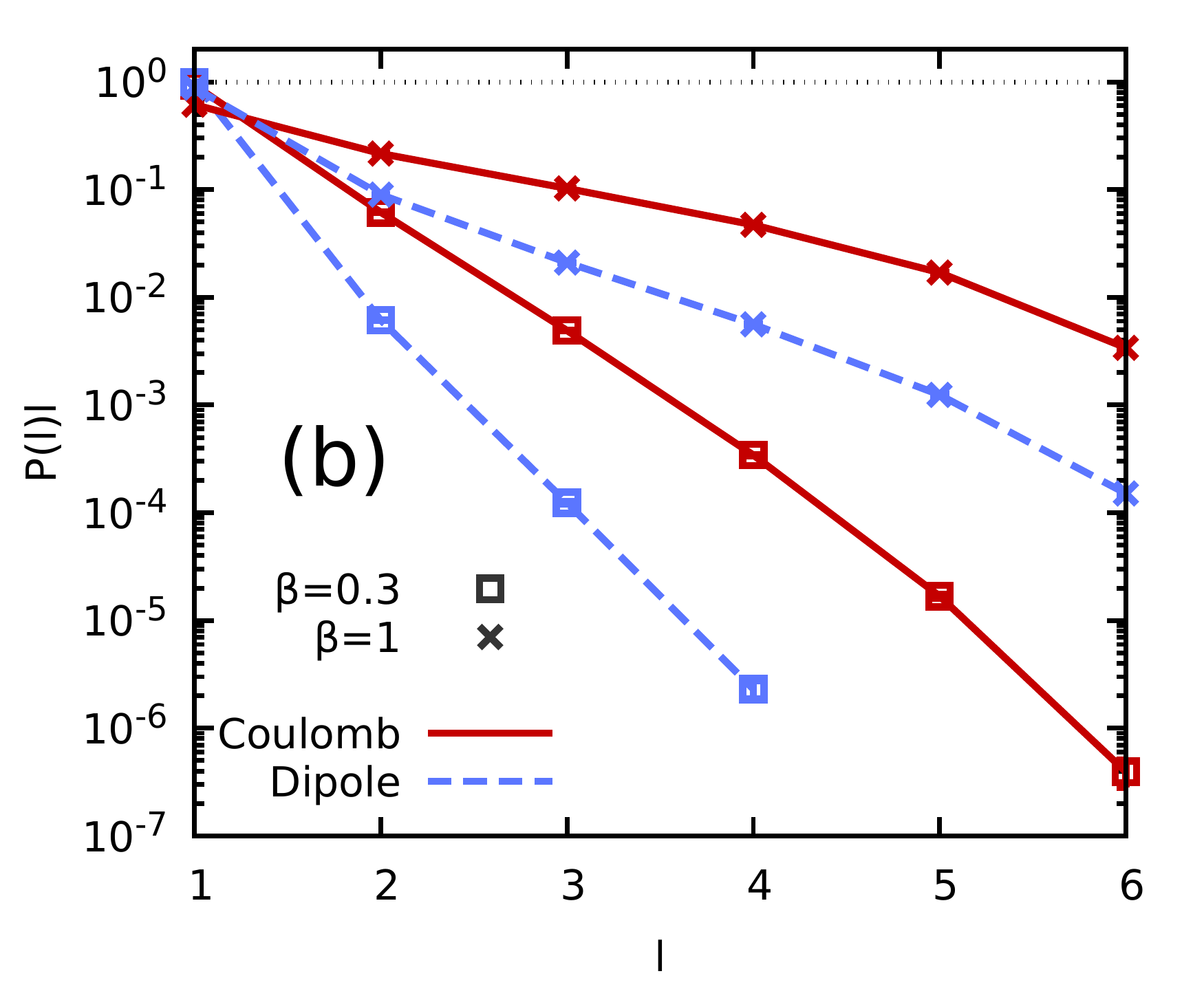}
\caption{\label{fig:Permutation_Cycles}
Permutation cycle distribution for $N=6$ spin-polarized fermions in a $2D$ harmonic trap: Panel (a) shows the probability of a single particle to be involved in a exchange-cycle of length $l$, $P(l)l$, for noninteracting fermions at $\beta=0.3$ (red), $\beta=1$ (blue), and $\beta=5$ (green). The dashed lines correspond to the theoretical result from Eq.~(\ref{eq:ideal_permutations}), and the points to our PIMC data.
Panel (b) shows the same information for electrons (Coulomb, solid red) and ultracold atoms (dipole interaction, dashed blue) with $\lambda=0.5$ for $\beta=0.3$ (squares) and $\beta=1$ (crosses).
}
\end{figure}

Let us conclude this section on the temperature dependence with a brief excursion to the distribution of permutation-cycles. In Fig.~\ref{fig:Permutation_Cycles}, we investigate the probability to find a particle involved in a permutation cycle of length $l$, $P(l)l$, see Ref.~\cite{dornheim_permutation_cycles} for a topical introduction and extensive discussion.
Panel (a) shows simulation results for $N=6$ noninteracting fermions in a $2D$ harmonic confinement at $\beta=0.3$ (red squares), $\beta=1$ (blue crosses), and $\beta=5$ (green circles). At the highest temperature, the paths resemble classical particles (see also Tab.~\ref{tab:SNAPSHOT_BOX}), pair-exchanges are quite improbable and approximately $90\%$ of particles are not involved in any exchange. Therefore, we find an average sign of $S\approx0.51$. At $\beta=1$, the situation has already drastically changed, and the distribution has become significantly flatter. Due to the resulting cancellation of positive and negative terms, the sign has decreased to $S\approx0.002$.
At the lowest temperature, $\beta=5$, the distribution has become almost completely flat and the sign vanishes within the given statistical uncertainty. In fact, it does hold $P(l)l=1/N$ in the zero temperature limit, which means that PIMC simulations are not possible in the ground state since the sign vanishes~\cite{krauth_book}.

To verify the correctness of our implementation, we compare our PIMC data to the theoretical result for $P(l)$, which can be phrased in terms of the noninteracting partition function at different temperature and system-size as~\cite{krauth_book,dornheim_permutation_cycles}
\begin{eqnarray}\label{eq:ideal_permutations}
P(l) = \frac{ Z'_1(l\beta) Z'_{N-l}(\beta) }{ l\ Z'_N(\beta) } \quad .
\end{eqnarray}
The corresponding dashed lines are in perfect agreement with our PIMC data for all temperatures $\beta$ and cycle-lengths $l$.

In Fig.~\ref{fig:Permutation_Cycles}~(b), we show results for $P(l)l$ for the same conditions as in panel (a), but with Coulomb (red) and dipole (blue) interaction and coupling strength $\lambda=0.5$. For $\beta=0.3$, we observe a qualitatively similar behavior as for the noninteracting case shown above. Still, the repulsion between the particles leads to a steeper decay of $P(l)l$ towards large $l$, which is even more pronounced in the case of dipoles. This is a direct consequence of the stronger repulsion at small distances in the latter case, which renders the formation of exchange-cycles within the simulation even more improbable, cf.~the discussion of Fig.~\ref{fig:2D_Coulomb_vs_2D_Dipole}. For $\beta=1$, the distribution is significantly less flat than in the noninteracting case, which is again more pronounced for the dipolar interaction.

\subsection{System-size dependence\label{sec:N}}

\begin{figure}
\includegraphics[width=0.41547\textwidth]{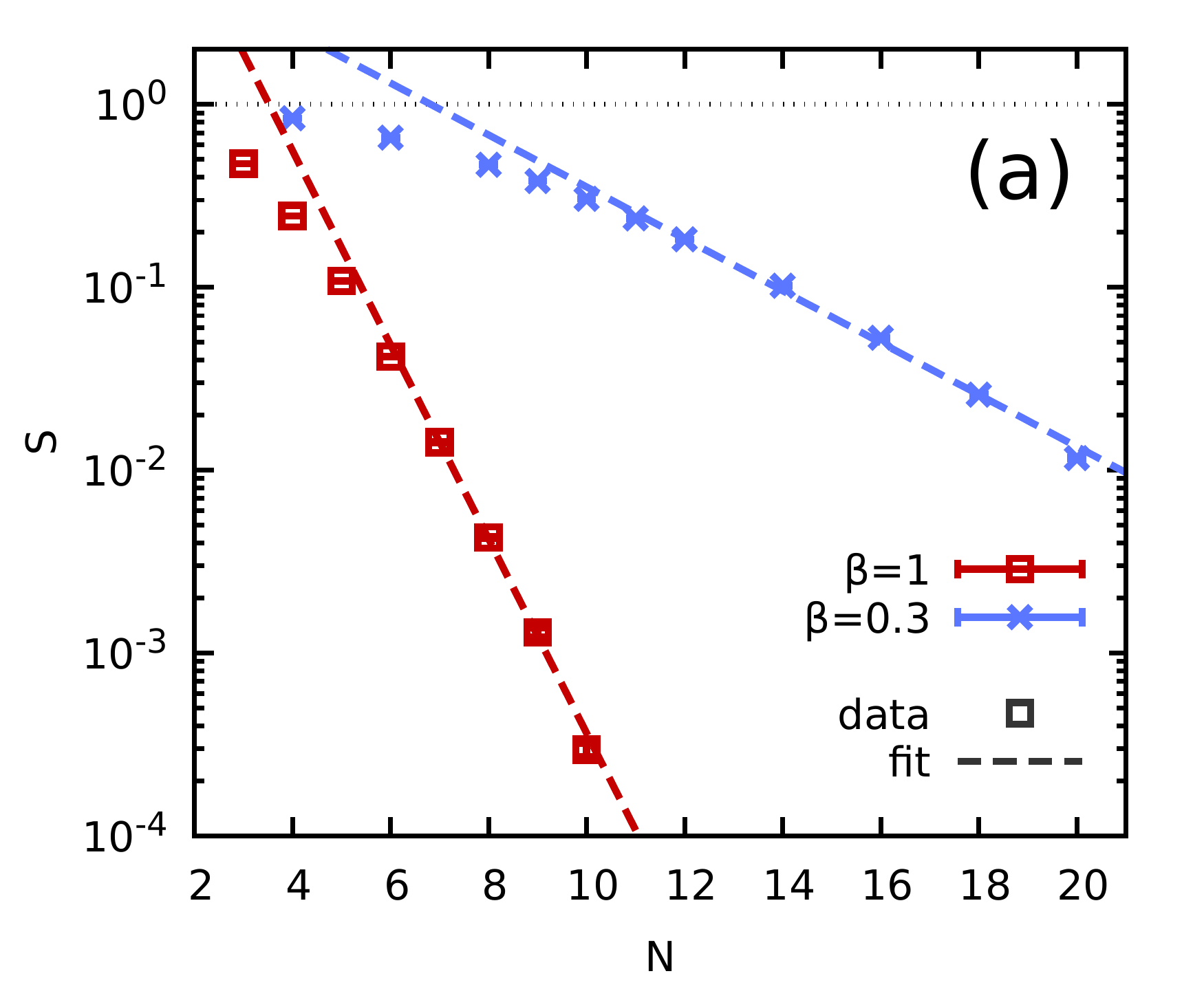}
\includegraphics[width=0.434\textwidth]{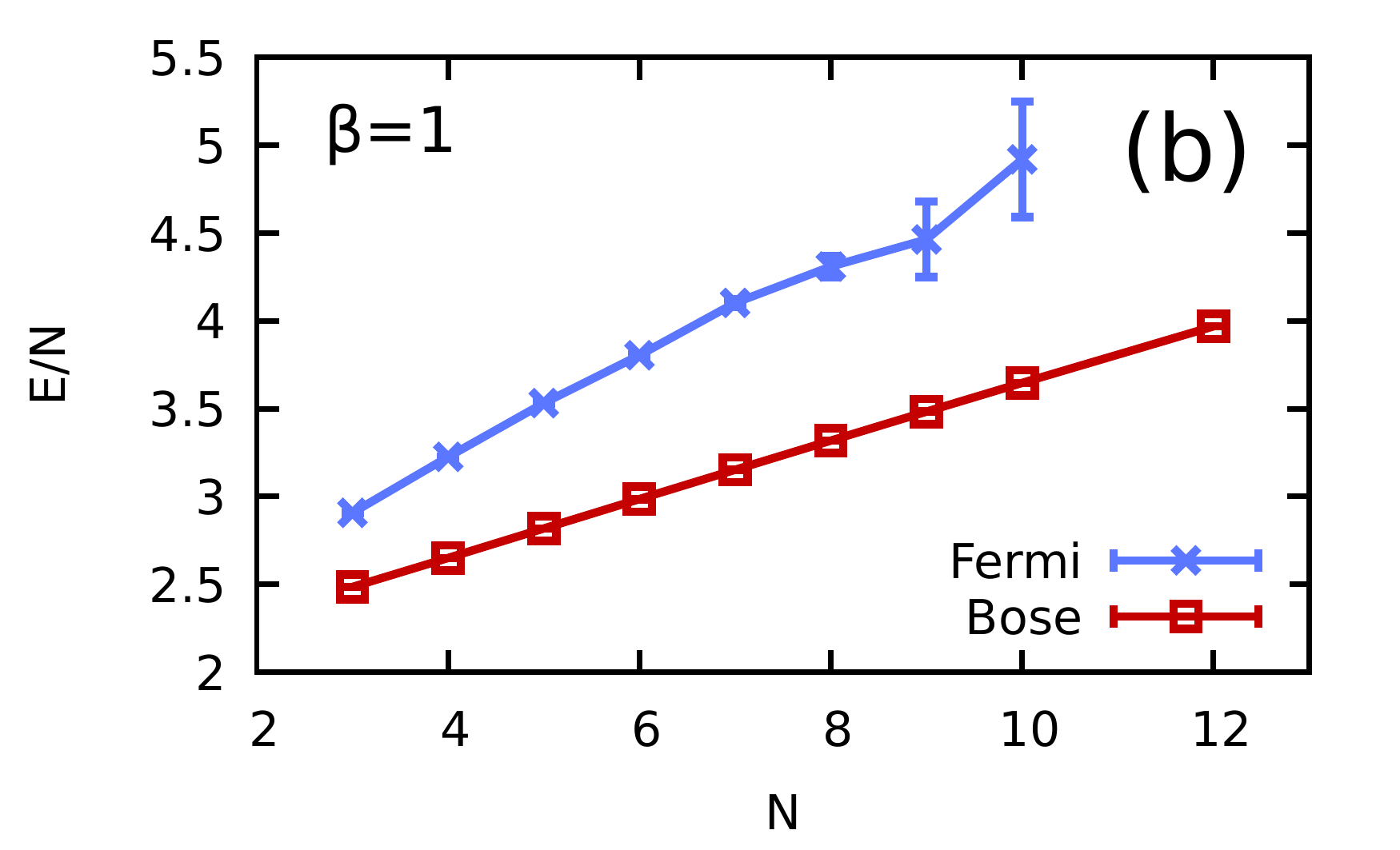}
\caption{\label{fig:2D_Coulomb_N_dependence}
System-size dependence of the fermion sign problem for electrons in a $2D$ harmonic trap: Panel (a) shows the average sign $S$ for $N=6$ spin-polarized electrons with $\lambda=0.5$ for $\beta=1$ (red) and $\beta=0.3$ (blue) with the points [dashed lines] depicting the PIMC data [a fit according to Eq.~(\ref{eq:N_fit})].
Panel (b) shows the corresponding results for the total energy per particle $E/N$ for Fermi (red squares) and Bose statistics (blue crosses). The PIMC results for $S$ and different energies are given in Tab.~\ref{tab:N_dependence_lambda0p5_Coulomb}.
}
\end{figure}

Another question that is of fundamental importance regarding fermionic PIMC simulations is the manifestation of the FSP with the system size. This topic is investigated in Fig.~\ref{fig:2D_Coulomb_N_dependence}~(a), where we show PIMC results for the average sign $S$ for $N=6$ electrons in a $2D$ harmonic trap with the coupling strength $\lambda=0.5$ and the inverse temperature $\beta=1$ (red squares), and $\beta=0.3$ (blue crosses).
Both data sets exhibit a steep decay with increasing $N$, which is significantly more pronounced for the lower temperature, as it is expected. To check the predicted exponential decay with $N$, we perform fits of the form
\begin{eqnarray}\label{eq:N_fit}
S_\beta(N)= a_\beta e^{-b_\beta N} \quad ,
\end{eqnarray}
with $a_\beta,b_\beta$ being the free parameters. The results for Eq.~(\ref{eq:N_fit}) are shown as the dashed lines, and are in qualitative agreement with the PIMC data. Still, the simulation results appear to exhibit an even faster decay than the exponential function from Eq.~(\ref{eq:N_fit}).

\begin{figure}
\includegraphics[width=0.41547\textwidth]{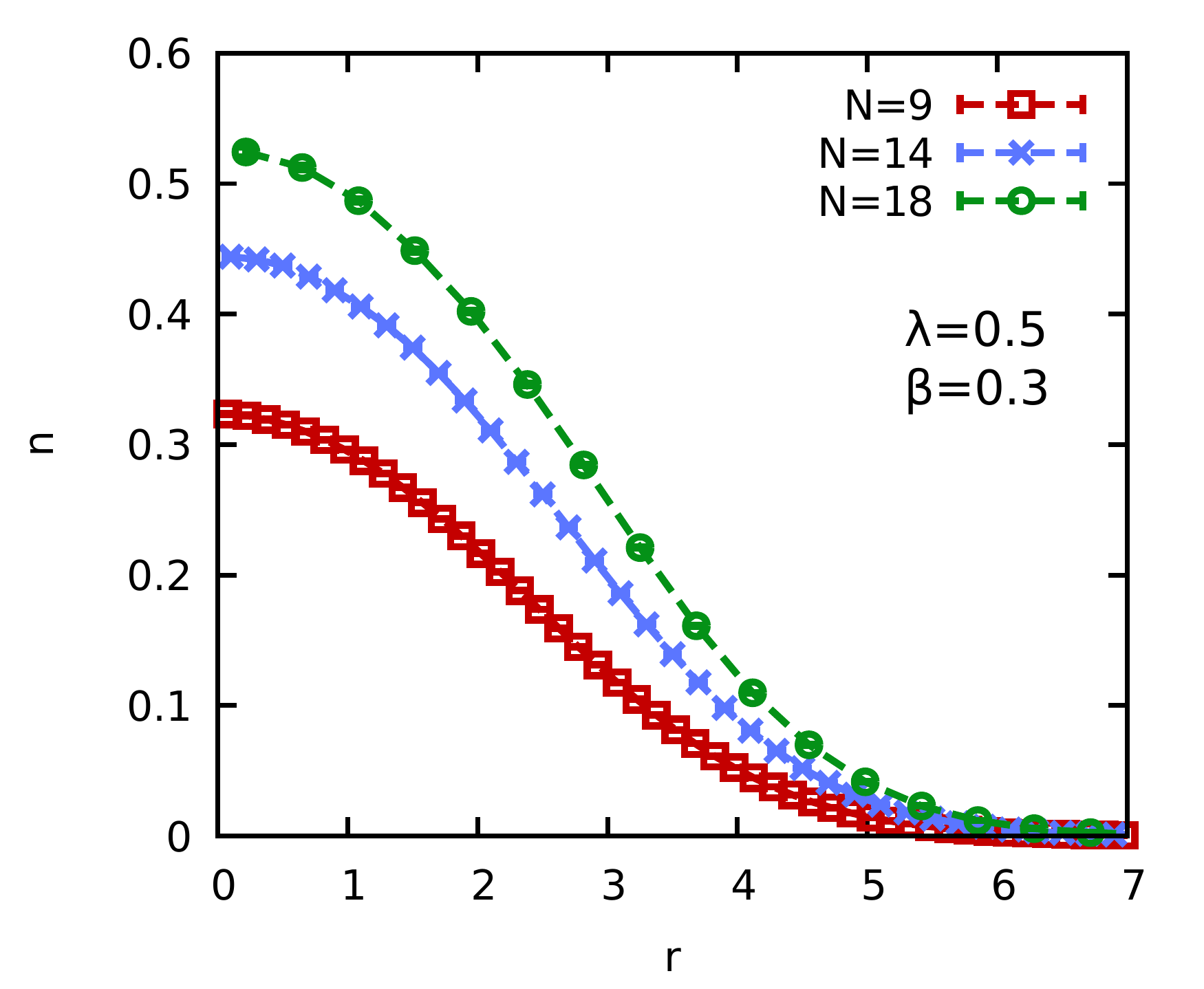}
\caption{\label{fig:2D_Coulomb_N_dependence_density}
System-size dependence of the radial density $n(r)$ for spin-polarized electrons (Coulomb) in a $2D$ harmonic confinement at $\lambda=0.5$ and $\beta=0.3$. The red squares, blue crosses, and green circles correspond to $N=9$, $N=14$, and $N=18$ electrons, respectively.
}
\end{figure}

To explain this finding, we plot the radial density $n(r)$ for $\beta=0.3$ and three different particle numbers in Fig.~\ref{fig:2D_Coulomb_N_dependence_density}.
Evidently, the addition of particles leads to an increased density, in particular around the center of the trap. Therefore, the system becomes more quantum degenerate, and the average sign decreases even faster than the exponential fit from Eq.~(\ref{eq:N_fit}).
It is important to note that the situation is entirely different for a uniform, periodic system like the UEG (see Sec.~\ref{sec:UEG}), where a change in system size does hardly affect the degree of degeneracy because the density remains constant. Therefore, one does indeed find an exponential decay of $S$ with $N$ in that case, cf.~Fig.~\ref{fig:YAKUB}.

Let us conclude this discussion of the system-size dependence of the FSP with the consideration of an observable. To this end, we show the $N$-dependence of the total energy per particle $E/N$ for the case of $\beta=1$ in Fig.~\ref{fig:2D_Coulomb_N_dependence}~(b)
both for Fermi (blue crosses) and Bose (red squares) statistics.
Evidently, the energy per particle does not remain constant for both cases, but increases, as it is expected. This trend is even more pronounced for the case of fermions, which are subject to the Pauli blocking. Thus, they get pushed away from the center of the trap, where the energy  
due to the external harmonic potential is large. 
In addition, we note the increasing error bars in the blue curve, which do not appear for bosons, and are a direct consequence of the corresponding decrease in $S$, cf.~Eq.~(\ref{eq:FSP}).

Extensive PIMC results for the $N$-dependence of spin-polarized electrons are given in Tab.~\ref{tab:N_dependence_lambda0p5_Coulomb}.

\subsection{Interaction and coupling-strength dependence\label{sec:coupling}}

\begin{figure}
\includegraphics[width=0.41547\textwidth]{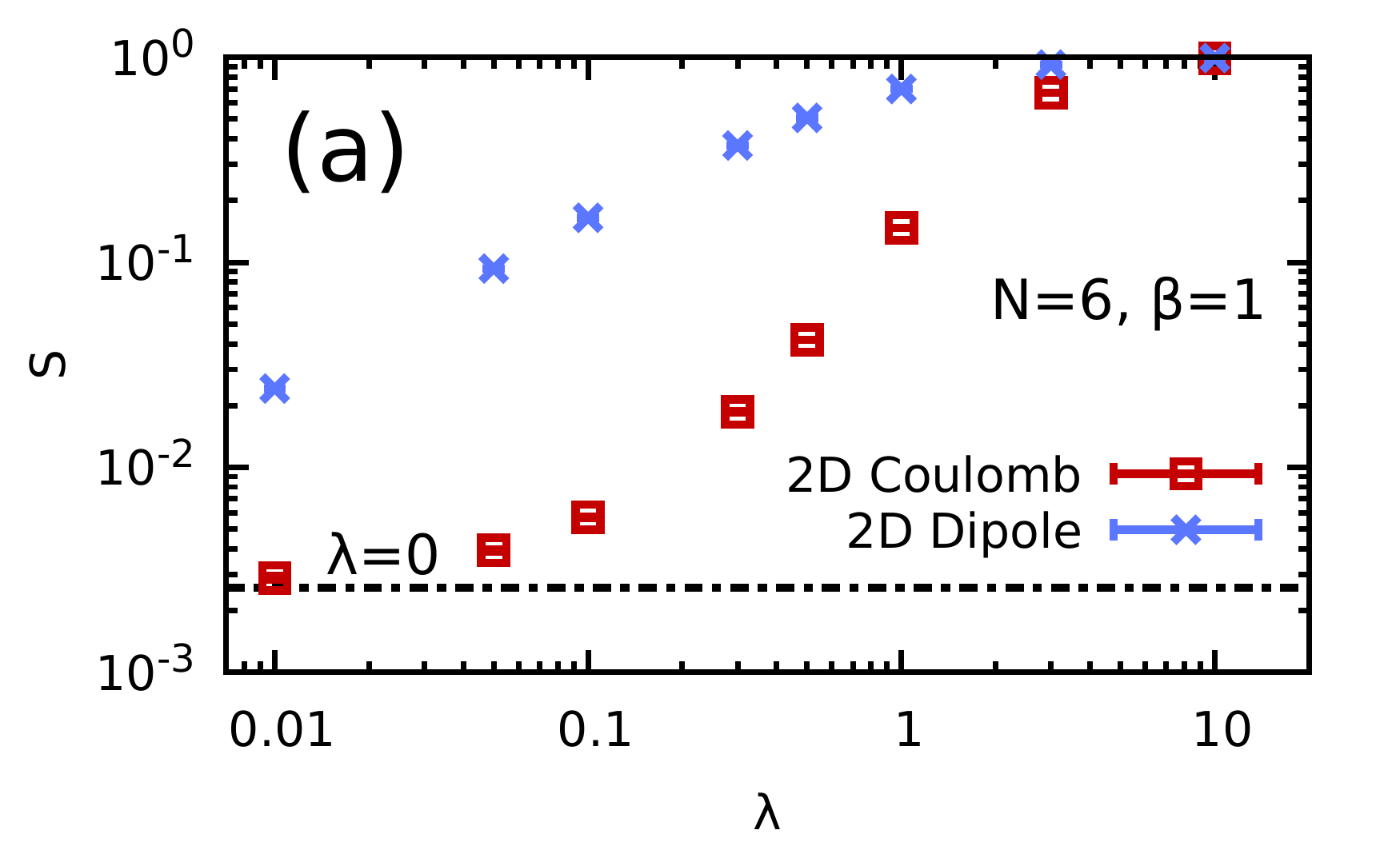}
\includegraphics[width=0.41547\textwidth]{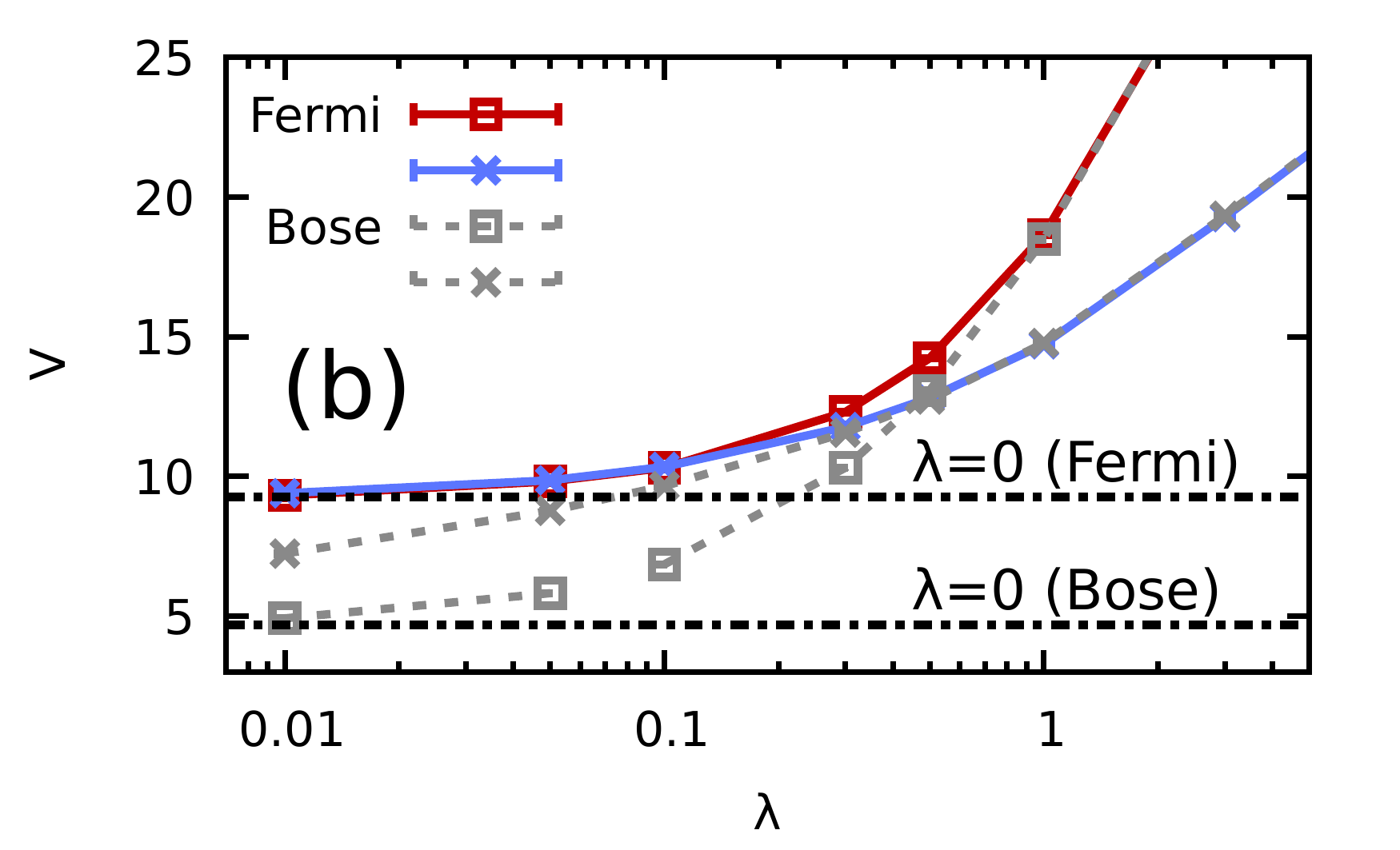}
\hspace*{-0.15cm}\includegraphics[width=0.42547\textwidth]{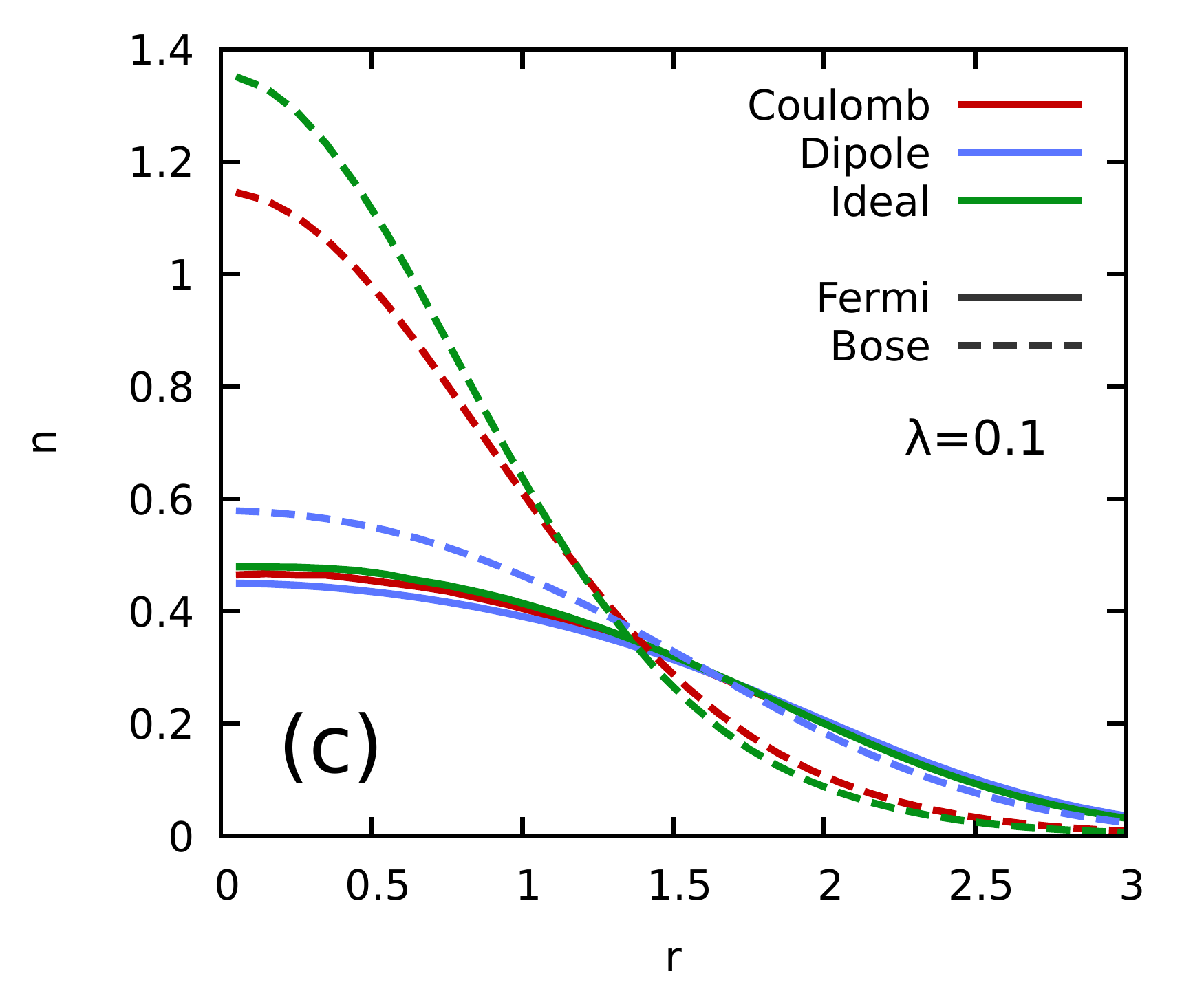}
\caption{\label{fig:2D_Coulomb_vs_2D_Dipole}
Impact of interaction-type on the fermion sign problem: Panel (a) shows the $\lambda$-dependence of the average sign $S$ for $N=6$ particles in a $2D$ harmonic trap at $\beta=1$ for Coulomb interaction (electrons, red squares) and dipole interaction (ultracold atoms, blue crosses). The dashed line depicts the noninteracting limit. Panel (b) depicts the total potential energy $V$, with the grey symbols corresponding to the bosonic expectation value. In panel (c), we show the radial density distributions $n(r)$ for $\lambda=0.1$ (and the noninteracting case, [top] green lines) for both Bose (dashed lines) and Fermi (solid lines) statistics. The PIMC results for $S$ and different energies are given in Tab.~\ref{tab:lambda_dependence_N6_beta1}.
}
\end{figure}

A somewhat less well understood question is the dependence of the FSP on the interaction-type and coupling strength. In Sec.~\ref{sec:temperature}, we have already seen that ultracold atoms with dipole interaction [$\alpha=3$, cf.~Eq.~(\ref{eq:Hamiltonian_trap})] exhibit a comparatively less severe sign problem than electrons at the same value of the coupling parameter $\lambda$, cf.~Fig.~\ref{fig:Permutation_Cycles}. In Fig.~\ref{fig:2D_Coulomb_vs_2D_Dipole}, we present a more systematic investigation of this issue by performing PIMC simulations of $N=6$ electrons (red squares) and ultracold atoms (blue crosses) at $\beta=1$.
Panel (a) shows the $\lambda$-dependence of the average sign $S$ over more than three orders of magnitude in the coupling strength. At $\lambda=10$, the particles are spatially separated by the strong repulsion for both types of interaction and fermionic exchange is suppressed. With decreasing $\lambda$, the particles get increasingly close to each other and the sign decreases for both data sets, although it does so significantly faster in the case of the Coulomb interaction. More specifically, the red curve has already almost attained the noninteracting limit ($\lambda=0$, dash-dotted black line) at $\lambda=0.01$, whereas the corresponding blue data point is still one order of magnitude larger. This is a direct consequence of the comparatively larger repulsion for the dipole-interaction at small distances, as we have already discussed in Sec.~\ref{sec:temperature}.

Let us next consider the corresponding $\lambda$-dependence of the potential energy $V$, which is shown in Fig.~\ref{fig:2D_Coulomb_vs_2D_Dipole}~(b). The squares and crosses depict data for Coulomb- and dipole-interaction, respectively, and the grey points show the corresponding results for Bose statistics. At strong coupling, quantum statistics are negligible, the grey and colored points are in perfect agreement, and the system resembles a semi-classical Coulomb- or dipole-system. With decreasing $\lambda$, there appears a transition region until eventually both the fermions and the bosons attain their respective noninteracting limit (dash-dotted black lines). Remarkably, this happens much faster for fermions, which are already in good agreement for both types of interaction at $\lambda=0.1$,
than for bosons, which still significantly deviate for $\lambda=0.01$.

The reason for this striking difference is illustrated in Fig.~\ref{fig:2D_Coulomb_vs_2D_Dipole}~(c), where we show the radial density $n(r)$ at $\lambda=0.1$ for Coulomb- (red), dipole- (blue), and no interaction (green) and for both Fermi (solid) and Bose (dashed) statistics. Let us first consider the three fermionic curves, which are in good agreement with each other, as it is by now expected from the observed corresponding agreement in $V$ [cf.~Fig.~\ref{fig:2D_Coulomb_vs_2D_Dipole}~(b)]. In stark contrast, the bosonic curve for the dipole-interaction significantly deviates from the other two, which explains the observed behavior in both $S$ and $V$: for Coulomb-interaction (or the noninteracting case), the paths that are sampled within our PIMC simulation are clustered around the center of the trap. The fermionic density, which remains large for much higher values of $r$, must subsequently be recovered by the cancellation and division by a small average sign $S$ according to Eq.~(\ref{eq:fermionic_expectation_value}).
For dipole-interaction, on the other hand, the strong repulsion at small distances has a very similar effect to the Pauli blocking, so that already the bosonic density is very close to its fermionic analogue. Consequently, the bosonic and fermionic configuration spaces and partition functions are almost equal, and the average sign $S\approx0.17$ is large.

In summary, we have found that the fermion sign problem is much less severe for interaction-types with a strong short-range repulsion. This makes the future systematic study of ultracold fermionic dipolar atoms~\cite{stuhler} (in the trap, in periodic boundary conditions, or in other geometries like bilayers~\cite{dynamic_alex2}) a promising project for future research.

Extensive PIMC data for the coupling-strength dependence of both electrons and ultracold atoms are given in Tab.~\ref{tab:lambda_dependence_N6_beta1}.

\begin{table*}
\caption{\label{tab:N_dependence_lambda0p5_Coulomb}System-size dependence:
\textit{Ab initio} path integral Monte Carlo results for spin-polarized electrons (Coulomb interaction) at $\beta=1$ (top half) and $\beta=0.3$ (bottom half) in a $2D$ harmonic trap with coupling strength $\lambda=0.5$. All results have been obtained  for $P=200$ imaginary-time propagators (see Appendix~\ref{sec:convergence} for details) and are given in oscillator units (i.e., energies in units of $E_0=\hbar\Omega$). Parts of the data set are shown in Fig.~\ref{fig:2D_Coulomb_N_dependence}.
}
\begin{ruledtabular}
\begin{tabular}{lllllll}
  $N$ & $S$ & $E_\textnormal{HO}$ & $V$ & $K$ & $K_\textnormal{vir}$ & $E$
\\ \hline$3$  &  $0.4746(2)$  &  $4.1585(9)$  &  $4.9640(8)$  &  $3.755(3)$ & $3.756(1)$  &  $8.719(3)$\\ 
$4$  &  $0.2453(1)$  &  $6.0593(9)$  &  $7.6279(8)$  &  $5.275(7)$ & $5.275(1)$  &  $12.903(7)$\\ 
$5$  &  $0.1085(1)$  &  $8.168(3)$  &  $10.721(2)$  &  $6.94(2)$ & $6.892(4)$  &  $17.66(2)$\\ 
$6$  &  $0.04184(8)$  &  $10.483(8)$  &  $14.234(5)$  &  $8.59(5)$ & $8.61(1)$  &  $22.82(5)$\\ 
$7$  &  $0.01425(9)$  &  $12.98(3)$  &  $18.13(1)$  &  $10.6(2)$ & $10.41(4)$  &  $28.7(2)$\\ 
$8$  &  $0.00428(6)$  &  $15.68(7)$  &  $22.40(3)$  &  $12.1(5)$ & $12.3(1)$  &  $34.5(5)$\\ 
$9$  &  $0.00130(8)$  &  $17.9(3)$  &  $26.7(2)$  &  $13(2)$ & $13.5(5)$  &  $40(2)$\\ 
$10$  &  $0.00030(3)$  &  $21.4(6)$  &  $31.8(3)$  &  $17(3)$ & $16.2(10)$  &  $49(3)$\\ 
 \hline
$4$  &  $0.8413(1)$  &  $14.392(5)$  &  $15.592(5)$  &  $13.782(6)$ & $13.792(9)$  &  $29.374(8)$\\ 
$6$  &  $0.6580(2)$  &  $22.492(8)$  &  $25.438(8)$  &  $21.01(1)$ & $21.02(1)$  &  $46.45(1)$\\ 
$8$  &  $0.4679(2)$  &  $31.118(10)$  &  $36.527(9)$  &  $28.44(2)$ & $28.41(2)$  &  $64.97(2)$\\ 
$9$  &  $0.3816(3)$  &  $35.65(2)$  &  $42.55(1)$  &  $32.24(2)$ & $32.20(2)$  &  $74.78(3)$\\ 
$10$  &  $0.3051(3)$  &  $40.32(2)$  &  $48.88(2)$  &  $36.04(4)$ & $36.04(3)$  &  $84.92(4)$\\ 
$11$  &  $0.2384(3)$  &  $45.06(2)$  &  $55.45(2)$  &  $39.86(5)$ & $39.87(4)$  &  $95.31(6)$\\ 
$12$  &  $0.1833(1)$  &  $49.98(1)$  &  $62.35(1)$  &  $43.85(5)$ & $43.80(2)$  &  $106.20(6)$\\ 
$14$  &  $0.1021(2)$  &  $60.11(3)$  &  $76.95(3)$  &  $51.7(2)$ & $51.70(5)$  &  $128.7(2)$\\ 
$16$  &  $0.0530(2)$  &  $70.75(6)$  &  $92.67(5)$  &  $59.7(3)$ & $59.79(9)$  &  $152.4(3)$\\ 
$18$  &  $0.0259(1)$  &  $81.89(8)$  &  $109.47(6)$  &  $69.6(6)$ & $68.1(1)$  &  $179.1(6)$\\ 
$20$  &  $0.01164(6)$  &  $93.3(1)$  &  $127.18(8)$  &  $75.8(10)$ & $76.4(2)$  &  $203.0(10)$\\ 
\end{tabular}
\end{ruledtabular}
\end{table*}

\begin{table*}
\caption{\label{tab:lambda_dependence_N6_beta1}Coupling strength dependence:
\textit{Ab initio} path integral Monte Carlo results for $N=6$ spin-polarized electrons (top half) and ultracold atoms with dipole interaction (bottom half) in a $2D$ harmonic trap at an inverse temperature $\beta=1$. All results have been obtained  for $P=200$ imaginary-time propagators (see Appendix~\ref{sec:convergence} for details) and are given in oscillator units (i.e., energies in units of $E_0=\hbar\Omega$). Parts of the data set are shown in Fig.~\ref{fig:2D_Coulomb_vs_2D_Dipole}.
}
\begin{ruledtabular}
\begin{tabular}{lllllll}
  $\lambda$ & $S$ & $E_\textnormal{HO}$ & $V$ & $K$ & $K_\textnormal{vir}$ & $E$
\\ \hline$0$  &  $0.00258(1)$  &  $9.27(2)$  &  $9.27(2)$  &  $9.3(1)$ & $9.27(3)$  &  $18.5(1)$\\ 
$0.01$  &  $0.00288(5)$  &  $9.24(8)$  &  $9.32(8)$  &  $9.2(4)$ & $9.2(1)$  &  $18.5(5)$\\ 
$0.05$  &  $0.00393(5)$  &  $9.42(6)$  &  $9.82(5)$  &  $9.5(3)$ & $9.22(9)$  &  $19.3(3)$\\ 
$0.1$  &  $0.00570(5)$  &  $9.51(4)$  &  $10.31(3)$  &  $8.7(2)$ & $9.11(6)$  &  $19.0(2)$\\ 
$0.3$  &  $0.01861(10)$  &  $9.98(2)$  &  $12.31(1)$  &  $8.7(1)$ & $8.81(2)$  &  $21.0(1)$\\ 
$0.5$  &  $0.04184(8)$  &  $10.483(8)$  &  $14.234(5)$  &  $8.59(5)$ & $8.61(1)$  &  $22.82(5)$\\ 
$1$  &  $0.1475(1)$  &  $11.657(3)$  &  $18.644(2)$  &  $8.16(2)$ & $8.164(5)$  &  $26.80(2)$\\ 
$3$  &  $0.6717(2)$  &  $15.859(2)$  &  $32.9762(9)$  &  $7.297(4)$ & $7.301(2)$  &  $40.273(4)$\\ 
$10$  &  $0.99069(4)$  &  $27.149(1)$  &  $67.7235(7)$  &  $6.864(2)$ & $6.862(2)$  &  $74.587(2)$\\ 
 \hline
$0.01$  &  $0.02419(4)$  &  $9.327(5)$  &  $9.401(4)$  &  $9.22(3)$ & $9.22(1)$  &  $18.62(3)$\\ 
$0.05$  &  $0.09311(7)$  &  $9.541(2)$  &  $9.865(2)$  &  $9.06(1)$ & $9.056(6)$  &  $18.93(1)$\\ 
$0.1$  &  $0.16501(6)$  &  $9.7708(10)$  &  $10.3445(8)$  &  $8.922(5)$ & $8.910(3)$  &  $19.266(5)$\\ 
$0.3$  &  $0.3721(1)$  &  $10.4931(8)$  &  $11.7701(7)$  &  $8.583(3)$ & $8.578(2)$  &  $20.353(3)$\\ 
$0.5$  &  $0.5070(1)$  &  $11.0464(7)$  &  $12.8143(6)$  &  $8.394(2)$ & $8.395(2)$  &  $21.208(2)$\\ 
$1$  &  $0.7015(2)$  &  $12.0977(10)$  &  $14.7323(8)$  &  $8.148(2)$ & $8.146(3)$  &  $22.880(3)$\\ 
$3$  &  $0.9228(2)$  &  $14.707(1)$  &  $19.2833(9)$  &  $7.842(2)$ & $7.842(3)$  &  $27.125(2)$\\ 
$10$  &  $0.99404(7)$  &  $19.4307(10)$  &  $27.2642(7)$  &  $7.680(2)$ & $7.680(3)$  &  $34.944(2)$\\ 
\end{tabular}
\end{ruledtabular}
\end{table*}

\subsection{Dimensionality versus Interaction-type, virial theorem\label{sec:dimensionality}}

\begin{figure}
\includegraphics[width=0.41547\textwidth]{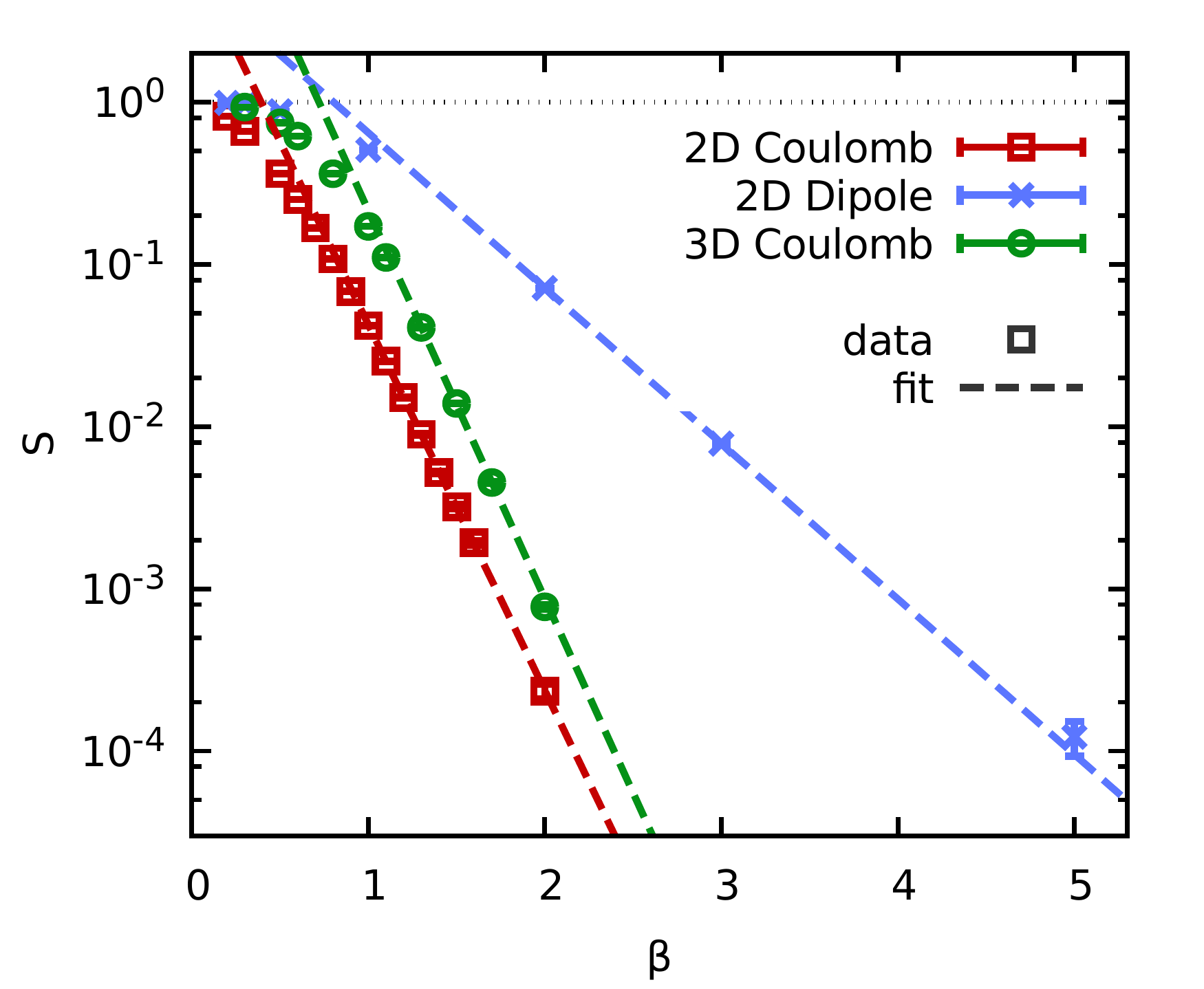}
\caption{\label{fig:big_comparison}
Dependence of the fermion sign problem on the inverse temperature $\beta$: Shown are PIMC results for $N=6$ spin-polarized fermions with $\lambda=0.5$ in $2D$ with Coulomb interaction (electrons, red squares), $2D$ with dipole interaction (ultracold atoms, blue crosses), and in $3D$ with Coulomb interaction (electrons, green circles). The dashed lines depict exponential fits according to Eq.~(\ref{eq:beta_fit}). All PIMC results for $S$ and different energies are given in Tab.~\ref{tab:beta_dependence_N6_lambda0.5}.
}
\end{figure}

The last question to be investigated in this work regarding fermions in a harmonic confinement is the impact of the dimensionality. 
In Fig.~\ref{fig:big_comparison}, we show the $\beta$-dependence of the average sign $S$ for $N=6$ and $\lambda=0.5$. The red squares, blue crosses, and green circles depict our PIMC results for $2D$ Coulomb, $2D$ dipoles, and $3D$ Coulomb, respectively. The corresponding dashed lines depict exponential fits according to Eq.~(\ref{eq:beta_fit}), which are in excellent agreement with the data for all types of systems. As usual, the dipole interaction leads to a significantly less steep decay of $S$ with $\beta$, cf.~Sec.~\ref{sec:coupling}.
In addition, we find that the Coulomb systems exhibit a very similar behavior of $S$, although the exponential decrease starts at somewhat lower temperatures in $3D$.
This is most likely due to the additional degree of freedom in this case, which allows the electrons to avoid each other more effectively.

In Tab.~\ref{tab:SNAPSHOT_BOX}, we compare snapshots from our PIMC simulation for all three kinds of system types at three different temperature regimes. For $\beta=0.3$ (left column), all systems exhibit a very similar behavior, with the extension of the paths, which is proportional to the thermal wave length $\lambda_\beta=\sqrt{2\pi\hbar^2\beta/m }$, being significantly smaller than the average inter-particle distance $\overline{r}$. At $\beta=1$ (center column), the paths are clustered more closely around the center of the trap in all three cases (the scale is equal for all three depicted values of $\beta$), and $\lambda_\beta$ is comparable to $\overline{r}$. In the case of dipole interaction (top row), the paths of individual particles are still mostly separated by the strong short-range repulsion (cf.~Sec.~\ref{sec:coupling}), and no exchange-cycle is present in the snapshot (the two particles in the front are close, but not connected). For $2D$ Coulomb (center row), on the other hand, fermionic exchange already plays a dominant role, and there appear two permutation-cycles with $N=3$ and $N=2$ particles in it. Going to $3D$ (bottom row), the situation looks qualitatively the same as in $2D$, and permutation-cycles are present, too. At low temperature, $\beta=5$, the thermal wavelength is larger than $\overline{r}$ in all three cases and the system is fully quantum degenerate. Yet, the dipole interaction manages to push the particles away from each other, and the corresponding average sign is several orders of magnitude larger than for Coulomb interaction, cf.~Fig.~\ref{fig:big_comparison}.
For Coulomb interaction in $2D$ and $3D$, the particles form an entangled knot of paths around the center of the trap, the probability to find an exchange cycle of length $l$ is almost constant (cf.~Fig.~\ref{fig:Permutation_Cycles}), and the average sign vanishes within the given statistical uncertainty.

\begin{figure}
\hspace*{-0.37cm}\includegraphics[width=0.42547\textwidth]{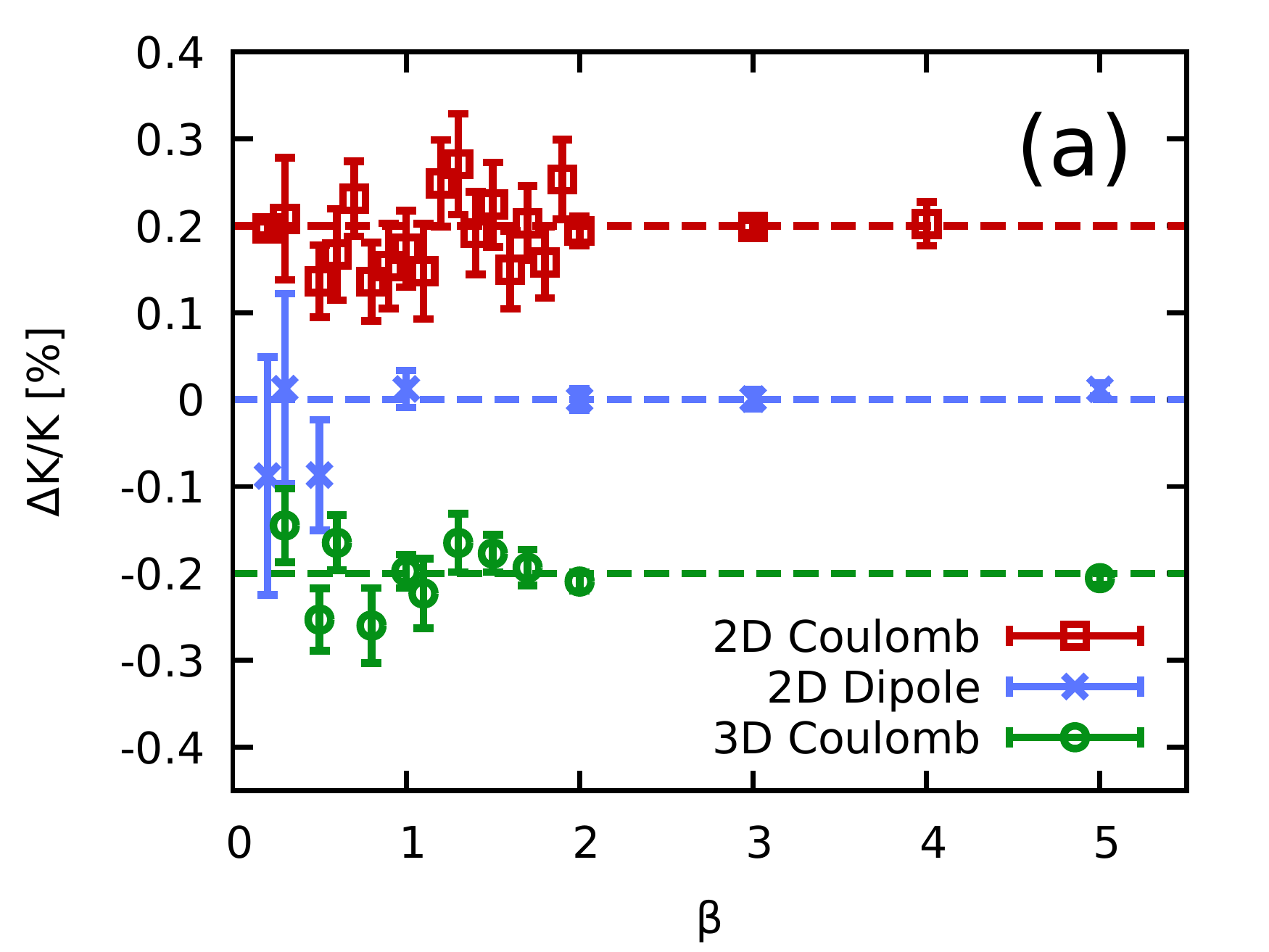}
\includegraphics[width=0.40\textwidth]{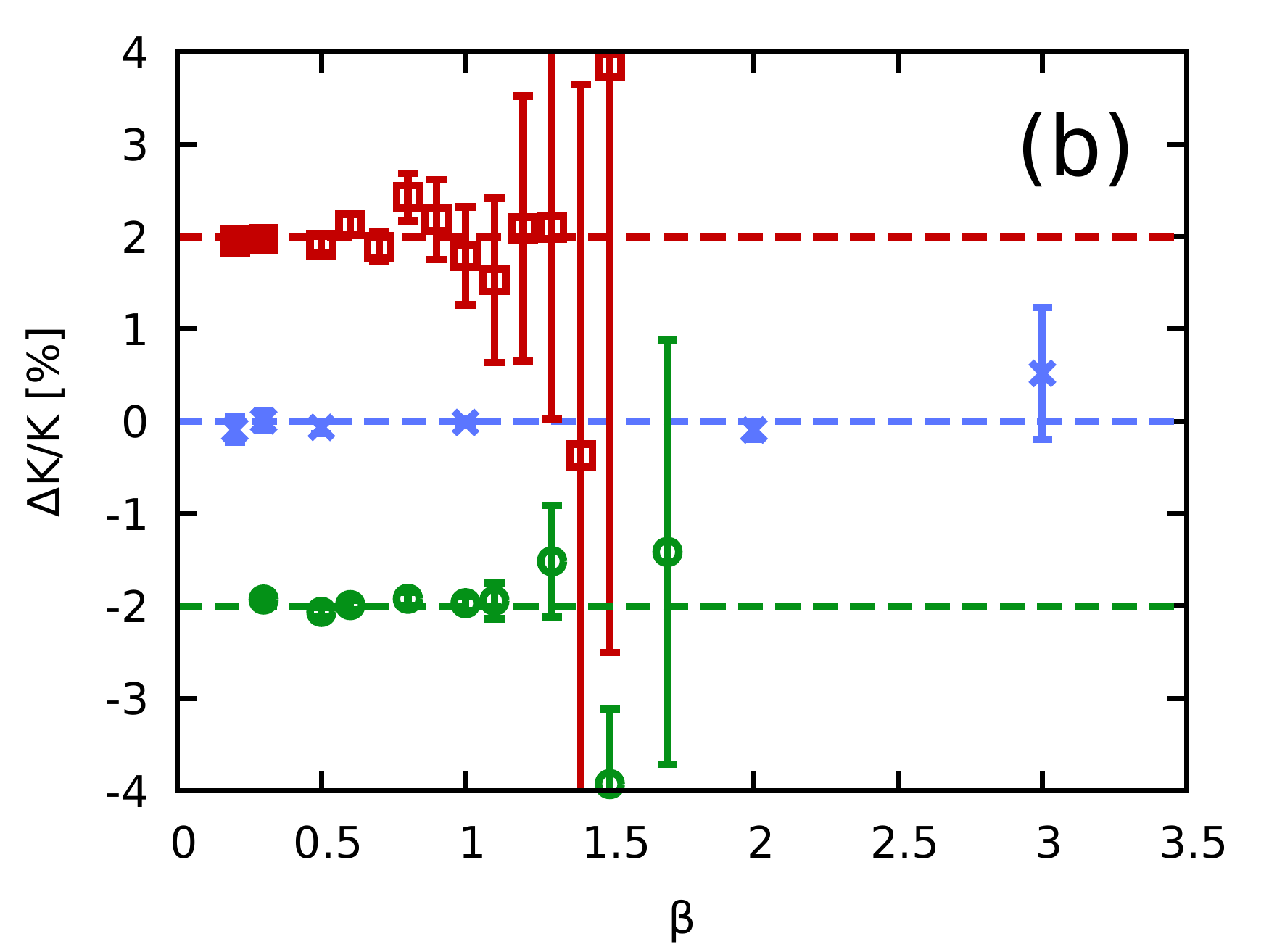}
\caption{\label{fig:Virial}
Verification of the virial theorem for $N=6$ spin-polarized fermions with $\lambda=0.5$: Relative difference (in $\%$) between the kinetic energy $K$ computed via the standard PIMC thermodynamic estimator and the virial theorem [cf.~Eq.~(\ref{eq:K_vir})]. The red squares, blue crosses, and green circles correspond to Coulomb interaction in $2D$ ($\alpha=1$), dipole interaction in $2D$ ($\alpha=3$), and Coulomb interaction in $3D$ ($\alpha=1$), respectively, and panels (a) and (b) show PIMC results for bosons and fermions.
The red and blue points have been shifted by $\pm0.2\%$ ($\pm2\%$) in the case of bosons (fermions) for better visibility.
The corresponding results for the average sign $S$ are shown in Fig.~\ref{fig:big_comparison}.
}
\end{figure}

Let us conclude this section by investigating the virial theorem~\cite{greiner_book}, which gives a relation between the different contributions to the total energy. For example, it holds
\begin{eqnarray}\label{eq:K_vir}
K = E_\textnormal{HO} - \alpha \frac{V - E_\textnormal{HO} }{2} \quad ,
\end{eqnarray}
with $V$ and $E_\textnormal{HO}$ being the total potential energy and the energy due to the external potential, respectively. Recall that $\alpha\in\{1,3\}$ distinguishes between Coulomb and dipolar interaction, cf.~Eq.~(\ref{eq:Hamiltonian_trap}). 

In Fig.~\ref{fig:Virial}, we show the relative difference between Eq.~(\ref{eq:K_vir}) and the kinetic energy as evaluated using the standard PIMC thermodynamic estimator (for an extensive discussion on energy estimation in PIMC simulations, see Ref.~\cite{janke}). Panels (a) and (b) show results for bosons and fermions, and the red squares, blue crosses, and green circles depict data for $2D$ Coulomb, $2D$ dipoles, and $3D$ Coulomb, respectively, with the red and green curves having been shifted for better visibility. Due to the absence of the FSP for Bose statistics, the statistical uncertainty is of the order of $\Delta K/K\sim10^{-4}$, and the difference between both results for $K$ vanishes within the error bars for all three data sets.
For fermions, the error eventually explodes with increasing $\beta$, but Eq.~(\ref{eq:K_vir}) still holds within the given uncertainty. Since Eq.~(\ref{eq:K_vir}) typically exhibits a smaller variance than the thermodynamic estimator for $K$, this route constitutes the method of choice and the results have been included as an extra column in all data tables as $K_\textnormal{vir}$.

\begin{table*}
\caption{\label{tab:beta_dependence_N6_lambda0.5}Temperature dependence:
\textit{Ab initio} path integral Monte Carlo results for $N=6$ spin-polarized electrons in a $2D$ (top half) and $3D$ (bottom half) harmonic trap with Coulomb interaction and coupling strength $\lambda=0.5$. All results have been obtained  for $P=200$ imaginary-time propagators (see Appendix~\ref{sec:convergence} for details) and are given in oscillator units (i.e., energies in units of $E_0=\hbar\Omega$). Parts of the data set are shown in Figs.~\ref{fig:2D_Coulomb_beta_dependence} and \ref{fig:big_comparison}.
}
\begin{ruledtabular}
\begin{tabular}{lllllll}
  $\beta$ & $S$ & $E_\textnormal{HO}$ & $V$ & $K$ & $K_\textnormal{vir}$ & $E$
\\ \hline
$0.3$  &  $0.6580(2)$  &  $22.492(8)$  &  $25.438(8)$  &  $21.01(1)$ & $21.02(1)$  &  $46.45(1)$\\ 
$0.5$  &  $0.3616(2)$  &  $15.246(4)$  &  $18.610(3)$  &  $13.55(1)$ & $13.564(6)$  &  $32.16(1)$\\ 
$0.6$  &  $0.2507(2)$  &  $13.552(4)$  &  $17.040(4)$  &  $11.82(1)$ & $11.807(6)$  &  $28.86(1)$\\ 
$0.8$  &  $0.1079(1)$  &  $11.557(5)$  &  $15.211(4)$  &  $9.77(2)$ & $9.730(8)$  &  $24.98(3)$\\ 
$1$  &  $0.04184(8)$  &  $10.483(8)$  &  $14.234(5)$  &  $8.59(5)$ & $8.61(1)$  &  $22.82(5)$\\ 
$1.1$  &  $0.02526(9)$  &  $10.12(1)$  &  $13.902(8)$  &  $8.19(7)$ & $8.23(2)$  &  $22.10(7)$\\ 
$1.3$  &  $0.00894(8)$  &  $9.62(3)$  &  $13.44(2)$  &  $7.7(2)$ & $7.71(5)$  &  $21.2(2)$\\ 
$1.5$  &  $0.00321(10)$  &  $9.15(9)$  &  $13.07(5)$  &  $7.3(5)$ & $7.2(1)$  &  $20.4(5)$\\ 
$1.7$  &  $0.00114(6)$  &  $8.7(2)$  &  $12.77(8)$  &  $8(1)$ & $6.7(3)$  &  $21(1)$\\ 
$2$  &  $0.00023(2)$  &  $8.5(3)$  &  $12.5(1)$  &  $4(1)$ & $6.5(5)$  &  $17(1)$\\ 
  \hline
$0.3$  &  $0.9298(1)$  &  $31.49(1)$  &  $33.68(1)$  &  $30.414(10)$ & $30.39(2)$  &  $64.09(2)$\\ 
$0.5$  &  $0.7434(2)$  &  $20.172(5)$  &  $22.833(5)$  &  $18.829(8)$ & $18.841(8)$  &  $41.66(1)$\\ 
$0.6$  &  $0.6163(2)$  &  $17.465(4)$  &  $20.286(4)$  &  $16.055(8)$ & $16.054(6)$  &  $36.341(9)$\\ 
$0.8$  &  $0.3610(2)$  &  $14.290(4)$  &  $17.336(3)$  &  $12.778(9)$ & $12.768(6)$  &  $30.114(9)$\\ 
$1$  &  $0.1712(1)$  &  $12.547(2)$  &  $15.736(2)$  &  $10.956(9)$ & $10.952(4)$  &  $26.692(9)$\\ 
$1.1$  &  $0.1102(2)$  &  $11.954(6)$  &  $15.196(5)$  &  $10.34(2)$ & $10.333(10)$  &  $25.54(2)$\\ 
$1.3$  &  $0.0408(2)$  &  $11.11(1)$  &  $14.429(8)$  &  $9.49(6)$ & $9.44(2)$  &  $23.92(6)$\\ 
$1.5$  &  $0.01388(8)$  &  $10.56(2)$  &  $13.93(1)$  &  $8.70(7)$ & $8.87(3)$  &  $22.63(7)$\\ 
$1.7$  &  $0.00452(7)$  &  $10.14(5)$  &  $13.56(3)$  &  $8.5(2)$ & $8.42(8)$  &  $22.0(2)$\\ 
$2$  &  $0.00077(3)$  &  $9.9(1)$  &  $13.28(6)$  &  $8.8(5)$ & $8.1(2)$  &  $22.1(5)$\\ 
\end{tabular}
\end{ruledtabular}
\end{table*}

\begin{table*}[]
    \centering
    \begin{tabular}{|m{1.72cm}| m{5.3cm}|m{4.9cm}|m{4.9cm}|} \hline\hline
    & \large \centering $\beta=0.3$ &\large \centering $\beta=1$ &\begin{center}\large $\beta=5$ \end{center}\\ \hline\hline
\large \begin{center}$2D$ Dipole\end{center}& \includegraphics[width=0.3\textwidth]{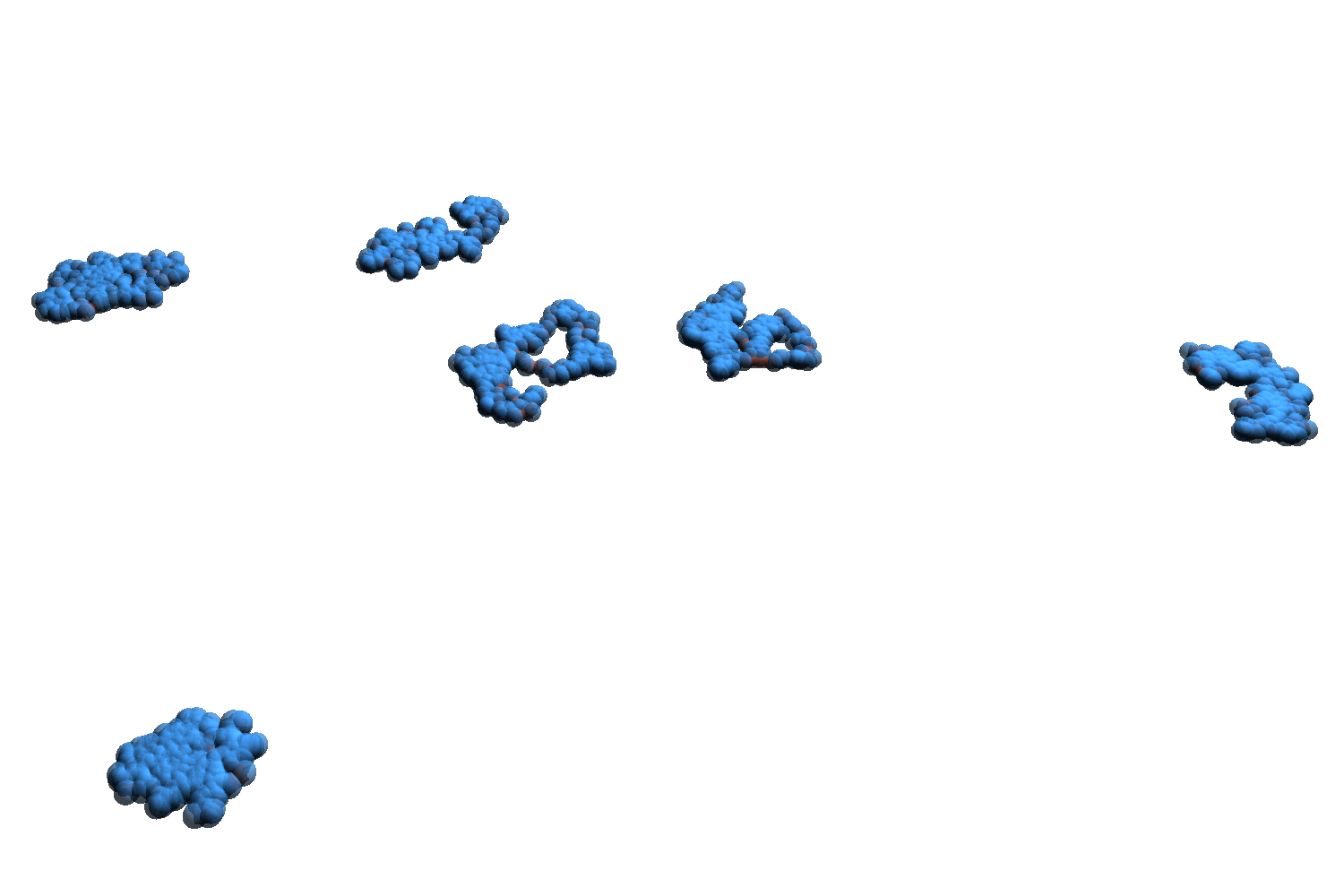} & 
\includegraphics[width=0.3\textwidth]{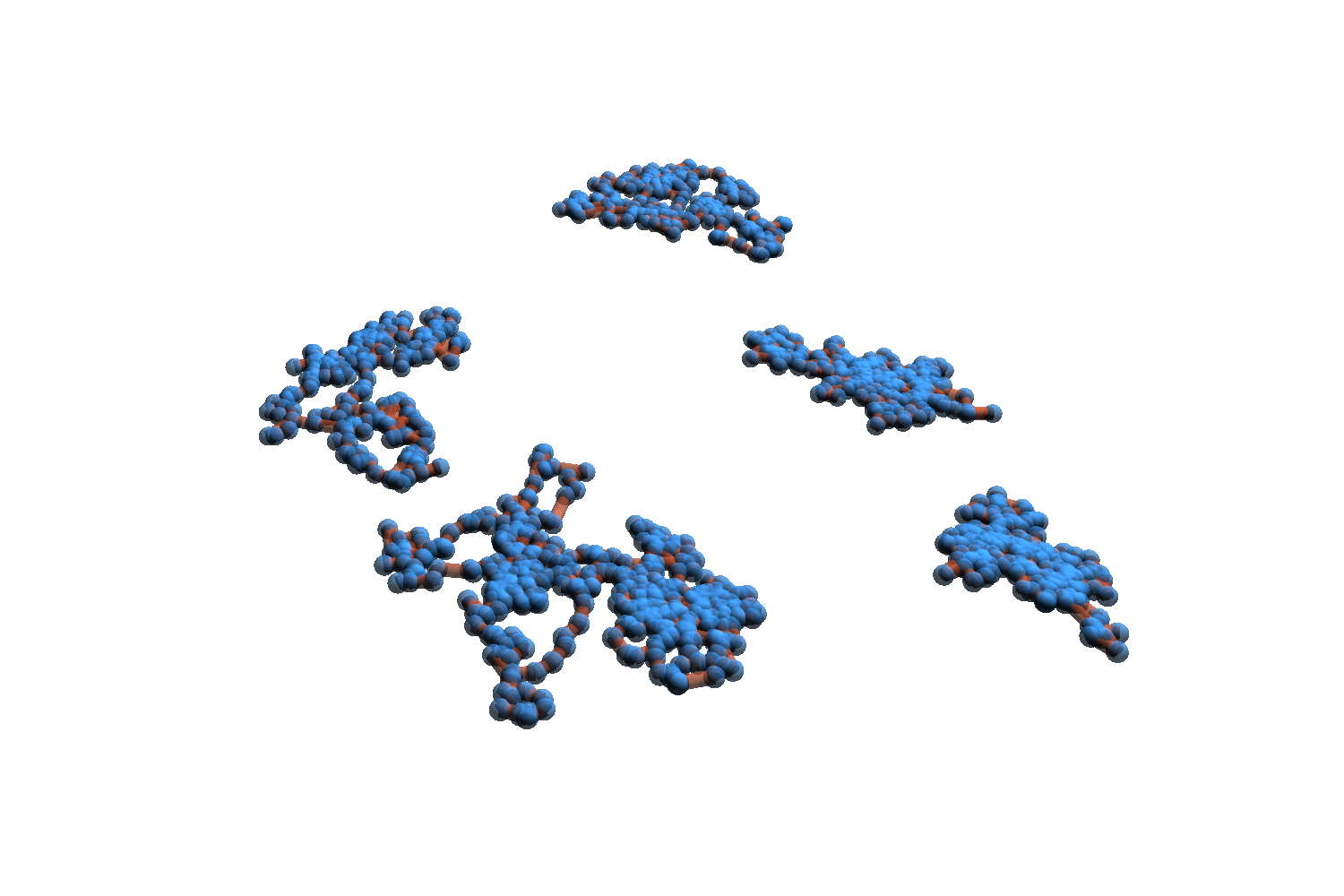} &
\includegraphics[width=0.3\textwidth]{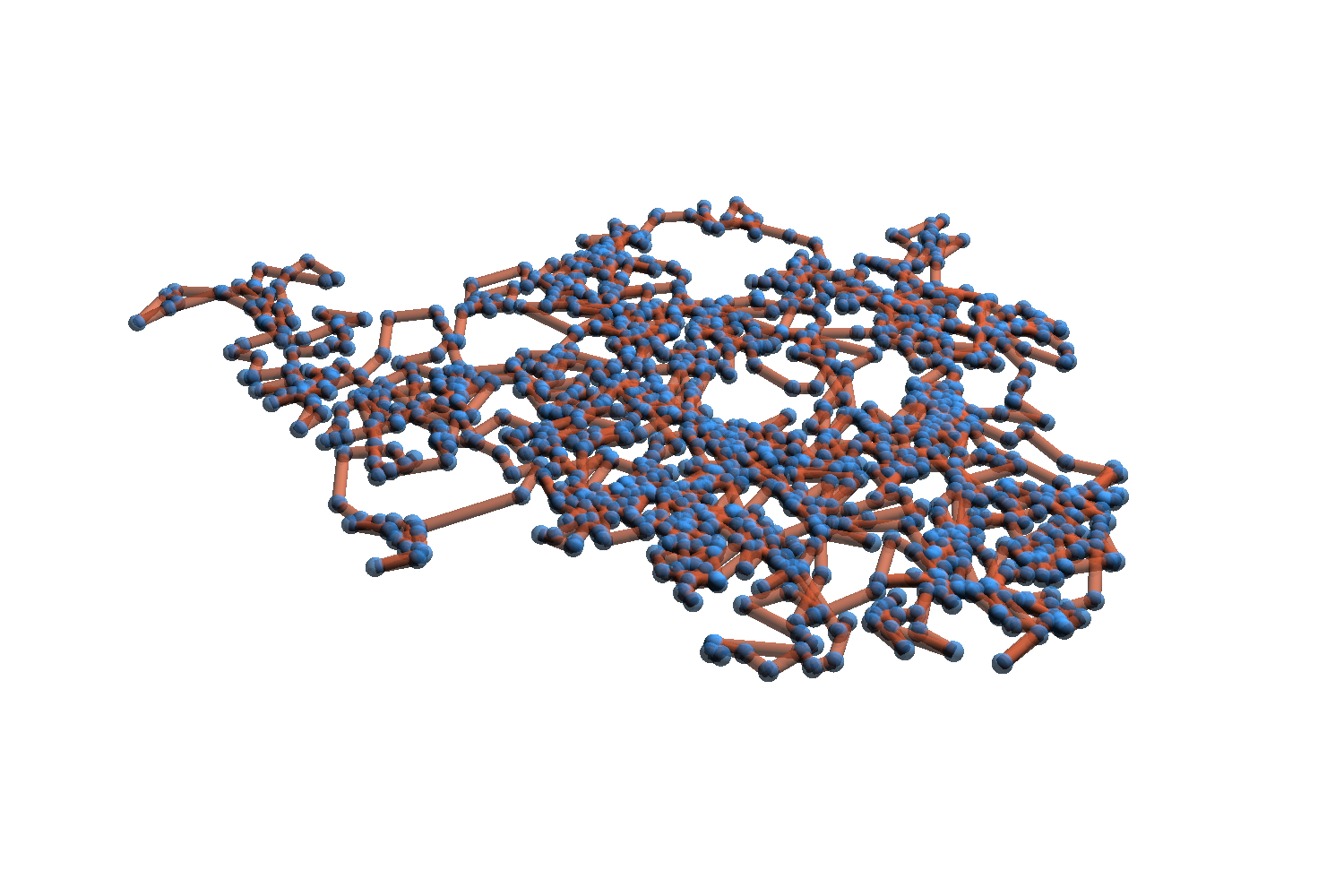} \\ \hline \large \begin{center}$2D$ Coulomb\end{center}&
\includegraphics[width=0.3\textwidth]{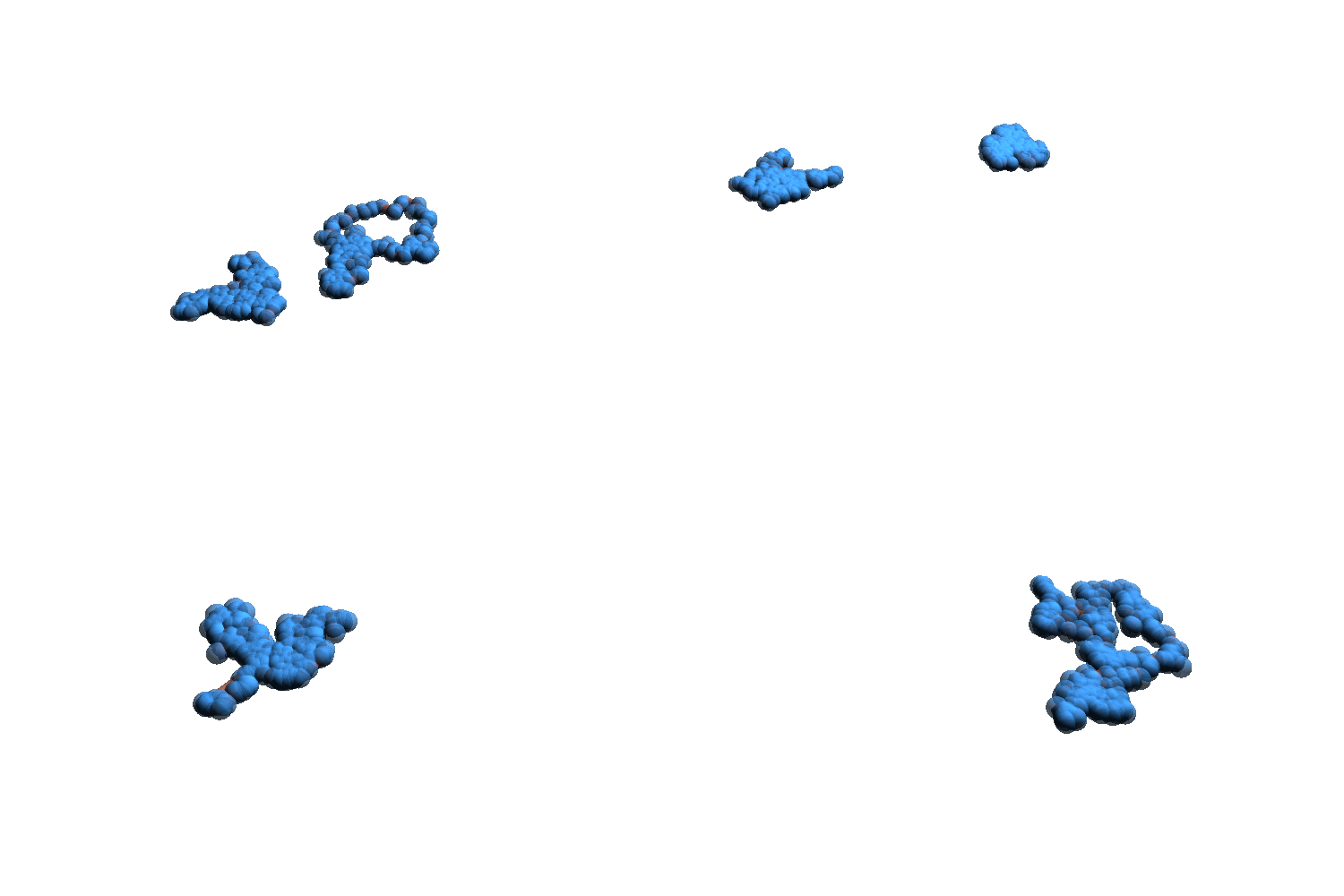} & 
\includegraphics[width=0.3\textwidth]{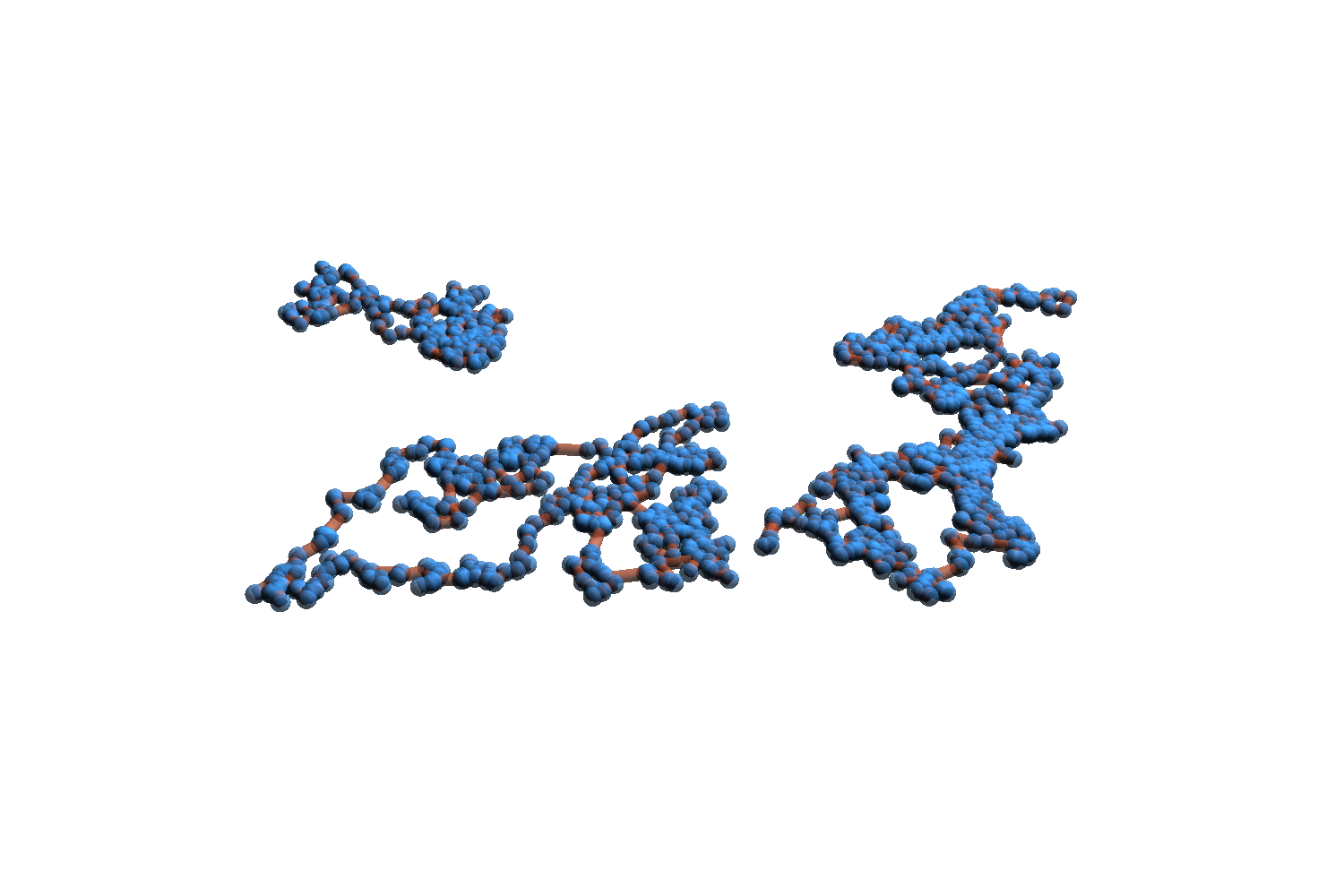} & 
\includegraphics[width=0.3\textwidth]{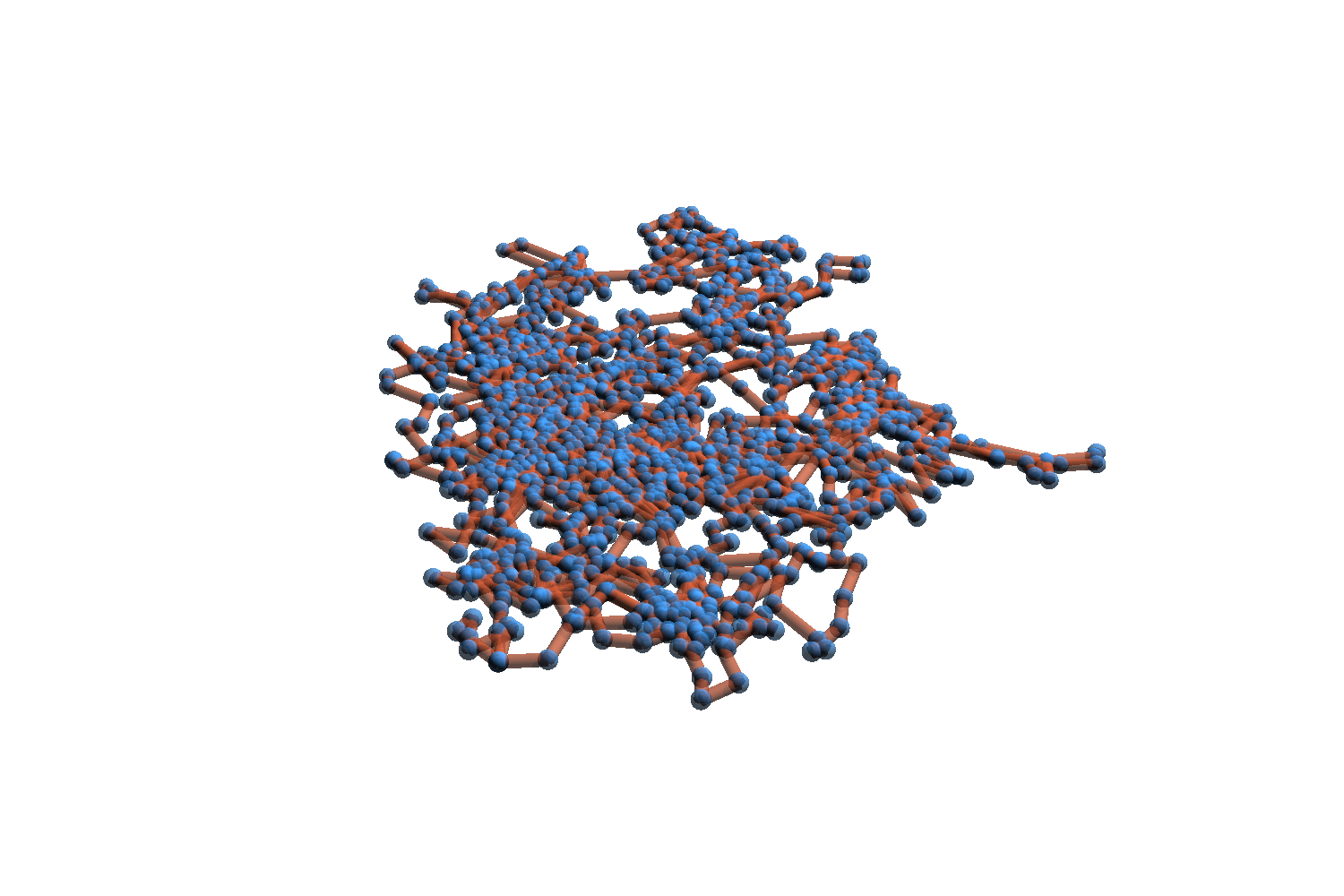} \\ \hline \large \begin{center}$3D$ Coulomb\end{center}&
\includegraphics[width=0.3\textwidth]{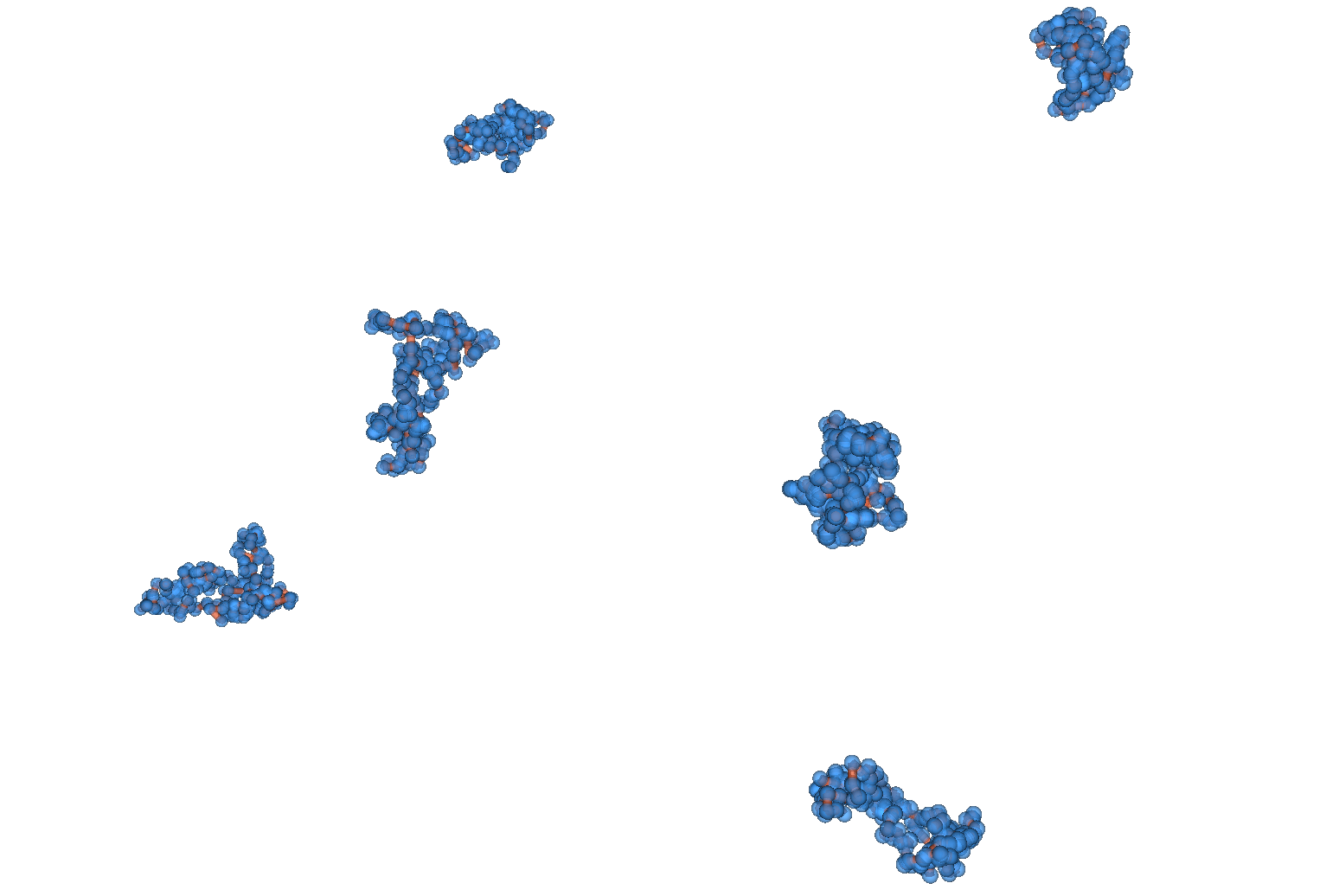} &
\includegraphics[width=0.3\textwidth]{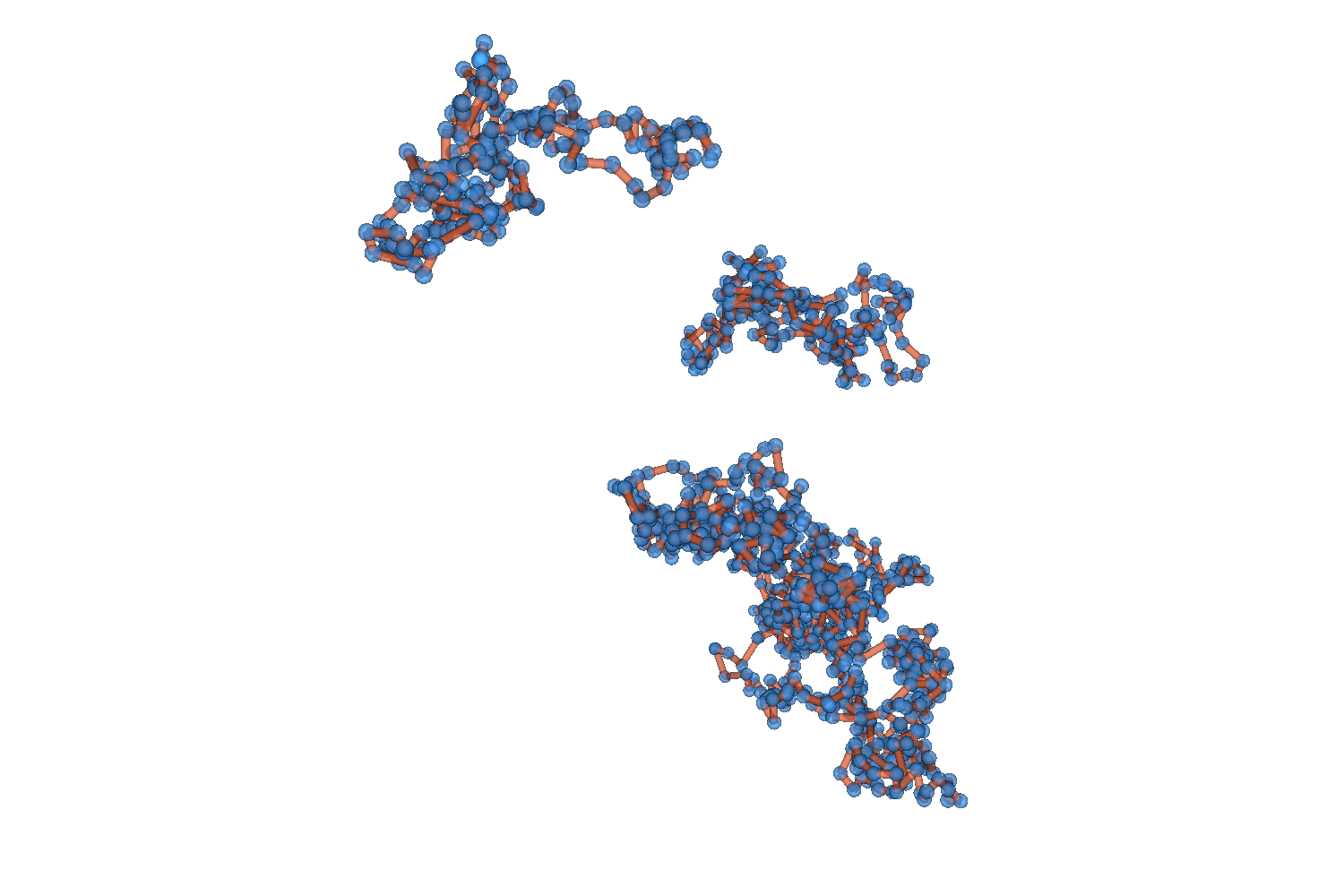} &
\includegraphics[width=0.3\textwidth]{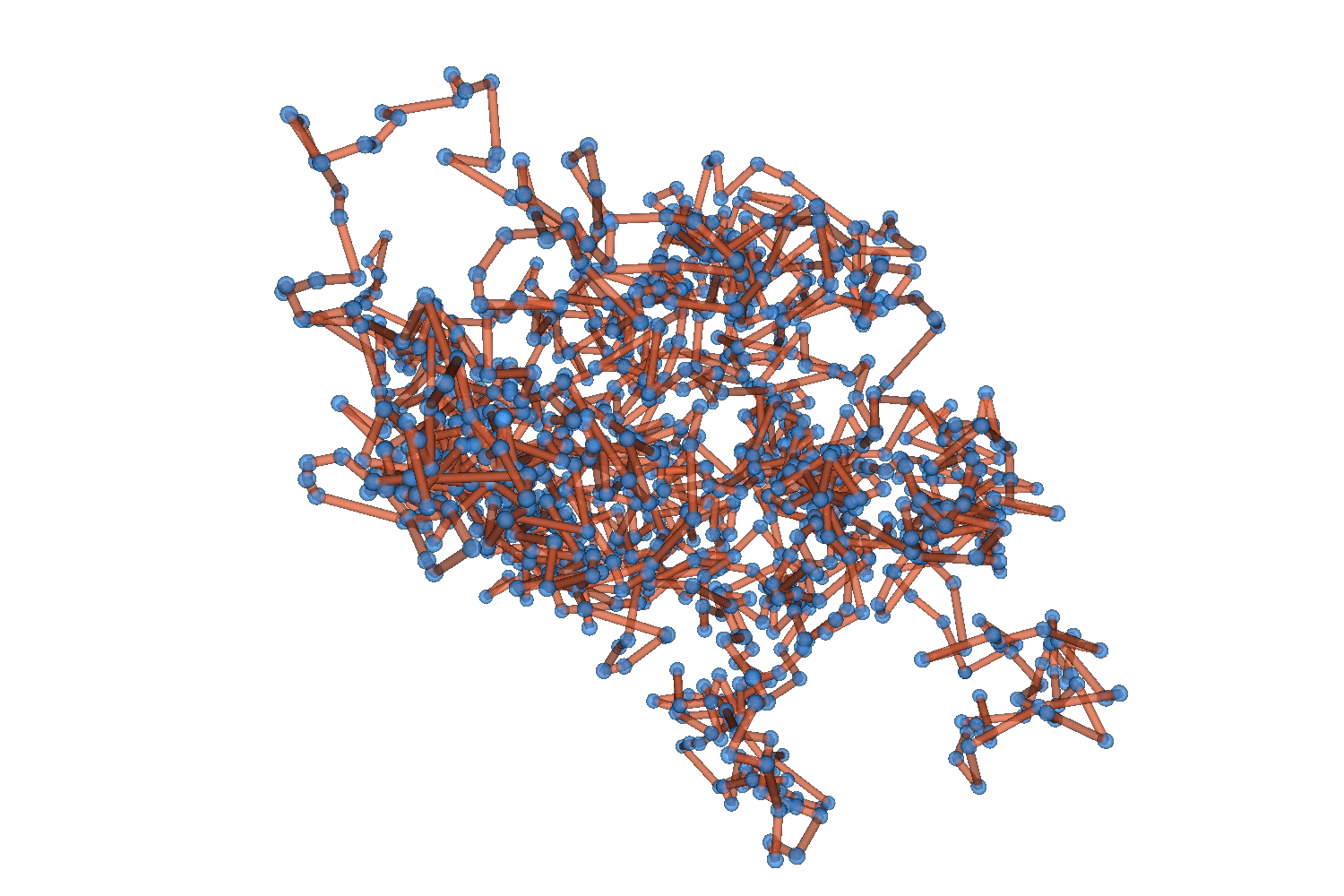} \\ \hline\hline
    \end{tabular}
    \caption{Snapshots from PIMC simulations of $N=6$ spin-polarized fermions with coupling strength $\lambda=0.5$ and $P=200$ imaginary-time slices. The left, center, and right column corresponds to the inverse temperature $\beta=0.3$, $\beta=1$, and $\beta=5$, respectively, and the top, center, and bottom row to a system of ultracold atoms with dipole interaction in $2D$, electrons with Coulomb interaction in a $2D$ harmonic trap, and electrons in a $3D$ harmonic trap. The corresponding full $\beta$-dependence of the average sign $S$ for all three systems in shown in Fig.~\ref{fig:big_comparison}.  }
    \label{tab:SNAPSHOT_BOX}
\end{table*}

\subsection{The uniform electron gas\label{sec:UEG}}

\begin{figure}
\includegraphics[width=0.41547\textwidth]{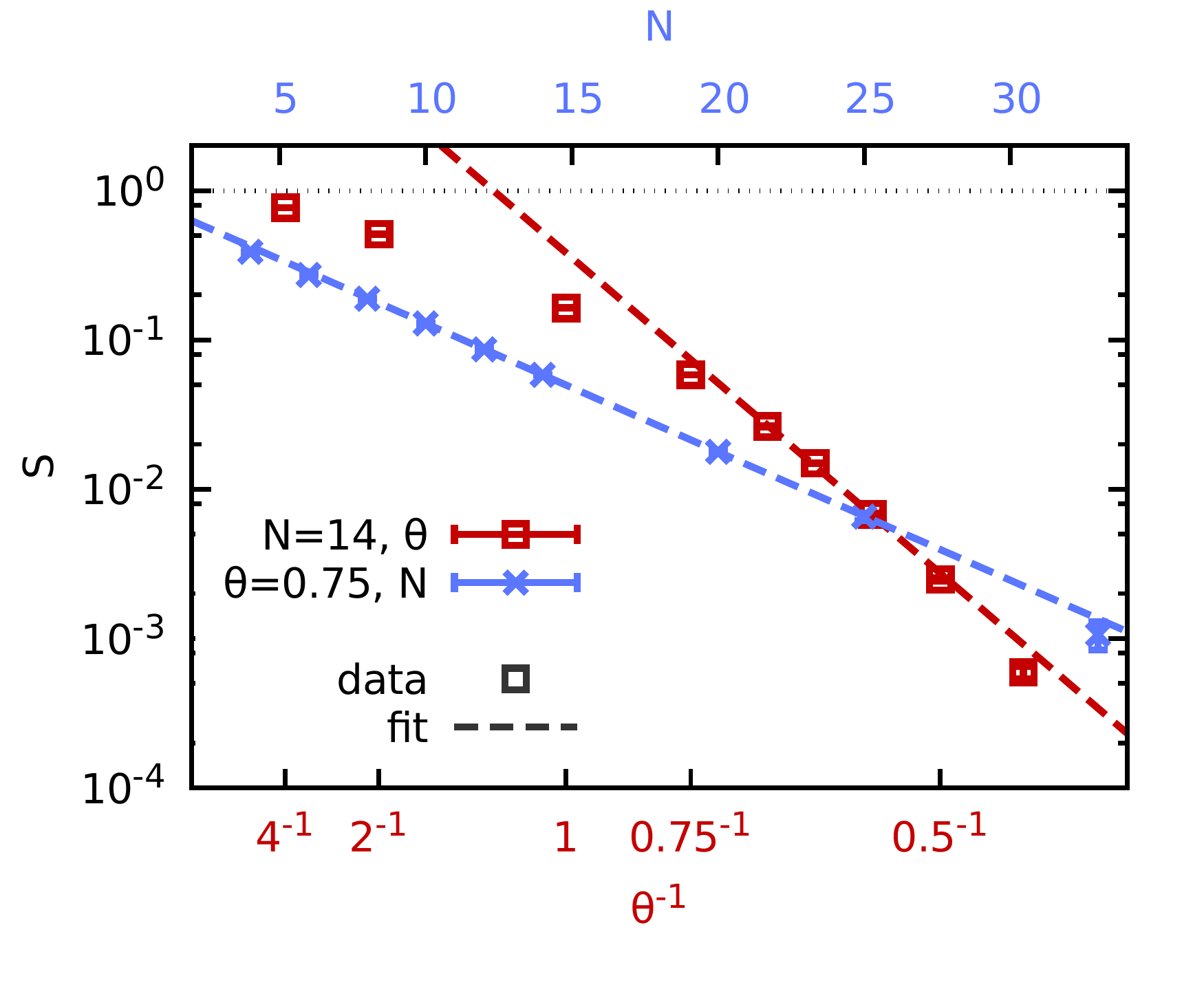}
\caption{\label{fig:YAKUB}
The fermion sign problem in PIMC simulation of the uniform electron gas at metallic density ($r_s=2$): Shown are the temperature dependence of the average sign $S$ for $N=14$ spin-polarized electrons (bottom $x$-axis) and the system-size dependence at $\theta=0.75$ (top $x$-axis). The symbols depict our PIMC data and the dashed lines correspond to exponential fits according to Eqs.~(\ref{eq:beta_fit}) and (\ref{eq:N_fit}).
}
\end{figure}

\begin{figure*}
\includegraphics[width=0.31547\textwidth]{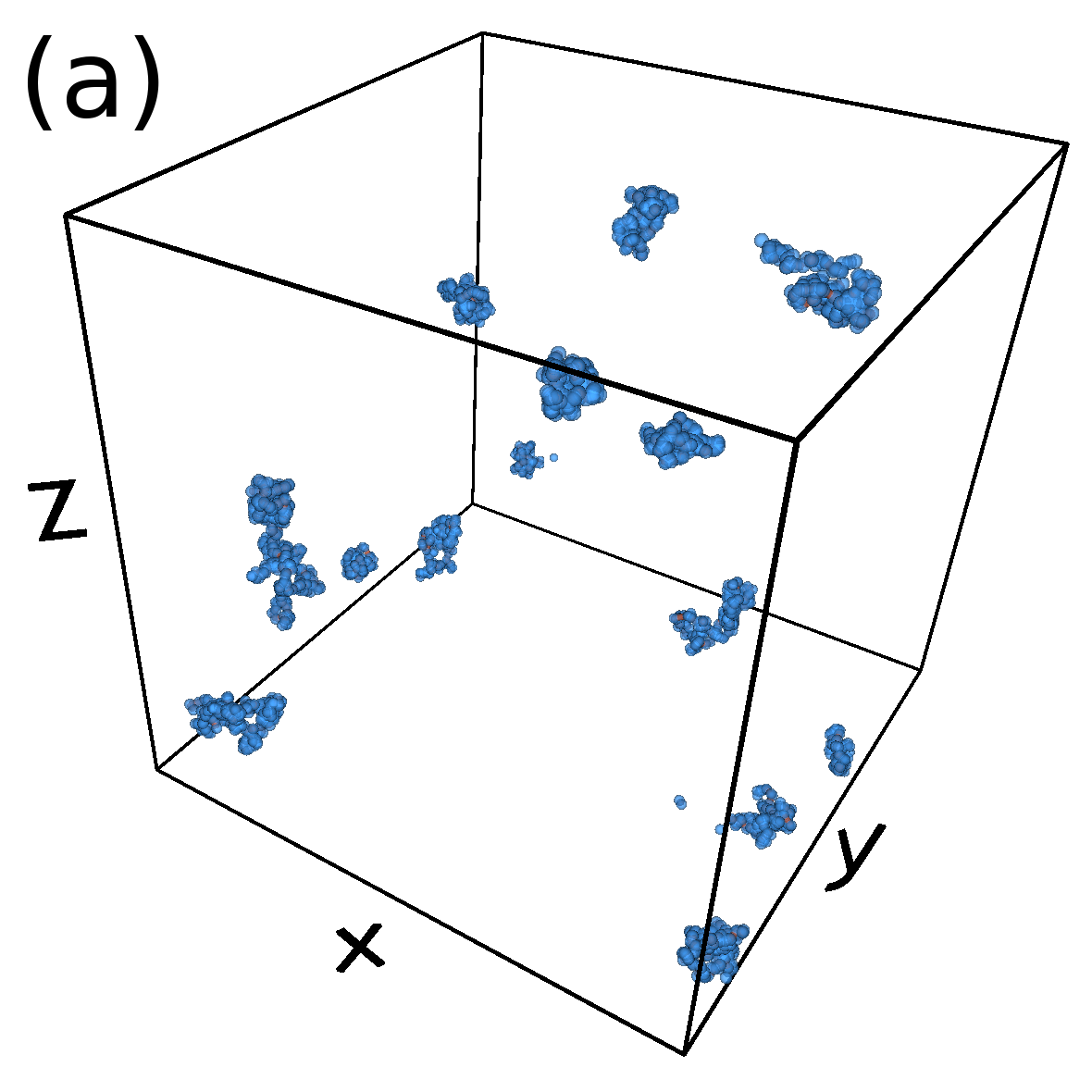}
\includegraphics[width=0.31547\textwidth]{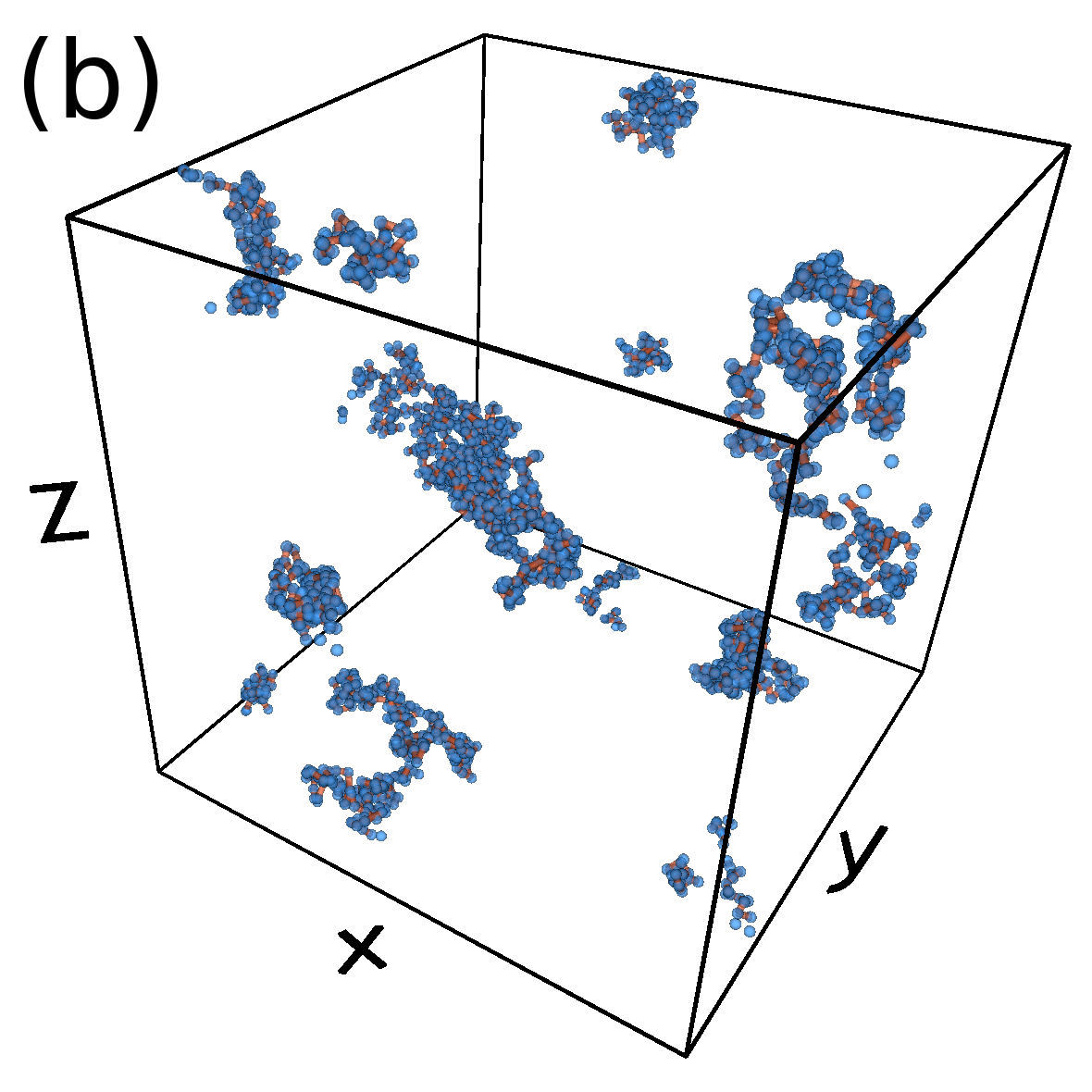}
\includegraphics[width=0.31547\textwidth]{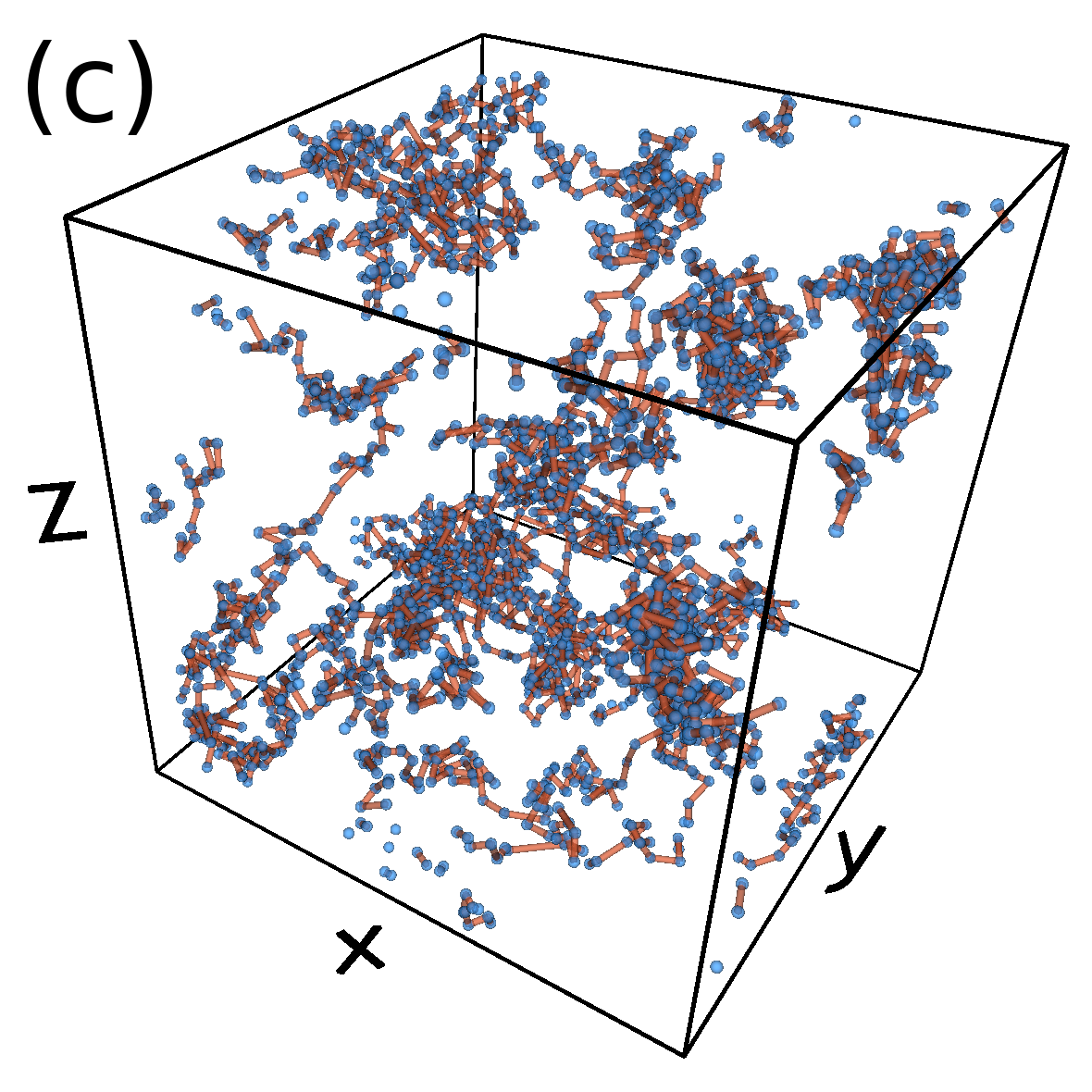}
\caption{\label{fig:YAKUB_BOXES}
Snapshots from a PIMC simulation of the spin-polarized UEG with $N=14$ electrons at $r_s=2$ with $P=200$ imaginary time slices. The temperature parameters are chosen as $\theta=4$ (a), $\theta=1$ (b), and $\theta=0.25$ (c).
}
\end{figure*}

Let us conclude this investigation of the fermion sign problem with a study of the uniform electron gas, which is shown in Fig.~\ref{fig:YAKUB}.
The top abscissa corresponds to the blue crosses, which depict the system-size dependence for the UEG at metallic density ($r_s=2$) in the warm dense matter regime~\cite{review}, $\theta=0.75$. Note that the density is kept constant by increasing the volume $V$
of the simulation cell when adding more electrons. Therefore, increasing $N$ only mitigates finite-size effects, but does not significantly affect the degree of quantum degeneracy. The dashed blue line depicts an exponential fit according to Eq.~(\ref{eq:N_fit}), which is in excellent agreement with our data points even for surprisingly small system size. Thus, the FSP does indeed constitute an exponential wall in terms of particle number $N$ for the UEG as predicted in Sec.~\ref{sec:FSP}, and the situation becomes only worse for the harmonic confinement. 
The red squares in the same plot show the decrease of $S$ with the inverse temperature for $N=14$ electrons at $r_s=2$ (bottom abscissa). Again, we find an exponential decay with $\beta\sim\theta^{-1}$, and simulations become unfeasible for $\theta\lesssim0.5$ even for such a comparatively small system size (a typical system size for the UEG are $N=33$ electrons~\cite{brown_ethan,groth,dornheim2}).


Lastly, we show snapshots from our PIMC simulation of the UEG in Fig.~\ref{fig:YAKUB_BOXES} for $N=14$ electrons at $r_s=2$ and $\theta=4$ (a), $\theta=1$ (b), and $\theta=0.25$ (c). At the highest temperature, the UEG resembles a semi-classical one-component plasma and the average sign $S\approx0.77$ is large. Panel (b) depicts a configuration from the interesting transition regime, where $\lambda_\beta$ becomes comparable to $\overline{r}$ and fermionic exchange-effects are important, but do not yet dominate. At $\theta=0.25$, the system is fully degenerate, the sign vanishes within the given statistical uncertainty, and standard PIMC simulations are unfeasible.

\section{Summary and discussion\label{sec:summary}}

In summary, we have presented a comprehensive, hands-on discussion of the fermion sign problem in path integral Monte Carlo simulations of degenerate Fermi systems. In particular, we have investigated the manifestation of the FSP regarding different parameters and have found the following: i) our PIMC data for the average sign $S$ are consistent with an exponential decrease in $S$ with increasing the inverse temperature $\beta$ for all considered system- and interaction-types; ii) while we do find an exponential decrease of $S$ with system size for the uniform electron gas, it decreases even faster for the case of the harmonic trap. This is explained by the increase in the radial density distribution $n(r)$ around the center of the trap, which leads to a higher degree of quantum degeneracy; iii) both the coupling strength $\lambda$ and the interaction-type have a large impact on the manifestation of the FSP. Firstly, there is a transition with decreasing $\lambda$ from the strongly coupled, quasi-classical regime (with $S\approx1$) to the respective noninteracting limit. Secondly, the short-range dipole interaction leads to a significantly less severe FSP compared to the long-range Coulomb repulsion, as the particles are effectively separated from each other within the PIMC simulation, which makes the formation of permutation-cycles less probable; iv) the increase of the dimensionality from $2D$ to $3D$ in the case of electrons in a harmonic confinement leads to a somewhat less severe FSP, although the scaling with $\beta$ is quite similar.

In addition, we have provided a practical example for the Monte-Carlo sampling of a fermionic observable, and have studied the probability distribution $P(VS/S)$ of a fermionic expectation value in the presence of the sign problem. In the case of a severe FSP, when the relative statistical uncertainty of $S$ is large, $P(VS/S)$ is given by a superposition of a Gaussian and a Lorentzian, which leads to a fat tail at large values and a divergence of the variance. For small errors in $S$, on the other hand, the distribution of the fermionic observable cannot be distinguished from a simple Gaussian, and the fermionic PIMC simulation is \emph{quasi-exact}.

We hope that our results---both regarding the manifestation of the FSP and the extensive data tables---will aid the future development of new simulation approaches for quantum degenerate, correlated Fermi systems. Moreover, the comparatively less severe manifestation of the FSP in the case of dipole interaction makes \textit{ab initio} PIMC simulations of ultracold dipolar atoms a promising project for future research, which could allow for unprecedented insights into, e.g., the emergence of pairing and fermionic superfluidity for a strongly correlated system.



 \section*{Acknowledgments}
 T.D.~wishes to thank Simon Groth and Jan Vorberger for their valuable feedback.

This work was partly funded by the Center of Advanced Systems Understanding (CASUS) which is financed by Germany's Federal Ministry of Education and Research (BMBF) and by the Saxon Ministry for Science and Art (SMWK) with tax funds on the basis of the budget approved by the Saxon State Parliament.

All calculations were carried out on the clusters \emph{hypnos} and \emph{hemera} at Helmholtz-Zentrum Dresden-Rossendorf (HZDR), and at the Norddeutscher Verbund f\"ur Hoch- und H\"ochstleistungsrechnen (HLRN) under grant shp00015.

\appendix

\renewcommand\thefigure{\thesection.\arabic{figure}}

\setcounter{figure}{0}  

\section{Convergence with imaginary-time slices $P$\label{sec:convergence}}

Since the operators for the kinetic and potential energy, $\hat K$ and $\hat V$, do not commute, the canonical density matrix within the PIMC formalism is typically decomposed using a suitable factorization scheme, see Refs.~\cite{brualla,sakkos} for a detailed discussion. In the present work, we restrict ourselves to the \textit{primitive factorization}
\begin{eqnarray}
\hat \rho_\epsilon = e^{-\epsilon(\hat K + \hat V)} = e^{-\epsilon\hat K} e^{-\epsilon \hat V} + \mathcal{O}\left(\epsilon^2\right)\quad , \label{eq:primitive}
\end{eqnarray}
with $\epsilon=\beta/P$ being the so-called imaginary-time step, which is justified by the Trotter formula~\cite{trotter}
\begin{eqnarray}
e^{-\beta(\hat K+ \hat V)} = \lim_{P\to\infty} \left( 
e^{-\epsilon \hat K} e^{-\epsilon\hat V}
\right)^P \quad .
\end{eqnarray}
Therefore, $P$ constitutes a convergence parameter within our simulations, and the factorization error in the expectation value of an observable $A$ due to Eq.~(\ref{eq:primitive}) scales as~\cite{brualla}
\begin{eqnarray}
A(P=\infty) = A(P) + \frac{\delta_A}{P^2} \quad . \label{eq:P_fit}
\end{eqnarray}
In the following, we will investigate the convergence with $P$ for a few representative cases.

\begin{figure}
\includegraphics[width=0.41\textwidth]{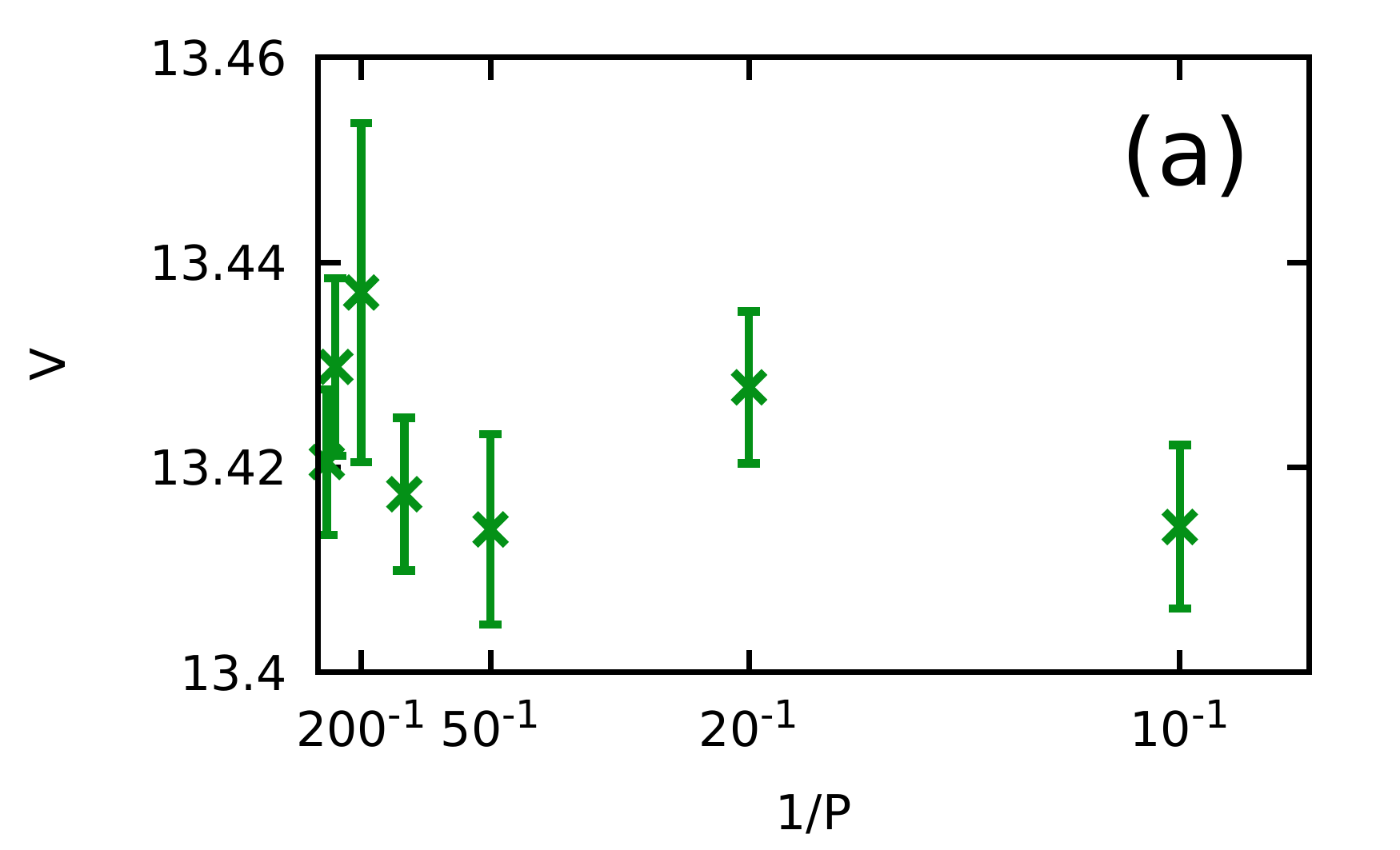}
\hspace*{0.3cm}\includegraphics[width=0.3875\textwidth]{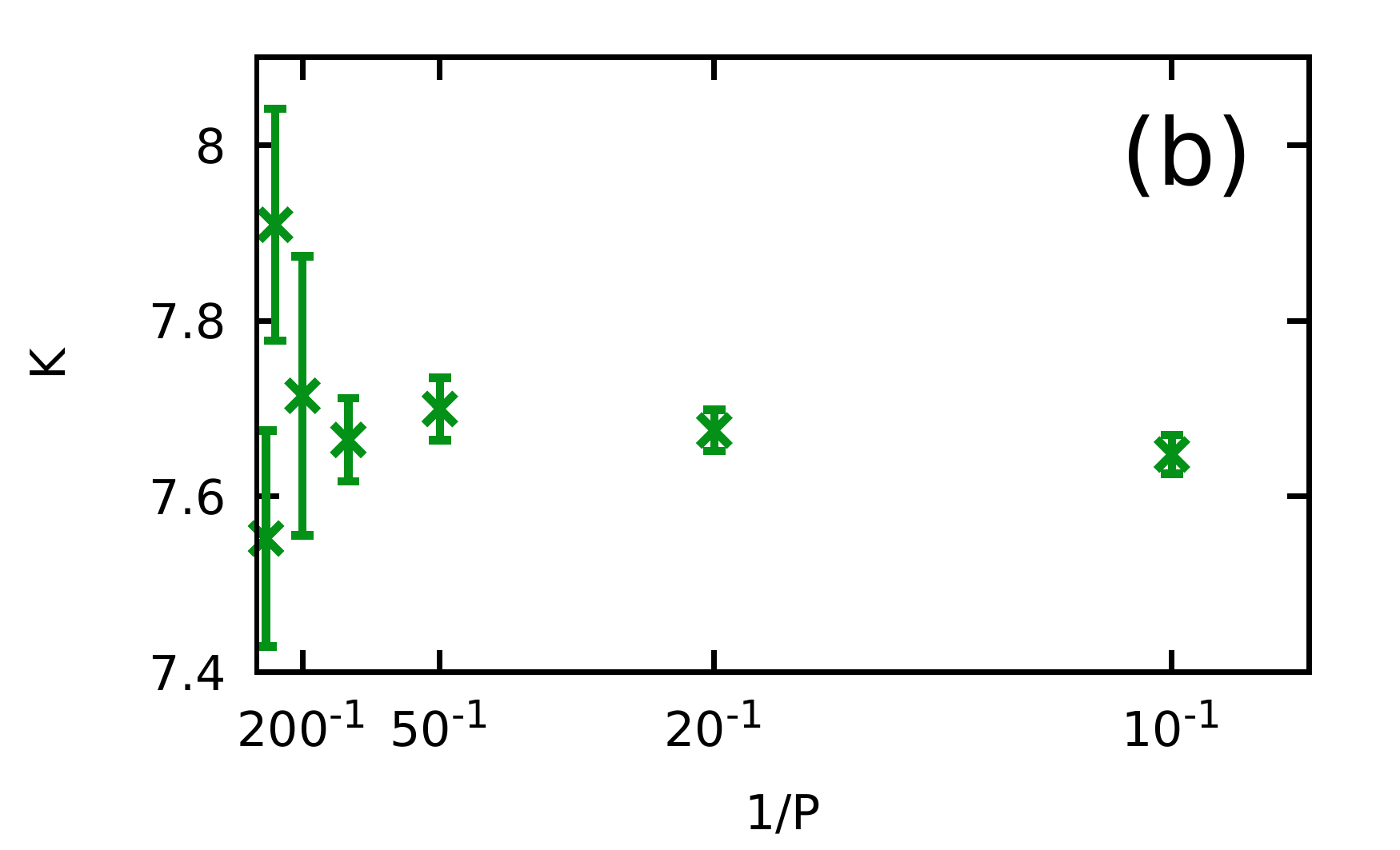}
\caption{\label{fig:Tconv}
Convergence with the number of primitive high-temperature factors $P$ for $N=6$ spin-polarized electrons at $\beta=1.3$ and $\lambda=0.5$ in a $2D$ harmonic trap. Panels (a) and (b) show PIMC results for the potential and kinetic energy, respectively.
}
\end{figure}

\begin{figure}
\includegraphics[width=0.41\textwidth]{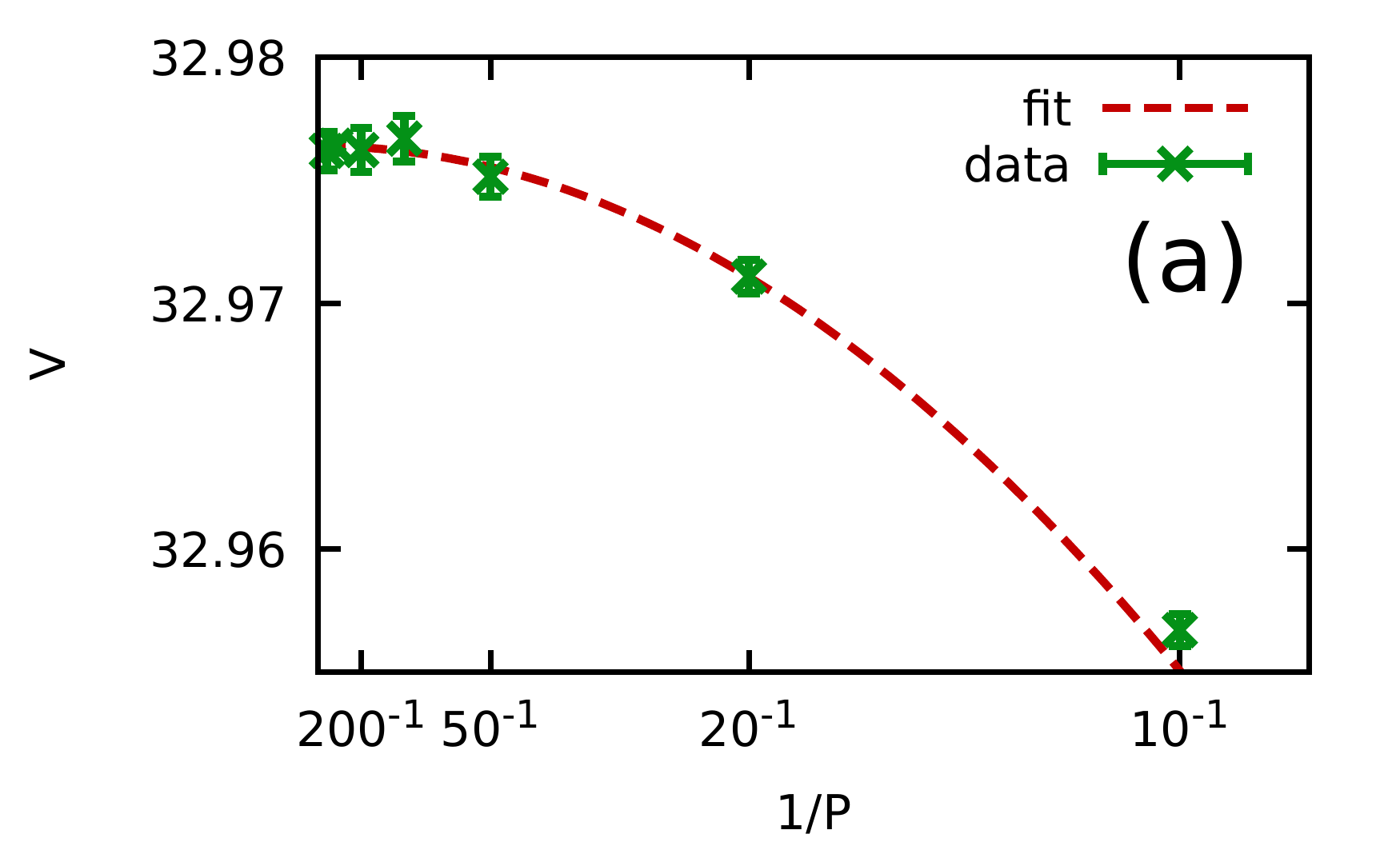}
\includegraphics[width=0.41\textwidth]{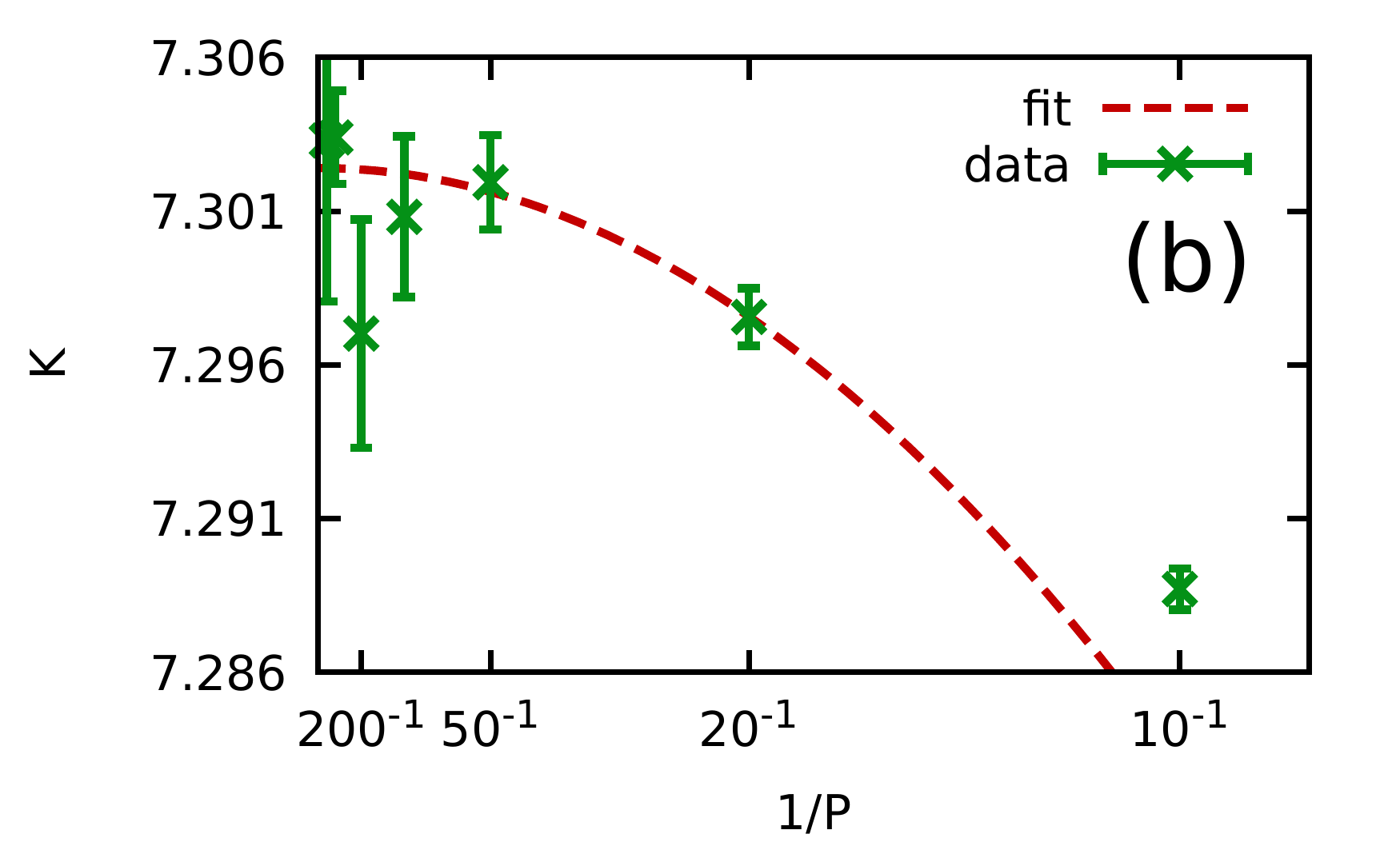}
\caption{\label{fig:l3conv}
Convergence with the number of primitive high-temperature factors $P$ for $N=6$ spin-polarized electrons at $\beta=1$ and $\lambda=3$ in a $2D$ harmonic trap. Panels (a) and (b) show PIMC results (green crosses) and a parabolic fit [cf.~Eq.~(\ref{eq:P_fit})] for the potential and kinetic energy, respectively.
}
\end{figure}

In Fig.~\ref{fig:Tconv}, we show the convergence of the potential energy (a) and kinetic energy (b) with $P$ for $N=6$ spin-polarized electrons with $\beta=1.3$ and $\lambda=0.5$ in a $2D$ harmonic trap, i.e., a data point from Tab.~\ref{tab:beta_dependence_N6_lambda0.5}. While 
this parameter combination does not constitute the lowest temperature considered in this work, it is still a good choice for this convergence study. For lower temperatures, the FSP leads to an exponentially increasing statistical uncertainty, and even large factorization errors cannot be resolved. Still, even at $\beta=1.3$ no factorization error can be resolved within the given error bars. For completeness, we mention that the increasing noise in $K$ towards large $P$ is a direct consequence of the utilized thermodynamic estimator, see Ref.~\cite{janke} for an extensive discussion.

A second degree of freedom worth considering is the interaction strength $\lambda$. In particular, one would expect that, for fixed temperature, the factorization error is most pronounced for intermediate coupling, as the system becomes effectively classical or noninteracting in the limits of $\lambda\gg 1$ and $\lambda\to0$, respectively.
To this end, we consider $N=6$ spin-polarized electrons in a $2D$ harmonic trap at $\lambda=3$ and $\beta=1$ (i.e., a parameter set from Tab.~\ref{tab:lambda_dependence_N6_beta1}) in Fig.~\ref{fig:l3conv}. The green crosses correspond to the PIMC results, and the dashed red lines to parabolic fits according to Eq.~(\ref{eq:P_fit}) for $20\leq P \leq 1000$. First and foremost, we do find a significant yet small dependence of our PIMC data on $P$, which is fully consistent with the expected factorization error. Moreover, we note that the data points for $P=100,200,500,$ and $1000$ cannot be distinguished within the given error bars, which means that the results for $P=200$ are indeed \textit{quasi-exact}.

\begin{figure}
\includegraphics[width=0.41\textwidth]{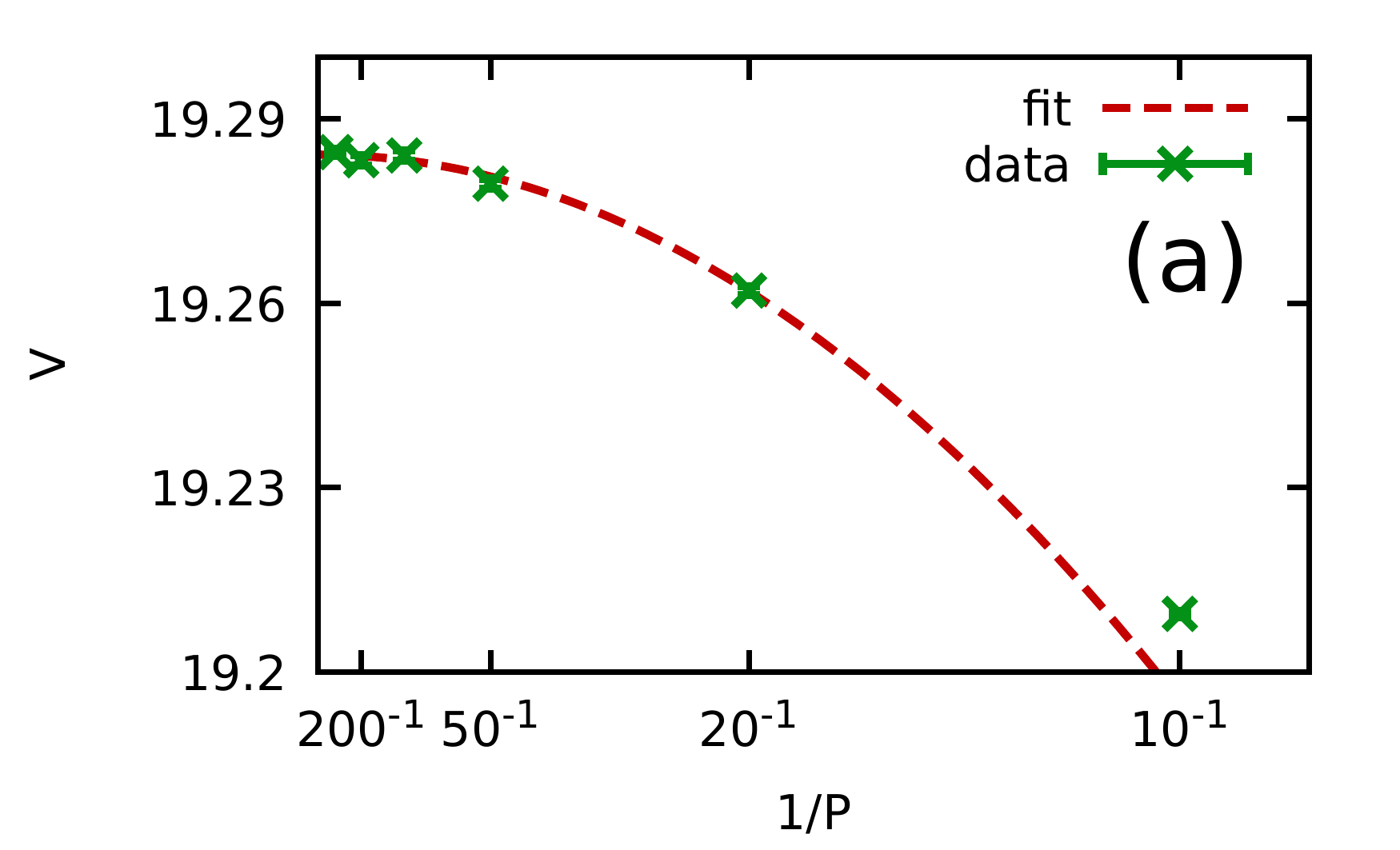}
\includegraphics[width=0.41\textwidth]{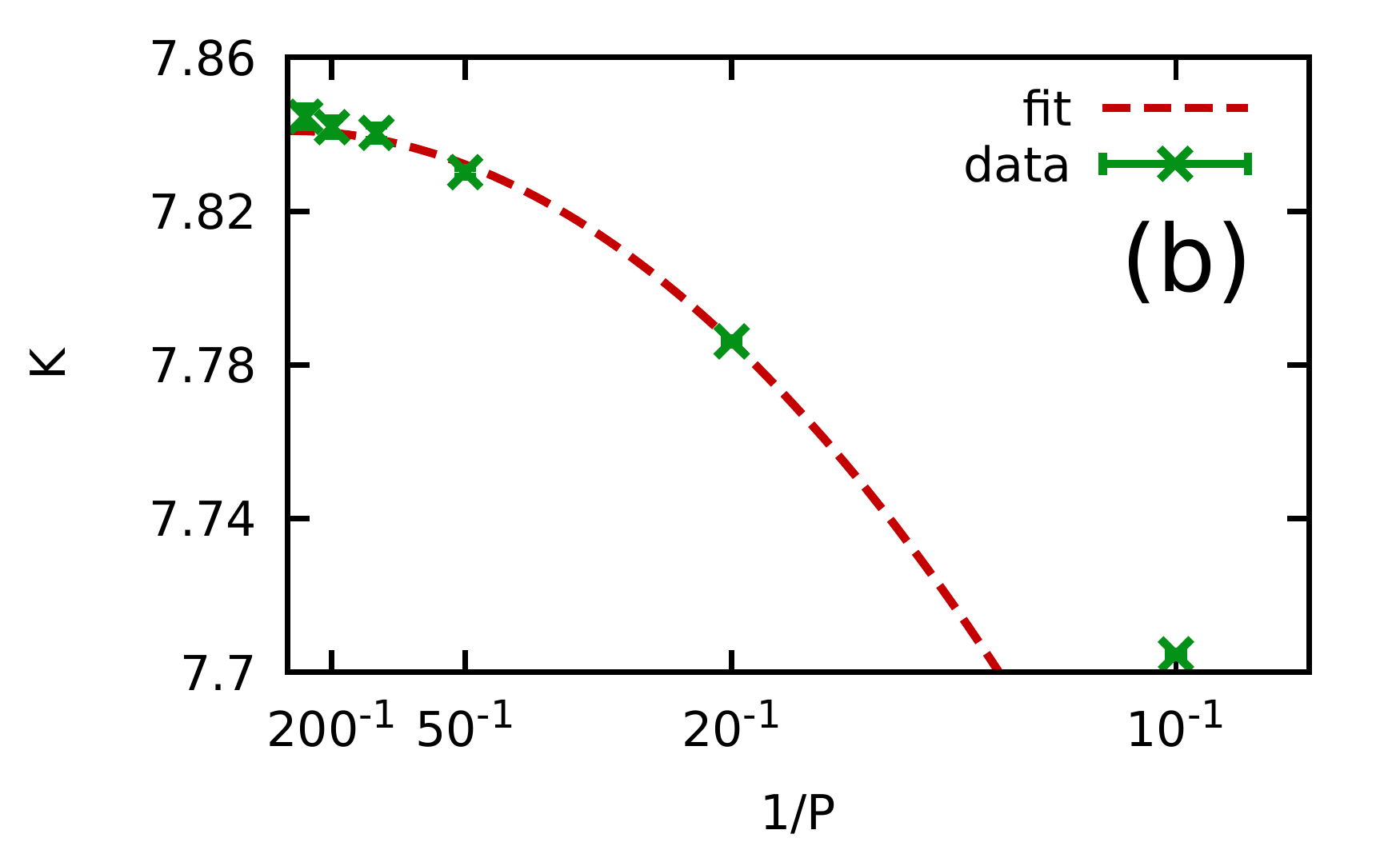}
\caption{\label{fig:l3conv_dip}
Convergence with the number of primitive high-temperature factors $P$ for $N=6$ ultracold atoms with dipole interaction at $\beta=1$ and $\lambda=3$ in a $2D$ harmonic trap. Panels (a) and (b) show PIMC results (green crosses) and a parabolic fit [cf.~Eq.~(\ref{eq:P_fit})] for the potential and kinetic energy, respectively.
}
\end{figure}

Lastly, we consider the case of dipole interaction (ultracold atoms) in Fig.~\ref{fig:l3conv_dip} for the same parameters as in Fig.~\ref{fig:l3conv}. Again, we find good agreement between the PIMC data and Eq.~(\ref{eq:P_fit}), and $P=200$ are converged within the given statistical uncertainty.


\newpage

\section*{References}
\begin{thebibliography}{10}


\bibitem{feynman} R.P.~Feynman and A.R.~Hibbs, Quantum Mechanics and Path Integrals, Dover Publications, New York (2010)


\bibitem{chandler} D.~Chandler and P.G.~Wolynes, Exploiting the isomorphism between quantum theory and classical statistical mechanics of polyatomic fluids, \href{https://aip.scitation.org/doi/abs/10.1063/1.441588}{\textit{J.~Chem.~Phys.}~\textbf{74}, 4078} (1981)


\bibitem{kleinert} H.~Kleinert, Path Integrals in Quantum Mechanics, Statistics, Polymer Physics, and Financial Markets, World Scientific, Singapore, 2009




\bibitem{berne3} M.F.~Herman, E.J.~Bruskin, and B.J.~Berne, On path integral Monte Carlo simulations, \href{https://aip.scitation.org/doi/abs/10.1063/1.442815}{\textit{J.~Chem.~Phys.}~\textbf{76}, 5150} (1982)

\bibitem{imada} M.~Takahashi and M.~Imada, Monte Carlo Calculation of Quantum Systems, \href{https://www.jstage.jst.go.jp/article/jpsj1946/53/3/53_3_963/_article/-char/ja/}{\textit{J.~Phys.~Soc.~Jpn.}~\textbf{53}, 963} (1984)

\bibitem{pollock} E.L.~Pollock and D.M.~Ceperley, Simulation of quantum many-body systems by path-integral methods, \href{https://journals.aps.org/prb/abstract/10.1103/PhysRevB.30.2555}{\textit{Phys.~Rev.~B} \textbf{30}, 2555} (1984)

\bibitem{cep} D.M.~Ceperley, Path integrals in the theory of condensed helium, \href{http://link.aps.org/doi/10.1103/RevModPhys.67.279}{\textit{Rev. Mod. Phys.} \textbf{67}, 279-355} (1995)


\bibitem{review} T.~Dornheim, S.~Groth, and M.~Bonitz, The uniform electron gas at warm dense matter conditions,
\href{https://www.sciencedirect.com/science/article/abs/pii/S0370157318300516}{\textit{Phys.~Reports}~\textbf{744}, 1--86} (2018)




\bibitem{metropolis} N.~Metropolis, A.W.~Rosenbluth, M.N.~Rosenbluth, A.H.~Teller, and E.~Teller, Equation of State Calculations by Fast Computing Machines, \href{https://aip.scitation.org/doi/abs/10.1063/1.1699114}{\textit{J.~Chem.~Phys.}~\textbf{21}, 1087} (1953)


\bibitem{curse} A.~Hinrichs, E.~Novak, M.~Ullrich and H.~Wozniakowski, The curse of dimensionality for numerical integration of smooth functions, \href{http://www.ams.org/journals/mcom/2014-83-290/S0025-5718-2014-02855-X/home.html}{\textit{Math.~Comp.}~\textbf{83}, 2853-2863} (2014)


\bibitem{filinov_chapter} A.~Filinov and M.~Bonitz, Classical and Quantum Monte Carlo, \textit{Introduction to Computational Methods in Many Body Physic}, M.~Bonitz and D.~Semkat (Eds.), Rinton Press, Paramus (New Jersey) (2006)

\bibitem{liu_monte_carlo} J.~Liu, Monte Carlo Strategies in Scientific Computing, Springer, New York (2013)


\bibitem{pimc_original} L.D.~Fosdick and H.F.~Jordan, Path-Integral Calculation of the Two-Particle Slater Sum for He$^4$, \href{https://journals.aps.org/pr/abstract/10.1103/PhysRev.143.58}{\textit{Phys.~Rev.}~\textbf{143}, 58} (1966)

\bibitem{pimc_original2} H.F.~Jordan and L.D.~Fosdick, Three-Particle Effects in the Pair Distribution Function for He$^4$ Gas, \href{https://journals.aps.org/pr/abstract/10.1103/PhysRev.171.128}{\textit{Phys.~Rev.}~\textbf{171}, 128} (1968)









\bibitem{ceperley_superfluid} E.L.~Pollock and D.M.~Ceperley, Path-integral computation of superfluid densities, \href{https://journals.aps.org/prb/abstract/10.1103/PhysRevB.36.8343}{\textit{Phys.~Rev.~B}~\textbf{36}, 8343} (1987)


\bibitem{sindzingre} P.~Sindzingre, M.L.~Klein, and D.M.~Ceperley, Path-integral Monte Carlo study of low-temperature $^4$He clusters, \href{https://journals.aps.org/prl/abstract/10.1103/PhysRevLett.63.1601}{\textit{Phys.~Rev.~Lett.}~\textbf{63}, 1601} (1989)




\bibitem{kwon} Y.~Kwon and K.B.~Whaley, Atomic-Scale Quantum Solvation Structure in Superfluid Helium-4 Clusters, \href{https://journals.aps.org/prl/abstract/10.1103/PhysRevLett.83.4108}{\textit{Phys.~Rev.~Lett.}~\textbf{83}, 4108} (1999)



\bibitem{dornheim_superfluid} T.~Dornheim, A.~Filinov, and M.~Bonitz, Superfluidity of strongly correlated bosons in two- and three-dimensional traps, \href{https://journals.aps.org/prb/abstract/10.1103/PhysRevB.91.054503}{\textit{Phys.~Rev.~B}~\textbf{91}, 054503} (2015)










\bibitem{BEC1} S.~Pilati, S.~Giorgini, M.~Modugno, and N.~Prokof'ev, Dilute Bose gas with correlated disorder: a path integral Monte Carlo study, \href{https://iopscience.iop.org/article/10.1088/1367-2630/12/7/073003/meta}{\textit{New J.~Phys.}~\textbf{12}, 073003} (2010)

\bibitem{BEC2} H.~Saito, Path-Integral Monte Carlo Study on a Droplet of a Dipolar Bose–Einstein Condensate Stabilized by Quantum Fluctuation, \href{https://journals.jps.jp/doi/full/10.7566/JPSJ.85.053001}{\textit{J.~Phys.~Soc.~Jpn.}~\textbf{85}, 053001} (2016)







\bibitem{jones_crystal} M.D.~Jones and D.M.~Ceperley, Crystallization of the one-component plasma at finite temperature, \href{https://journals.aps.org/prl/abstract/10.1103/PhysRevLett.76.4572}{\textit{Phys.~Rev.~Lett.}~\textbf{76}, 4572} (1996)


\bibitem{filinov_PRL} A.V.~Filinov, M.~Bonitz, and Yu.E.~Lozovik, Wigner Crystallization in Mesoscopic 2D Electron Systems, \href{https://journals.aps.org/prl/abstract/10.1103/PhysRevLett.86.3851}{\textit{Phys.~Rev.~Lett.}~\textbf{86}, 3851} (2001)


\bibitem{clark_casula} B.K.~Clark, M.~Casula, and D.M.~Ceperley, Hexatic and Mesoscopic Phases in a 2D Quantum Coulomb System, \href{https://journals.aps.org/prl/abstract/10.1103/PhysRevLett.103.055701}{\textit{Phys.~Rev.~Lett.}~\textbf{103}, 055701} (2009)







\bibitem{berne1} D.~Thirumalai and B.J.~Berne, On the calculation of time correlation functions in quantum systems: Path integral techniques, \href{https://aip.scitation.org/doi/abs/10.1063/1.445597}{\textit{J.~Chem.~Phys.}~\textbf{79}, 5029} (1983)


\bibitem{berne2} E.~Gallicchio and B.J.~Berne, The absorption spectrum of the solvated electron in fluid helium by maximum entropy inversion of imaginary time correlation functions from path integral Monte Carlo simulations, \href{https://aip.scitation.org/doi/abs/10.1063/1.467892}{\textit{J.~Chem.~Phys.}~\textbf{101}, 9909} (1994)


\bibitem{dynamic_folgepaper} S.~Groth, T.~Dornheim, and J.~Vorberger, \textit{Ab Initio} Path Integral Monte Carlo Approach to the Static and Dynamic Density Response of the Uniform Electron Gas, \href{https://link.aps.org/doi/10.1103/PhysRevB.99.235122}{\textit{Phys.~Rev.~B} \textbf{99}, 235122} (2019)





\bibitem{jarrell} M.~Jarrell and J.E.~Gubernatis, Bayesian inference and the analytic continuation of imaginary-time quantum Monte Carlo data, \href{https://www.sciencedirect.com/science/article/abs/pii/0370157395000747}{\textit{Phys.~Reports}~\textbf{269}, 133-195} (1996)


\bibitem{schoett} J.~Sch\"ott, E.G.C.P.~van Loon, I.L.M.~Locht, M.I.~Katsnelson, and I.~Di Marco, Comparison between methods of analytical continuation for bosonic functions, \href{https://journals.aps.org/prb/abstract/10.1103/PhysRevB.94.245140}{\textit{Phys.~Rev.~B}~\textbf{94}, 245140} (2016)

















\bibitem{dornheim_dynamic} T.~Dornheim, S.~Groth, J.~Vorberger, and M.~Bonitz, \textit{Ab initio} Path Integral Monte Carlo Results for the Dynamic Structure Factor of Correlated Electrons: From the Electron Liquid to Warm Dense Matter, \href{https://journals.aps.org/prl/abstract/10.1103/PhysRevLett.121.255001}{\textit{Phys.~Rev.~Lett.}~\textbf{121}, 255001} (2018)



\bibitem{vitali} E.~Vitali, M.~Rossi, L.~Reatto, and D.E.~Galli, \textit{Ab initio} low-energy dynamics of superfluid and solid $^4$He, \href{https://journals.aps.org/prb/abstract/10.1103/PhysRevB.82.174510}{\textit{Phys.~Rev.~B}~\textbf{82}, 174510} (2010)


\bibitem{supersolid} S.~Saccani, S.~Moroni, and M.~Boninsegni, 
Excitation Spectrum of a Supersolid, \href{https://journals.aps.org/prl/abstract/10.1103/PhysRevLett.108.175301}{\textit{Phys.~Rev.~Lett.}~\textbf{108}, 175301} (2012)












\bibitem{boninsegni1} M.~Boninsegni, N.V.~Prokofev, and B.V.~Svistunov, Worm algorithm and diagrammatic Monte Carlo: A new approach to continuous-space path integral Monte Carlo simulations, \href{https://journals.aps.org/pre/abstract/10.1103/PhysRevE.74.036701}{\textit{Phys.~Rev.~E}~\textbf{74}, 036701} (2006)

\bibitem{boninsegni2} M.~Boninsegni, N.V.~Prokofev, and B.V.~Svistunov, Worm Algorithm for Continuous-Space Path Integral Monte Carlo Simulations, \href{https://journals.aps.org/prl/abstract/10.1103/PhysRevLett.96.070601}{\textit{Phys.~Rev.~Lett.}~\textbf{96}, 070601} (2006)

















\bibitem{ceperley_fermions} D.M.~Ceperley, Path Integral Monte Carlo Methods for Fermions, \textit{Monte Carlo and Molecular Dynamics of Condensed Matter Systems}, K.~Binder and G.~Ciccotti (Eds.), Bologna (Italy) (1996)













\bibitem{loh} E.Y.~Loh, J.E.~Gubernatis, R.T.~Scalettar, S.R.~White, D.J.~Scalapino and R.L.~Sugar, Sign problem in the numerical simulation of many-electron systems, \href{http://link.aps.org/doi/10.1103/PhysRevB.41.9301}{\textit{Phys. Rev. B} \textbf{41}, 9301-9307} (1990)


\bibitem{troyer} M.~Troyer and U.J.~Wiese, Computational Complexity and Fundamental Limitations to Fermionic Quantum Monte Carlo Simulations, \href{http://link.aps.org/doi/10.1103/PhysRevLett.94.170201}{\textit{Phys. Rev. Lett.} \textbf{94}, 170201} (2005)


\bibitem{lyubartsev} A.P.~Lyubartsev, 
Simulation of excited states and the sign problem in the path integral Monte Carlo method, \href{https://iopscience.iop.org/article/10.1088/0305-4470/38/30/003/meta}{\textit{J.~Phys.~A: Math.~Gen.}~\textbf{38}, 6659-6674} (2005)

\bibitem{vozn} M.A.~Voznesenskiy, P.N.~Vorontsov-Velyaminov, and A.P.~Lyubartsev, Path-integral–expanded-ensemble Monte Carlo method in treatment of the sign problem for fermions, \href{https://journals.aps.org/pre/abstract/10.1103/PhysRevE.80.066702}{\textit{Phys.~Rev.~E}~\textbf{80}, 066702} (2009)











\bibitem{fsp_note} For completeness, we note that quantum Monte Carlo simulations without a sign problem are possible for a few special cases, see, e.g., Refs.~\cite{ch1,ch2,ch3}.

\bibitem{ch1} Z.-X.~Li, Y.~F. Jiang, and H.~Yao, Solving the fermion sign problem in quantum Monte Carlo simulations by Majorana representation, \href{https://journals.aps.org/prb/abstract/10.1103/PhysRevB.91.241117}{\textit{Phys.~Rev.~B} \textbf{91}, 241117(R)} (2015)

\bibitem{ch2} Z.-X.~Li, Y.~F. Jiang, and H.~Yao, Majorana-Time-Reversal Symmetries: A Fundamental Principle for Sign-Problem-Free Quantum Monte Carlo Simulations, \href{ https://journals.aps.org/prl/abstract/10.1103/PhysRevLett.117.267002}{ \textit{Phys.~Rev.~Lett.}~\textbf{117}, 267002} (2016)

\bibitem{ch3} Z.-X.~Li and H.~Yao, Sign-Problem-Free Fermionic Quantum Monte Carlo: Developments and Applications, \href{ https://www.annualreviews.org/doi/full/10.1146/annurev-conmatphys-033117-054307}{ \textit{Ann.~Rev.~Cond.~Mat.~Phys.}~\textbf{10}, 337-356} (2019)










\bibitem{fermi1} Q.~Chen, J.~Stajic, S.~Tan, and K.~Levin, BCS–BEC crossover: From high temperature superconductors to ultracold superfluids, \href{https://www.sciencedirect.com/science/article/abs/pii/S0370157305001067}{\textit{Phys.~Reports}~\textbf{412}, 1-88} (2005)




\bibitem{fermi2} A.~Bulgac, J.E.~Drut, and P.~Magierski, Quantum Monte Carlo simulations of the BCS-BEC crossover at finite temperature, \href{https://journals.aps.org/pra/abstract/10.1103/PhysRevA.78.023625}{\textit{Phys.~Rev.~A}~\textbf{78}, 023625} (2008)

\bibitem{fermi3} E.~Burovski, E.~Kozik, N.~Prokof'ev, B.~Svistunov, and M.~Troyer, Critical Temperature Curve in BEC-BCS Crossover, \href{https://journals.aps.org/prl/abstract/10.1103/PhysRevLett.101.090402}{\textit{Phys.~Rev.~Lett.}~\textbf{101}, 090402} (2008)






\bibitem{wigner_molecule1} S.M.~Reimann, M.~Koskinen, and M.~Manninen, Formation of Wigner molecules in small quantum dots, \href{https://journals.aps.org/prb/abstract/10.1103/PhysRevB.62.8108}{\textit{Phys.~Rev.~B}~\textbf{62}, 8108} (2000)

\bibitem{wigner_molecule2} B.~Reusch, W.~H\"ausler, and H.~Grabert, Wigner molecules in quantum dots, \href{https://journals.aps.org/prb/abstract/10.1103/PhysRevB.63.113313}{\textit{Phys.~Rev.~B}~\textbf{63}, 113313} (2001)

\bibitem{wigner_molecule3} E.~R\"as\"anen, H.~Saarikoski, M.J.~Puska, and R.M.~Nieminen, Wigner molecules in polygonal quantum dots: A density-functional study, \href{https://journals.aps.org/prb/abstract/10.1103/PhysRevB.67.035326}{\textit{Phys.~Rev.~B}~\textbf{87}, 035326} (2003)




\bibitem{ross} M.~Ross, Matter under extreme conditions of temperature and pressure, \href{https://doi.org/10.1088\%2F0034-4885\%2F48\%2F1\%2F001}{\textit{Rep.~Prog.~Phys.}~\textbf{48}, 1--52} (1985)

\bibitem{koenig} M.~Koenig \textit{et al.}, Progress in the study of warm dense matter, \href{https://doi.org/10.1088\%2F0741-3335\%2F47\%2F12b\%2Fs31}{\textit{Plasma Phys.~Control.~Fusion} \textbf{47}, B441--B449} (2005)



\bibitem{fortov_review} V.E.~Fortov, Extreme states of matter on Earth and in space, \href{https://www.turpion.org/php/paper.phtml?journal_id=pu&paper_id=6821}{\textit{Phys.-Usp.}~\textbf{52}, 615--647} (2009)






\bibitem{militzer1} J.~Vorberger, I.~Tamblyn, B.~Militzer, and S.A.~Bonev, Hydrogen-helium mixtures in the interiors of giant planets, \href{ https://journals.aps.org/prb/abstract/10.1103/PhysRevB.75.024206 }{
\textit{Phys.~Rev.~B} \textbf{75}, 024206} (2007)

\bibitem{militzer2} B.~Militzer, W.B.~Hubbard, J.~Vorberger, I.~Tamblyn, and S.A.~Bonev, A Massive Core in Jupiter Predicted
from First-Principles Simulations, \href{https://iopscience.iop.org/article/10.1086/594364/meta}{ \textit{Astrophys.~J.~Lett.}~\textbf{688}, L45} (2008)

 \bibitem{knudson} M.D.~Knudson, M.P.~Desjarlais, R.W.~Lemke, T.R.~Mattsson, M.~French, N.~Nettelmann, and R.~Redmer, Probing the Interiors of the Ice Giants: Shock Compression of Water to $700$ GPa and $3.8$ g/cm$^3$, \href{https://journals.aps.org/prl/abstract/10.1103/PhysRevLett.108.091102}{ \textit{Phys.~Rev.~Lett.}~\textbf{108}, 091102} (2012) 







\bibitem{saumon1} D.~Saumon, W.B.~Hubbard, G.~Chabrier, and H.M.~van~Horn, The role of the
molecular-metallic transition of hydrogen in the evolution of Jupiter, Saturn, and
brown dwarfs, \href{http://adsabs.harvard.edu/full/1992ApJ...391..827S}{ \textit{Astrophys.~J.}~\textbf{391}, 827-831} (1992)

\bibitem{saumon2} W.B.~Hubbard, T.~Guillot, J.I.~Lunine, A.~Burrows, D.~Saumon, M.S.~Marley, and R.~S. Freedman, Liquid
metallic hydrogen and the structure of brown dwarfs and giant planets, \href{https://aip.scitation.org/doi/abs/10.1063/1.872570}{ \textit{Phys.~Plasmas} \textbf{4}, 2011–2015} (1997) 


\bibitem{becker} A.~Becker, W.~Lorenzen, J.J.~Fortney, N.~Nettelmann, M.~Sch\"ottler, and R.~Redmer, Ab initio equations of state for
hydrogen (H-REOS.3) and helium (He-REOS.3) and their
implications for the interior of brown dwarfs, \href{https://iopscience.iop.org/article/10.1088/0067-0049/215/2/21/meta}{ \textit{Astrophys.~J.~Suppl.~Ser.}~\textbf{215}, 21} (2014)









\bibitem{hu1} S.X.~Hu, B.~Militzer, V.N.~Goncharov, and S.~Skupsky, Strong Coupling and Degeneracy Effects in Inertial Confinement Fusion Implosions, \href{https://journals.aps.org/prl/abstract/10.1103/PhysRevLett.104.235003}{\textit{Phys.~Rev.~Lett.}~\textbf{104}, 235003} (2010)


\bibitem{hu2} S.X.~Hu, B.~Militzer, V.N.~Goncharov, and S.~Skupsky, First-principles equation-of-state table of deuterium for inertial confinement fusion applications, \href{https://journals.aps.org/prb/abstract/10.1103/PhysRevB.84.224109}{\textit{Phys.~Rev.~B} \textbf{84}, 224109} (2011)




\bibitem{falk_wdm} K.~Falk, Experimental methods for warm dense matter research, \href{https://www.cambridge.org/core/journals/high-power-laser-science-and-engineering/article/experimental-methods-for-warm-dense-matter-research/7205AE1029BEA0061044F84875F1CEDB}{\textit{High Power Laser Sci. Eng.}~\textbf{6}, e59} (2018)









\bibitem{wdm_book} F.~Graziani, M.P.~Desjarlais, R.~Redmer, and S.B.~Trickey (eds.), Frontiers and Challenges in Warm Dense Matter, Springer International Publishing (2014)





\bibitem{torben_eur} T.~Ott, H.~Thomsen, J.W.~Abraham, T.~Dornheim, and M.~Bonitz, Recent progress in the theory and simulation of strongly correlated plasmas: phase transitions, transport, quantum, and magnetic field effects, \href{https://link.springer.com/article/10.1140/epjd/e2018-80385-7}{\textit{Eur.~Phys.~J.~D} \textbf{72}, 84} (2018)


\bibitem{quantum_theory} G.~Giuliani and G.~Vignale, Quantum Theory of the Electron Liquid, Cambridge University Press (2008)







\bibitem{brown_chapter} E.~Brown, M.A.~Morales, C.~Pierleoni, and D.~Ceperley, Quantum Monte Carlo Techniques and Applications for Warm Dense
Matter, in F. Graziani, M. P. Desjarlais, R. Redmer, and S. B.
Trickey (Eds.), \textit{Frontiers and Challenges in Warm Dense Matter},
Springer International Publishing (2014)












\bibitem{cpimc_original} T.~Schoof, M.~Bonitz, A.~Filinov, D.~Hochstuhl, and J.W.~Dufty, Configuration Path Integral Monte Carlo, \href{https://onlinelibrary.wiley.com/doi/abs/10.1002/ctpp.201100012}{\textit{Contrib.~Plasma Phys.}~\textbf{51}, 687-697} (2011)


\bibitem{brown_ethan} E.W.~Brown, B.K.~Clark, J.L.~DuBois, and D.M.~Ceperley, Path-Integral Monte Carlo Simulation of the Warm Dense Homogeneous Electron Gas, \href{https://journals.aps.org/prl/abstract/10.1103/PhysRevLett.110.146405}{\textit{Phys.~Rev.~Lett.}~\textbf{110}, 146405} (2013)

\bibitem{blunt1} N.S.~Blunt, T.W.~Rogers, J.S.~Spencer, and W.M.C.~Foulkes, Density-matrix quantum Monte Carlo method, \href{https://journals.aps.org/prb/abstract/10.1103/PhysRevB.89.245124}{\textit{Phys.~Rev.~B}~\textbf{89}, 245124} (2014)

\bibitem{schoof_prl} T.~Schoof, S.~Groth, J.~Vorberger, and M.~Bonitz, \textit{Ab Initio} Thermodynamic Results for the Degenerate Electron Gas at Finite Temperature, \href{https://journals.aps.org/prl/abstract/10.1103/PhysRevLett.115.130402}{\textit{Phys.~Rev.~Lett.}~\textbf{115}, 130402} (2015)

\bibitem{malone1} F.D.~Malone, N.S.~Blunt, J.J.~Shepherd, D.K.K.~Lee, J.S.~Spencer, and W.M.C.~Foulkes, Interaction picture density matrix quantum Monte Carlo, \href{https://aip.scitation.org/doi/abs/10.1063/1.4927434}{\textit{J.~Chem.~Phys.}~\textbf{143}, 044116} (2015)

\bibitem{blunt2} N.S.~Blunt, A.~Alavi, and G.H.~Booth, Krylov-Projected Quantum Monte Carlo Method, \href{https://journals.aps.org/prl/abstract/10.1103/PhysRevLett.115.050603}{\textit{Phys.~Rev.~Lett.}~\textbf{115}, 050603} (2015)

\bibitem{malone2} F.D.~Malone, N.S.~Blunt, E.W.~Brown, D.K.K.~Lee, J.S.~Spencer, W.M.C.~Foulkes, and J.J.~Shepherd, Accurate Exchange-Correlation Energies for the Warm Dense Electron Gas, \href{https://journals.aps.org/prl/abstract/10.1103/PhysRevLett.117.115701}{\textit{Phys.~Rev.~Lett.}~\textbf{117}, 115701} (2016)


\bibitem{dornheim} T.~Dornheim, S.~Groth, A.~Filinov and M.~Bonitz, Permutation blocking path integral Monte Carlo: a highly efficient approach to the simulation of strongly degenerate non-ideal fermions, \href{ http://iopscience.iop.org/1367-2630/17/7/073017 }{ \textit{New J. Phys.} \textbf{17}, 073017} (2015)

\bibitem{dornheim2} T.~Dornheim, T.~Schoof, S.~Groth, A.~Filinov, and M.~Bonitz, Permutation Blocking Path Integral Monte Carlo Approach to the Uniform Electron Gas at Finite Temperature, \href{ http://scitation.aip.org/content/aip/journal/jcp/143/20/10.1063/1.4936145 }{ \textit{ J. Chem. Phys.} \textbf{143}, 204101} (2015)


\bibitem{vladimir_UEG} V.S.~Filinov, V.E.~Fortov, M.~Bonitz, and Zh.A.~Moldabekov, Fermionic path-integral Monte Carlo results for the uniform electron gas at finite temperature, \href{https://journals.aps.org/pre/abstract/10.1103/PhysRevE.91.033108}{\textit{Phys.~Rev.~E} \textbf{91}, 033108} (2015)


\bibitem{groth} S.~Groth, T.~Schoof, T.~Dornheim, and M.~Bonitz, \textit{Ab Initio} Quantum Monte Carlo Simulations of the Uniform Electron Gas without Fixed Nodes, \href{ http://link.aps.org/doi/10.1103/PhysRevB.93.085102 }{ \textit{Phys.~Rev.~B} \textbf{93}, 085102} (2016)


\bibitem{dornheim3} T.~Dornheim, S.~Groth, T.~Schoof, C.~Hann, and M.~Bonitz, \textit{Ab initio} quantum Monte Carlo simulations of the Uniform electron gas without fixed nodes: The unpolarized case, \href{ http://link.aps.org/doi/10.1103/PhysRevB.93.205134 }{ \textit{Phys.~Rev.~B} \textbf{93}, 205134 } (2016)


\bibitem{dornheim_prl} T.~Dornheim, S.~Groth, T.~Sjostrom, F.D.~Malone, W.M.C.~Foulkes, and M.~Bonitz, \textit{Ab Initio}   Quantum Monte Carlo Simulation of the Warm Dense Electron Gas in the Thermodynamic Limit, \href{ http://link.aps.org/doi/10.1103/PhysRevLett.117.156403}{\textit{Phys.~Rev.~Lett.}~\textbf{117}, 156403} (2016)



\bibitem{dornheim_pop} T.~Dornheim, S.~Groth, F.D.~Malone, T. Schoof, T.~Sjostrom, W.M.C.~Foulkes, and M.~Bonitz, \textit{Ab Initio}   Quantum Monte Carlo Simulation of the Warm Dense Electron Gas, \href{ http://aip.scitation.org/doi/full/10.1063/1.4977920}{ \textit{Phys.~Plasmas}  {\bf 24}, 056303} (2017)


\bibitem{groth_prl} S.~Groth, T.~Dornheim, T.~Sjostrom, F.D.~Malone, W.M.C.~Foulkes, and M.~Bonitz, \textit{Ab initio} Exchange--Correlation Free Energy of the Uniform Electron Gas at Warm Dense Matter Conditions, \href{https://journals.aps.org/prl/abstract/10.1103/PhysRevLett.119.135001}{\textit{Phys.~Rev.~Lett.}~\textbf{119}, 135001} (2017)


\bibitem{dubois} J.L.~DuBois,  E.W.~Brown,  and  B.J.~Alder,  Overcoming the
Fermion Sign Problem in Homogeneous Systems, in E.~Schwegler,
B.M.~Rubenstein, and S.B.~Libby (Eds.),
Advances in the Computational Sciences-Symposium in Honor of Dr Berni Alder's
90th Birthday, World Scientific, Singapore (2017)


\bibitem{dornheim_pre} T.~Dornheim, S.~Groth, J.~Vorberger, and M.~Bonitz, Permutation Blocking Path Integral Monte Carlo approach to the Static Density Response of the Warm Dense Electron Gas, \href{https://journals.aps.org/pre/abstract/10.1103/PhysRevE.96.023203}{\textit{Phys.~Rev.~E}~\textbf{96}, 023203} (2017)

\bibitem{groth_jcp} S.~Groth, T.~Dornheim, and M.~Bonitz, Configuration Path Integral Monte Carlo approach to the Static Density Response of the Warm Dense Electron Gas, \href{https://aip.scitation.org/doi/abs/10.1063/1.4999907}{\textit{J.~Chem.~Phys.}~\textbf{147}, 164108} (2017)


\bibitem{claes} J.~Claes and B.K.~Clark, Finite-temperature properties of strongly correlated systems via variational Monte Carlo, \href{https://journals.aps.org/prb/abstract/10.1103/PhysRevB.95.205109}{\textit{Phys.~Rev.~B} \textbf{95}, 205109} (2017)




\bibitem{dornheim_cpp} T.~Dornheim, S.~Groth, and M.~Bonitz, Ab initio results for the Static Structure Factor of the Warm Dense Electron Gas, \href{https://onlinelibrary.wiley.com/doi/full/10.1002/ctpp.201700096}{\textit{Contrib.~Plasma Phys.}~\textbf{57}, 468-478} (2017)





\bibitem{brenda} Y.~Liu, M.~Cho, and B.~Rubenstein, \textit{Ab Initio} Finite Temperature Auxiliary Field Quantum Monte Carlo, \href{https://pubs.acs.org/doi/abs/10.1021/acs.jctc.8b00569}{\textit{J.~Chem.~Theory~Comput.}~\textbf{14}, 4722-4732} (2018)


\bibitem{universe} V.~Filinov and A.~Larkin, Quantum Dynamics of Charged Fermions in the Wigner Formulation of Quantum Mechanics, \href{https://www.mdpi.com/2218-1997/4/12/133/htm}{\textit{Universe} \textbf{4}, 133} (2018)

\bibitem{dornheim_neu} T.~Dornheim, S.~Groth, and M.~Bonitz, Permutation Blocking Path Integral Monte Carlo Simulations of Degenerate Electrons at Finite Temperature, \href{https://onlinelibrary.wiley.com/doi/full/10.1002/ctpp.201800157}{\textit{Contrib.~Plasma Phys.}~\textbf{59}, e201800157} (2019)








\bibitem{dornheim_permutation_cycles} T.~Dornheim, S.~Groth, A.~Filinov, and M.~Bonitz, Path Integral Monte Carlo Simulation of Degenerate Electrons: Permutation-Cycle Properties, \href{https://aip.scitation.org/doi/10.1063/1.5093171}{\textit{J.~Chem.~Phys.}~\textbf{151}, 014108} (2019)











\bibitem{trotter} H.~De Raedt and B.~De Raedt, Applications of the generalized Trotter formula, \href{https://journals.aps.org/pra/abstract/10.1103/PhysRevA.28.3575}{\textit{Phys.~Rev.~A} \textbf{28}, 3575} (1983)






\bibitem{fermion_nodes} D.M.~Ceperley, Fermion nodes, \href{https://link.springer.com/article/10.1007/BF01030009}{\textit{J.~Stat.~Phys.}~\textbf{63}, 1237-1267} (1991)



\bibitem{node_note} Fermionic quantum Monte-Carlo simulations based on the fixed-node approximation (e.g., Refs.~\cite{gs2,spink,brown_ethan}) are not afflicted by the FSP and, therefore, do not suffer from the associated exponential increase in computation time. In the ground state, the nodes are variational with respect to the energy, and are even exactly known in some cases~\cite{loos_paper_1,loos_paper_2}. In contrast, relatively little is known about the nodal structure at finite temperature.





\bibitem{gs2} D.M.~Ceperley and B.J.~Alder, Ground State of the Electron Gas by a Stochastic Method, \href{ http://link.aps.org/doi/10.1103/PhysRevLett.45.566}{ \textit{Phys. Rev. Lett.} \textbf{45}, 566} (1980)


\bibitem{spink} G.G.~Spink, R.J.~Needs, and N.D.~Drummond, Quantum Monte Carlo study of the three-dimensional spin-polarized homogeneous electron gas, \href{https://journals.aps.org/prb/abstract/10.1103/PhysRevB.88.085121}{\textit{Phys.~Rev.~B}~\textbf{88}, 085121} (2013)




\bibitem{loos_paper_1} P.-F.~Loos and P.M.W.~Gill, Exact Wave Functions of Two-Electron Quantum Rings, \href{https://journals.aps.org/prl/abstract/10.1103/PhysRevLett.108.083002}{\textit{Phys.~Rev.~Lett.}~\textbf{108}, 083002} (2012)

\bibitem{loos_paper_2} P.-F.~Loss and P.M.W.~Gill, Uniform electron gases. I. Electrons on a ring, \href{https://aip.scitation.org/doi/abs/10.1063/1.4802589}{\textit{J.~Chem.~Phys.}~\textbf{138}, 164124} (2013)










\bibitem{dornheim_analyzing} T.~Dornheim, H.~Thomsen, P.~Ludwig, A.~Filinov, and M.~Bonitz, Analyzing Quantum Correlations Made Simple, \href{https://onlinelibrary.wiley.com/doi/abs/10.1002/ctpp.201500120}{\textit{Contrib.~Plasma Phys.}~\textbf{56}, 371-379} (2016)






\bibitem{reimann} S.M.~Reimann and M.~Manninen, Electronic structure of quantum dots, \href{https://journals.aps.org/rmp/abstract/10.1103/RevModPhys.74.1283}{\textit{Rev.~Mod.~Phys.}~\textbf{74}, 1283} (2002)





















\bibitem{stuhler} J. Stuhler, A. Griesmaier, T. Koch, M. Fattori, T. Pfau, S. Giovanazzi, P. Pedri, and L. Santos, Observation of Dipole-Dipole Interaction in a Degenerate Quantum Gas, \href{https://journals.aps.org/prl/abstract/10.1103/PhysRevLett.95.150406}{\textit{Phys.~Rev.~Lett.}~\textbf{95}, 150406} (2005)

\bibitem{dynamic_alex1} A.~Filinov and M.~Bonitz, Collective and single-particle excitations in two-dimensional dipolar Bose gases, \href{https://journals.aps.org/pra/abstract/10.1103/PhysRevA.86.043628}{\textit{Phys.~Rev.~A} \textbf{86}, 043628} (2012)





\bibitem{jan_willem} J.W.~Abraham and M.~Bonitz, Quantum Breathing Mode of Trapped Particles: From Nanoplasmas to Ultracold Gases, \href{https://onlinelibrary.wiley.com/doi/abs/10.1002/ctpp.201300066}{\textit{Contrib.~Plasma Phys.}~\textbf{54}, 27-99} (2014)

























\bibitem{loos} P.-F.~Loos and P.M.W.~Gill, The uniform electron gas, \href{http://onlinelibrary.wiley.com/doi/10.1002/wcms.1257/abstract}{ \textit{Comput.~Mol.~Sci.}~\textbf{6}, 410-429 } (2016)
















\bibitem{fraser} L.M.~Fraser, W.M.C.~Foulkes, G.~Rajagopal, R.J.~Needs, S.D.~Kenny, and A.J.~Williamson, Finite-size effects and Coulomb interactions in quantum Monte Carlo calculations for homogeneous systems with periodic boundary conditions, \href{https://journals.aps.org/prb/abstract/10.1103/PhysRevB.53.1814}{\textit{Phys.~Rev.~B} \textbf{53}, 1814} (1996)



\bibitem{yakub1} E.~Yakub and C.~Ronchi, An efficient method for computation of
long-ranged Coulomb forces in computer
simulation of ionic fluids, \href{https://aip.scitation.org/doi/abs/10.1063/1.1624364}{\textit{J.~Chem.~Phys.}~\textbf{119}, 11556} (2003)

\bibitem{yakub2} E.~Yakub and C.~Ronchi, A New Method for Computation of Long Ranged
Coulomb Forces in Computer Simulation of
Disordered Systems, \href{https://link.springer.com/article/10.1007/s10909-005-5451-5}{\textit{J.~Low Temp.~Phys.}~\textbf{139}, 633-643} (2005)




\bibitem{mezza} F.~Mezzacapo and M.~Boninsegni, Structure, superfluidity, and quantum melting of hydrogen clusters, \href{https://journals.aps.org/pra/abstract/10.1103/PhysRevA.75.033201}{\textit{Phys.~Rev.~A} \textbf{75}, 033201} (2007)





\bibitem{brualla} L.~Brualla, K.~Sakkos, J.~Boronat, and J.~Casulleras, Higher order and infinite Trotter-number extrapolationsin path integral Monte Carlo, \href{https://aip.scitation.org/doi/abs/10.1063/1.1760512}{\textit{J.~Chem.~Phys.}~\textbf{121}, 636} (2004)

\bibitem{sakkos} K.~Sakkos, J.~Casulleras, and J.~Boronat, High order Chin actions in path integral Monte Carlo, \href{https://aip.scitation.org/doi/abs/10.1063/1.3143522}{\textit{J.~Chem.~Phys.}~\textbf{130}, 204109} (2009)










\bibitem{monte_carlo_book} G.~Fishman, Monte Carlo, \textit{Springer Series in Operations Research and Financial Engineering}, Springer, New York (1996)





\bibitem{error_note} Strictly speaking, Eq.~(\ref{eq:naive_error}) only holds for $M$ independent Monte Carlo samples. In practice, however, adjacent elements in the Markov chain are correlated leading to an increased statistical uncertainty. This must be taken into account via a so-called autocorrelation time, which is explained in detail in, e.g., Ref.~\cite{error_cite}. In contrast, all error bars given in Tabs.~\ref{tab:N_dependence_lambda0p5_Coulomb}, \ref{tab:lambda_dependence_N6_beta1}, and \ref{tab:beta_dependence_N6_lambda0.5} have been obtained by evaluating Eq.~(\ref{eq:naive_error}) for $N_\textnormal{s}$ statistically independent random seeds (instead of $M$ correlated measurements).




\bibitem{error_cite} V.~Ambegaokar and M.~Troyer, Estimating errors reliably in Monte Carlo simulations of the Ehrenfestmodel, \href{https://aapt.scitation.org/doi/abs/10.1119/1.3247985}{\textit{Am.~J.~Phys.}~\textbf{78}, 150} (2010)


\bibitem{hatano} N.~Hatano, Data Analysis for Quantum Monte Carlo Simulations with the Negative-Sign Problem, \href{https://journals.jps.jp/doi/abs/10.1143/JPSJ.63.1691}{\textit{J.~Phys.~Soc.~Jpn.}~\textbf{63}, 1691-1697} (1993)

































\bibitem{mandelbrot} B.B.~Mandelbrot, Fractals and Scaling in Finance, Springer Science+Business Media, New York (1997)


\bibitem{taleb} N.N.~Taleb, The Black Swan, Random House (2009)












\bibitem{krauth_book} W.~Krauth,
Statistical  Mechanics: Algorithms  and  Computations, \textit{Oxford Master Series in Physics}, Oxford University Press, 2006











\bibitem{dynamic_alex2} A.~Filinov, Correlation effects and collective excitations in bosonic bilayers: Role of quantum statistics, superfluidity, and the dimerization transition, \href{https://journals.aps.org/pra/abstract/10.1103/PhysRevA.94.013603}{\textit{Phys.~Rev.~A} \textbf{94}, 013603} (2016)













\bibitem{greiner_book} W.~Greiner, D.~Rischke, L.~Neise, and H.~St\"ocker, Thermodynamics and Statistical Mechanics, Springer, New York (2012)


\bibitem{janke} W.~Janke and T.~Sauer, Optimal energy estimation in path-integral Monte Carlo simulations, \href{https://aip.scitation.org/doi/abs/10.1063/1.474309}{\textit{J.~Chem.~Phys.}~\textbf{107}, 5821} (1997)





\end{thebibliography}
\end{document}